# AI for DevSecOps: A Landscape and Future Opportunities

MICHAEL FU, Monash University, Australia
JIRAT PASUKSMIT, Atlassian, Australia
CHAKKRIT TANTITHAMTHAVORN, Monash University, Australia

DevOps has emerged as one of the most rapidly evolving software development paradigms. With the growing concerns surrounding security in software systems, the DevSecOps paradigm has gained prominence, urging practitioners to incorporate security practices seamlessly into the DevOps workflow. However, integrating security into the DevOps workflow can impact agility and impede delivery speed. Recently, the advancement of artificial intelligence (AI) has revolutionized automation in various software domains, including software security. AI-driven security approaches, particularly those leveraging machine learning or deep learning, hold promise in automating security workflows. They reduce manual efforts, which can be integrated into DevOps to ensure uninterrupted delivery speed and align with the DevSecOps paradigm simultaneously. This paper seeks to contribute to the critical intersection of AI and DevSecOps by presenting a comprehensive landscape of AI-driven security techniques applicable to DevOps and identifying avenues for enhancing security, trust, and efficiency in software development processes. We analyzed 99 research papers spanning from 2017 to 2023. Specifically, we address two key research questions (RQs). In RQ1, we identified 12 security tasks associated with the DevSecOps process and reviewed existing AI-driven security approaches, the problems they addressed, and the 65 benchmarks used to evaluate those approaches. Drawing insights from our findings, in RQ2, we discussed state-of-the-art AI-driven security approaches, highlighted 15 challenges in existing research, and proposed 15 corresponding avenues for future opportunities.

## 1 INTRODUCTION

The traditional software development lifecycle (SDLC) adopts a sequential and siloed approach, with distinct phases executed in a linear fashion, resulting in limited collaboration, slow feedback loops, increased risk of defects, difficulty in managing changes, lack of agility, and increased costs and time-to-market. To address these limitations, DevOps emerged over a decade ago, combining development (Dev) and operations (Ops) to integrate individuals, processes, and technology across application planning, development, delivery, and operations. Moreover, DevOps facilitates coordination and collaboration among previously segregated roles like development, quality engineering,

Authors' addresses: Michael Fu, Monash University, Clayton, Australia, yeh.fu@monash.edu; Jirat Pasuksmit, Atlassian, Sydney, Australia, jpasuksmit@atlassian.com; Chakkrit Tantithamthavorn, Monash University, Clayton, Australia, chakkrit@monash.edu.

and security [148]. DevOps has been a widely adopted SDLC [133], with organizations generally expressing satisfaction and positivity regarding their transition to DevOps practices [57].

While DevOps improves collaboration, automation, and agility in software development and operations, it often overlooks security considerations until later stages of the development process [154]. This delayed focus on security can lead to vulnerabilities and risks being introduced into the software, potentially exposing organizations to cyber threats and compliance issues [103]. To overcome these security limitations and ensure the secure delivery of software products while preserving DevOps agility, the concept of DevSecOps (Development, Security, and Operations) emerged. DevSecOps involves considering application and infrastructure security from the outset and automating certain security gates to prevent slowdowns in the DevOps workflow [265]. However, incorporating security practices into DevOps to achieve DevSecOps poses several challenges. Common security practices, such as security code review [89], often demand significant manual effort, potentially hindering the agility of DevOps. While automated tools like static application security testing (SAST) can assist in identifying vulnerabilities during code review, they still face limitations in practical effectiveness, especially in real-world scenarios [26]. Additionally, recent research advocates for further automation of traditionally manual security practices to better align with rapid software deployment cycles [177].

In recent years, the rapid advancement of artificial intelligence (AI) technologies has significantly impacted software development and cybersecurity. As organizations increasingly adopt DevSecOps practices [73], integrating security into every phase of the software development lifecycle becomes crucial. AI-driven security approaches are gaining traction in research and are seen as vital tools in this integration. According to Google's 2023 State of DevOps Report, DevOps teams recognize AI's potential in enhancing data analysis, automating security tasks, and improving bug identification [71]. This growing emphasis on AI within the DevSecOps framework reflects the industry's commitment to embedding security more deeply into DevOps processes, ensuring that security measures are proactive and continuous throughout the development lifecycle.

We observe that several systematic literature reviews (SLRs) have explored the DevSecOps domain from diverse perspectives. As presented in Table 1, one SLR focused on AI-driven approaches (i.e., machine learning and deep learning) specifically for the operation and monitoring step in DevSecOps [10], while other studies, not primarily focused on AI, covered all steps of DevSecOps [5, 114, 135, 154, 155, 170, 177, 223]. However, none of the previous studies reviewed AI-driven approaches for all steps in DevSecOps comprehensively Despite the growing intersection of AI and DevSecOps, our analysis reveals a notable absence of an SLR specifically examining literature concerning AI-driven methodologies and tools aimed at automating and enhancing the security aspect of DevOps that helps achieve the DevSecOps paradigm. To bridge this gap, our research aims to contribute to this critical intersection by providing a comprehensive landscape of AI techniques applicable to DevSecOps and identifying future opportunities for enhancing security, trust, and efficiency in software development processes.

In this paper, *we define AI approaches as those employing machine learning or deep learning algorithms.* We iteratively defined our search string and searched papers from top-tier software engineering and security conferences and journals, focusing on those published between 2017 and 2023. This timeframe was chosen due to the significant advancements in AI and Large Language Models (LLMs) since the proposal of the transformer architecture in 2017. Our automated literature-searching process yielded a collection of 1,683 papers, from which we manually reviewed and filtered out 1,595 papers based on our selection criteria. Moreover, our snowballing search identified 11 additional papers. We systematically analyzed the collected 99 papers and presented two key aspects: Firstly, we identified existing AI-driven security methods that can be integrated into the DevOps workflow. Secondly, we delved into the challenges and future research opportunities arising from the current landscape of AI-driven security methods. Based on our SLR, this paper presents the following contributions:

- **Landscape of AI-Driven Security Approaches for DevSecOps**: We comprehensively reviewed 99 AI-driven security approaches relevant to DevSecOps, introducing the specific problems they address, detailing the proposed AI methodologies, and presenting the benchmarks used for their evaluation. Notably, we identified and categorized 65 benchmarks, including both synthetic and real-world datasets, which are crucial for assessing the effectiveness of these AI-driven approaches within the DevSecOps paradigm.
- **Identification of 15 Challenges and 15 Future Research Opportunities**: We systematically identified 15 key challenges faced by state-of-the-art AI-driven security approaches, drawing insights from both AI and software engineering perspectives. From an AI standpoint, challenges such as data quality and model explainability are emphasized, while from the software engineering perspective, issues like the complexity of securing Cyber-Physical Systems (CPS), particularly the need to address multiple simultaneous threats across various CPS layers, are highlighted. Correspondingly, we derived 15 future research directions that aim to address these challenges and enhance AI-driven security approaches within the DevSecOps paradigm.

## 2 BACKGROUND AND RELATED WORK

In this section, we define the DevOps process of this study and present an overview of each step in DevOps. We then compare our study with existing reviews on DevSecOps.

### 2.1 DevOps

To identify the five steps in the DevOps workflow used in this paper, the first author initially conducted a review of the relevant literature to explore various definitions of the DevOps process from both industry and academic articles. The first author then discussed the initial findings with the other two authors to incorporate their insights: the second author contributed industry experience, while the third author provided an academic perspective. After synthesizing these inputs, the authors collectively decided to adopt the definition used by Microsoft Microsoft [146], which is widely recognized and implemented in the industry. This choice was made to ensure alignment with widely accepted industry practices and to adopt a well-established framework that effectively captures the essential stages of the DevOps process.

To identify the 12 security tasks within the DevOps workflow, we relied on the collective expertise of the three authors in AI and software security. The first author was responsible for providing an initial overview of potential security tasks based on existing knowledge and relevant software engineering and software security literature. This overview was then discussed and refined through iterative brainstorming sessions and online meetings among the three authors. Through these discussions, we ensured that each task was relevant to specific security aspects of individual DevOps steps. This iterative and collaborative approach was essential for mitigating potential biases. The authors reached a consensus on the 12 identified tasks before proceeding with the literature search and further review, ensuring that the identified tasks were well-founded and relevant to our study.

For consistency, we followed the definition provided by Microsoft [146] throughout the paper and determined DevOps as a five-step workflow (depicted in Figure 1): (1) Plan, (2) Development, (3) Code Commit, (4) Build, Test and Deployment, and (5) Operation and Monitoring. Below, we introduce each step in detail.

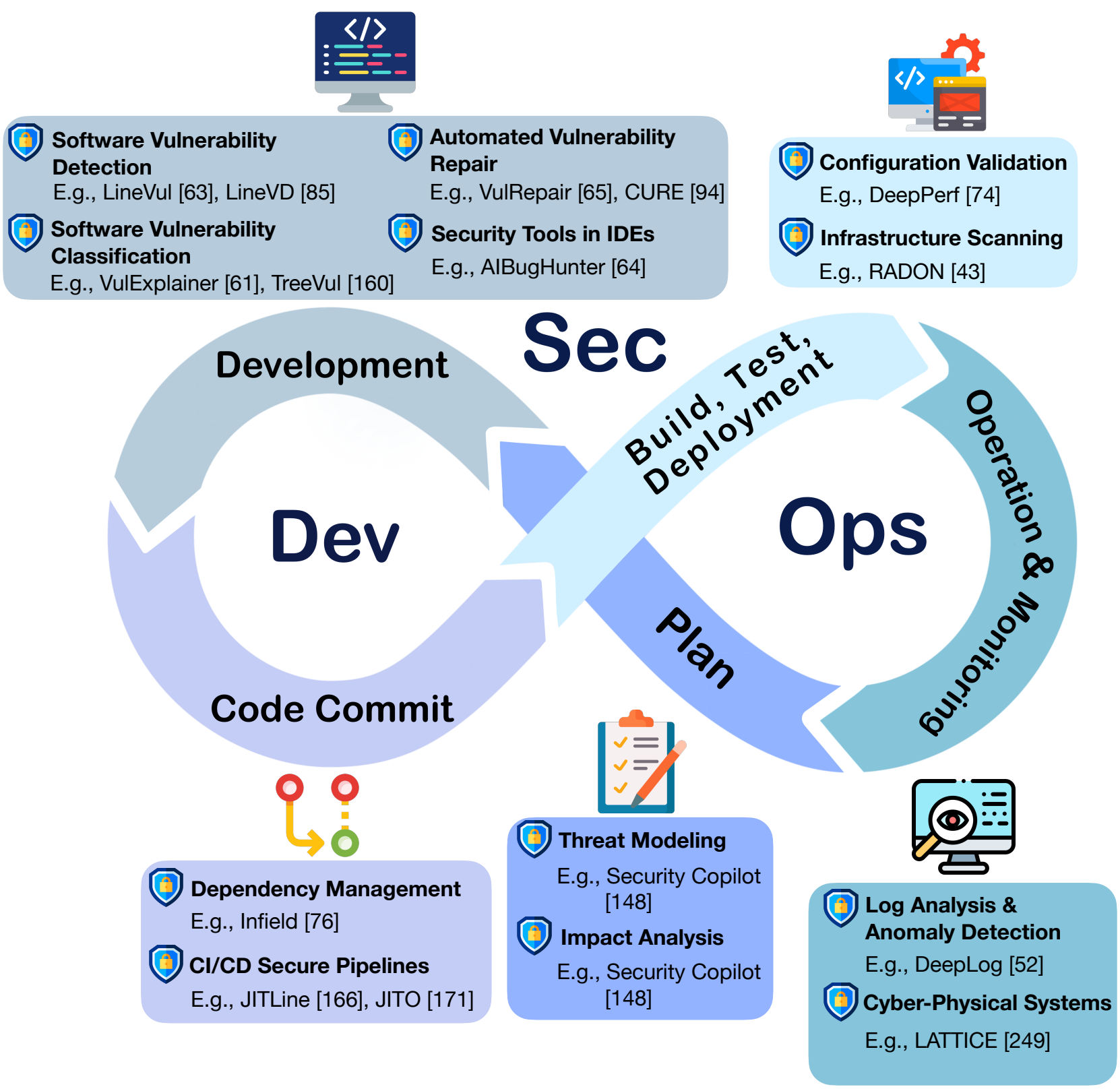

Fig. 1. The overview of the DevOps workflow and the identified security tasks relevant under each step in the DevOps process.

**Plan**. In the "Plan" step of DevOps, teams define project goals, requirements, and timelines. This phase establishes the initial roadmap for the project, involving activities such as gathering user stories, prioritizing features, and assigning tasks. During this phase, the team identifies project objectives and security needs, engaging in threat modeling to grasp security vulnerabilities and plan security measures accordingly [197]. Furthermore, software impact analysis is conducted to identify entities directly or indirectly influenced by a change [9]. This involves the process of assessing and estimating the potential consequences before implementing a modification in the deployed product [105].

**Development**. In the "Development" step of DevOps, software engineers are responsible for implementing the features and functionalities outlined in the requirements and specifications established during the planning phase. In this step, developers operate within an Integrated Development Environment (IDE) where they use static analysis tools (e.g., Checkmarx [28], Flawfinder [239], and Snyk [202]) to scan for potential errors and vulnerabilities before compiling the code.

**Code Commit**. In the "Code Commit" step of DevOps, developers use version control systems like Git to commit their code changes, facilitating collaboration and tracking changes. This step is integral to the Continuous Integration/Continuous Deployment (CI/CD) pipeline, where CI involves the regular integration of developers' work into the main branch of the version control system [205], while CD automates the deployment of software changes to production without human intervention [194]. Furthermore, dependency management plays a critical role in this step, as modern software often relies heavily on third-party code, such as external libraries, to streamline development processes [115]. However, this practice can introduce dependency vulnerabilities [169], underscoring the importance of effectively managing external libraries and dependencies to ensure the reliability and security of the software product. For instance, Dependabot [69] in GitHub helps users monitor their software dependencies and issues security alerts to users if finding vulnerable dependencies.

**Build, Test, and Deployment**. In the "Build and Test" step of DevOps, software code undergoes compilation and rigorous testing to ensure functionality and reliability. Depending on the organization's infrastructure and DevOps practices, these processes may occur either on-premises or in the cloud. In software systems, various configurations are used to control features, endpoints (e.g., cache server addresses), security, fault tolerance, tunable behaviors (e.g., timeouts, throttling limits) and so on [92]. Thus, configuration validation tools are used to ensure the proper configuration of cloud environments. Infrastructure as Code (IaC) simplifies infrastructure management by provisioning consistent environments through machine-readable code, eliminating the need for manual provisioning and management of servers and other components during application development and deployment [178]. Various tools and platforms support IaC, including Terraform, Cloudify, Docker Swarm, Kubernetes, Packer, and Chef, Ansible, Puppet for configuration management [74]. Then, the validated software code is deployed to the production environment to make it available for end-users, involving customization, configuration, and installation [24]. This phase involves automating the deployment process to ensure consistency and reliability [93].

**Operation and Monitoring**. In the "Operation and Monitoring" step of DevOps, the focus shifts to maintaining and monitoring the deployed software to ensure optimal performance and security. This phase involves leveraging actionable intelligence and employing data-driven, event-driven processes to promptly identify, evaluate, and respond to potential risks [146]. Log analysis is commonly used to detect and diagnose abnormal behavior, enhancing system reliability using data from application logs and runtime environments [22]. Additionally, anomaly detection involves identifying uncommon and unexpected occurrences over time [16]. Hagemann and Katsarou [76] categorize methods for detecting anomalies into three groups: machine learning, deep learning, and statistical approaches. Finally, feedback loops are established to gather insights from the aforementioned monitoring activities, enabling continuous improvement and refinement of the DevOps pipeline and security measures.

### 2.2 Other Reviews in DevSecOps

We are aware that there are existing literature reviews of the DevSecOps process, and we have identified 9 related reviews, which are summarized in Table 1. Myrbakken and Colomo-Palacios [154] conducted one of the early literature reviews of DevSecOps, focusing on providing an overview and defining the DevSecOps process, considering DevSecOps as a relatively new concept at the time. Prates et al. [170] reviewed DevSecOps metrics that can be used to monitor the process. For example, defect density and defect burn rate [36] can be employed in the "Development" step to monitor the quality of code being produced and the efficiency of defect resolution over time.

Recent studies have identified several challenges when implementing the DevSecOps paradigm [5, 114]. For instance, Akbar et al. [5] pointed out challenges such as the use of immature automated deployment tools and a lack of software security awareness. Furthermore, Rajapakse et al. [177] organized challenges and solutions in their SLR based on existing literature, presenting future research opportunities in DevSecOps for unresolved problems. Additionally, Valdés-Rodríguez et al. [223] discussed current trends in existing methods for software development involving security.

Table 1. Comparison with other related reviews focusing on DevSecOps. **#P** - total number of reviewed papers, **SLR** - Is the study a systematic literature review?, **A** - Does the study focus on the machine learning and deep learning approaches for DevSecOps?, **B** - Does the study encompass all steps in DevSecOps?

| Reference | Year | Focus | Time | #P | SLR | A | B |
|---|---|---|---|---|---|---|---|
| Myrbakken and Colomo-Palacios [154] | 2017 | This study explores the definition, characteristics, benefits, and evolution of DevSecOps while also pointing out the challenges in its adoption. | 2014-2017 | 52 | | | ✓ |
| Prates et al. [170] | 2019 | This study reviews important metrics for monitoring the DevSecOps process. | 2013-2019 | 13 | | | |
| Mao et al. [135] | 2020 | This study reports the state-of-the-practice of DevSecOps and calls for academia to pay more attention to DevSecOps. | 2013-2019 | 141 | | | ✓ |
| Bahaa et al. [10] | 2021 | This study reviews machine learning approaches on the detection of IoT attacks. | 2016-2020 | 49 | ✓ | ✓ | |
| Leppänen et al. [114] | 2022 | This study reviews security challenges and practices for DevOps software development. | 2015-2019 | 18 | ✓ | | ✓ |
| Akbar et al. [5] | 2022 | This study identifies and prioritizes the challenges associated with implementing the DevSecOps process. | -2022 | 46 | | | ✓ |
| Naidoo and Möller [155] | 2022 | This study presents a socio-technical framework of DevSecOps based on a systematic literature review. | 2016-2020 | 26 | ✓ | | ✓ |
| Rajapakse et al. [177] | 2022 | This study reviews the challenges faced by practitioners in DevSecOps adoption, assesses solutions in the literature, and highlights areas requiring future research. | 2011-2020 | 54 | ✓ | | ✓ |
| Valdés-Rodríguez et al. [223] | 2023 | The study reviews methods or models suitable for integrating security into the agile software development life cycle. | 2018-2023 | 39 | ✓ | | ✓ |
| **This study** | 2024 | Our study introduces a landscape of state-of-the-art AI-driven security methodologies and tools tailored for DevSecOps, while also pinpointing the challenges faced by these AI-driven approaches and deriving potential avenues for future research. | 2017-2023 | 99 | ✓ | ✓ | ✓ |

On the other hand, Bahaa et al. [10] concentrated on machine learning/deep learning approaches for Internet of Things (IoT) attacks, while Naidoo and Möller [155] delved into the socio-technical perspective of DevSecOps.

In contrast to previous reviews, our SLR specifically focuses on the emerging trend of AI-driven security approaches within the DevSecOps landscape. While prior works such as Rajapakse et al. [177] have thoroughly categorized challenges related to People, Practices, Tools, and Infrastructure in DevSecOps, our objective is distinct: to categorize AI-driven security approaches, pinpoint the challenges they present, and uncover future research prospects specifically related to AI-driven methods in DevSecOps. To our knowledge, this paper represents one of the first SLRs that focuses on AI-driven security approaches across every step of the DevSecOps process.

## 3 APPROACH

This systematic literature review (SLR) follows the principles outlined by Keele et al. [101] and Kitchenham et al. [106], a framework widely adopted in the context of DevSecOps-related SLRs [5, 173, 177, 187]. Our methodology encompasses three stages: (1) formulation of the review plan, (2) execution of the review, and (3) comprehensive examination of the review outcomes. In the following sections, we introduce each of these steps in detail.

### 3.1 Research Questions

For a thorough understanding of AI-driven security approaches in DevSecOps, it is crucial to explore the existing AI techniques applicable to each DevSecOps step. Additionally, identifying challenges within these AI-based security techniques is essential for deriving future research directions aimed at further enhancing these techniques. Thus, this systematic literature review aims to address the following research questions:

- **RQ1: What are the existing AI methods and tools employed at each stage of the DevSecOps process, and what specific security challenges do they address?**
- **RQ2: What challenges and future research opportunities exist for AI-driven DevSecOps?**

Table 2. The overview of the selected software engineering and security conferences and journals.

| **Software Engineering Conference** | **Acronym** | **CORE Rank** | **#P** |
|---|---|---|---|
| International Conference on Software Engineering | ICSE | **A*** | 12 |
| Foundations of Software Engineering | FSE | **A*** | 12 |
| Automated Software Engineering Conference | ASE | **A*** | 14 |
| Mining Software Repositories | MSR | **A** | 5 |
| Software Analysis, Evolution and Reengineering | SANER | **A** | 5 |
| International Symposium on Software Testing and Analysis | ISSTA | **A** | 1 |
| International Conference on Software Maintenance and Evolution | ICSME | **A** | 1 |
| Evaluation and Assessment in Software Engineering | EASE | **A** | 0 |
| Empirical Software Engineering and Measurement | ESEM | **A** | 1 |
| **Software Engineering Journal** | **Acronym** | **Impact Factor** | **#P** |
| Transactions on Software Engineering | TSE | **7.4** | 11 |
| Transactions on Software Engineering and Methodology | TOSEM | **4.4** | 4 |
| Information and Software Technology | IST | **3.9** | 5 |
| Empirical Software Engineering | EMSE | **3.8** | 5 |
| Journal of Systems and Software | JSS | **3.5** | 5 |
| **Security Conference** | **Acronym** | **CORE Rank** | **#P** |
| Network and Distributed System Security Symposium | NDSS | **A*** | 2 |
| Symposium on Security and Privacy | SP | **A*** | 1 |
| Computer and Communications Security | CCS | **A*** | 5 |
| USENIX Security Symposium | USENIX | **A*** | 0 |
| **Security Journal** | **Acronym** | **Impact Factor** | **#P** |
| Transactions on Dependable and Secure Computing | TDSC | **7.3** | 8 |
| Transactions on Information Forensics and Security | TIFS | **7.2** | 2 |

### 3.2 Literature Search Strategy

We follow the iterative approach outlined by Kitchenham et al. [106] to develop the search string for this study. For each of the five steps in DevOps, as defined in Section 2.1, we first identify the common security processes associated with that step. Next, we combine these security processes with AI-related terms to form the search string for AI-driven security approaches for that step. In particular, the full list of security tasks at each step of DevOps, which this study concentrates on, is presented in Figure 1. For instance, in the planning step, common security processes include threat modeling and software impact analysis. We extract the keywords "threat modeling" and "software impact analysis" with AI-related terms, resulting in a search string such as *(threat modeling OR software impact analysis) AND (AI-related terms)*. This iterative process is repeated for each step in the DevOps lifecycle. This process will result in a set of DevSecOps activity keywords and a set of AI-related keywords. The complete set of search keywords is as follows:

- *Keywords related to DevSecOps activities: Threat Modeling, Software Impact Analysis, Static Application Security Testing, SAST, Software Vulnerability Detection, Software Vulnerability Prediction, Software Vulnerability Classification, Automated Vulnerability Repair, Automated Program Repair, Dependency Management, Dependency Vulnerability, Package Management, CI/CD Secure Pipeline, Software Defect Prediction, Defect Prediction, SDP, Continuous Integration, Continuous Deployment, Configuration Validation, Infrastructure Scanning, Infrastructure as Code, IaC, Log Analysis, Anomaly Detection, Cyber-Physical Systems*
- *Keywords related to AI: Artificial Intelligence, AI, Deep Learning, DL, Machine Learning, ML, LLM, Large Language Model, Language Model, LM, Natural Language Processing, NLP, Transformer, Supervised Learning, Semi-supervised Learning, Unsupervised Learning*

After an iterative refinement process, our final search string is as follows:

"([Threat Modeling] OR [Software Impact Analysis] OR [Static Application Security Testing] OR [SAST] OR [Software Vulnerability Detection] OR [Software Vulnerability Prediction] OR [Software Vulnerability Classification] OR [Automated Vulnerability Repair] OR [Automated Program Repair] OR [Dependency Management] OR [Dependency Vulnerability] OR [Package Management] OR [CI/CD Secure Pipeline] OR [Software Defect Prediction] OR [Defect Prediction] OR [SDP] OR [Continuous Integration] OR [Continuous Deployment] OR [Configuration Validation] OR [Infrastructure Scanning] OR [Infrastructure as Code] OR [IaC] OR [Log Analysis] OR [Anomaly Detection] OR [Cyber-Physical Systems]) **AND** ([Artificial Intelligence] OR [AI] OR [Deep Learning] OR [DL] OR [Machine Learning] OR [ML] OR [LLM] OR [Large Language Model] OR [Language Model] OR [LM] OR [Natural Language Processing] OR [NLP] OR [Transformer] OR [Supervised Learning] OR [Semi-supervised Learning] OR [Unsupervised Learning])"

We use Harzing's Publish or Perish software for our automated search process [84] with the Google Scholar search engine, aiming to gather high-quality and impactful research papers for our review. To achieve this, we target both top-tier software engineering (SE) conferences/journals and reputable software security venues. The targeted venues are summarized in Table 2. Specifically, we focus on 9 SE conferences, all of which are ranked either CORE A* or CORE A according to International CORE Conference Rankings (ICORE) [39], along with 5 SE journals with impact factors (IFs) ranging from 3.5 to 7.4. Additionally, we include 4 security conferences ranked CORE A* and 2 security journals with IFs of 7.2 and 7.3. We incorporate SE and security journals with IFs spanning from 3.5 to 7.4 to strike a balance between diversity and reputation. While journals with a higher IF enhance visibility and credibility, those with a lower IF broaden our literature search.

It is worth noting that we deliberately chose the top-tier SE and security conferences and journals to ensure the quality and impact of the studies included in our review. Our comprehensive search process led to a collection of 1,683 unique papers across 20 distinguished venues, ultimately identifying 88 studies that met our rigorous criteria and were deemed suitable for inclusion. This substantial number of papers demonstrates the thoroughness of our review and compares favorably with similar SLRs outlined in Table 1. While we acknowledge the potential for overlooking relevant studies by excluding lower-ranking venues, our deliberate emphasis on esteemed SE and security venues enables us to capture emerging trends and valuable insights from reputable sources. Notably, this approach aligns with the strategies adopted by other SLRs in the software engineering field such as [235].

This paper reviews AI-based security approaches that can be integrated into the DevOps process to achieve the DevSecOps paradigm. It is worth noting that the recent advancements of AI such as language models (LM) and large language models (LLM) stem from the transformer architecture published in 2017 by Vaswani et al. [225]. Thus, we focus our search on papers published from 2017 until the end of 2023 to examine the state-of-the-art AI-based security methods.

### 3.3 Literature Selection: Inclusion-Exclusion Criteria and Quality Assessment Criteria

As presented in Table 3, we have formulated three inclusion criteria (IC) and five exclusion criteria (EC) to ensure that the papers selected are qualified and highly relevant to this study. In addition, a well-crafted quality assessment can help prevent biases introduced by low-quality studies [259] and Kitchenham et al. [106] also suggested that such a process should be considered mandatory in any systematic review to avoid research bias. Thus, in Table 3, we further outline five quality assessment criteria (QAC) aimed at assessing the relevance, clarity, validity, and significance of included papers.

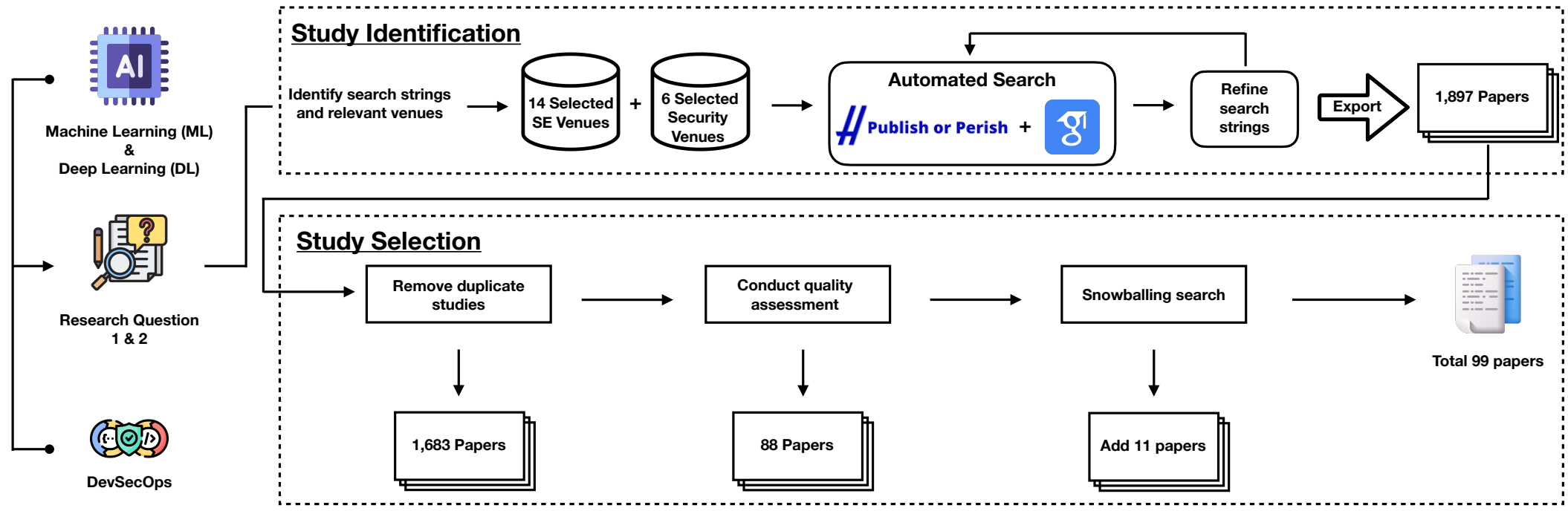

Fig. 2. The overview of our study identification and selection process.

Table 3. The inclusion-exclusion criteria and quality assessment criteria.

| | Inclusion Criteria |
|---|---|
| IC-1 | The paper must be peer-reviewed and published at a journal, conference, or workshop |
| IC-2 | The paper focuses on machine learning/deep learning for security purposes for any of the step included in DevOps |
| IC-3 | The paper with accessible full text |
| | **Exclusion Criteria** |
| EC-1 | The paper is a duplicate or continuation of another study already included in the review |
| EC-2 | The paper written in languages other than English |
| EC-3 | The paper is a literature review |
| EC-4 | The paper is a replication study |
| EC-5 | Studies from sources such as books, theses, technical reports, monographs, keynotes, panels, or venues that do not undergo a full peer-review process |
| | **Quality Assessment Criteria** |
| QAC-1 | The paper presents empirical results or case studies demonstrating the effectiveness of machine learning/deep learning techniques in improving security practices within DevSecOps |
| QAC-2 | The research objective is described |
| QAC-3 | The paper describes techniques or methodologies |
| QAC-4 | The paper describes evaluation or validation methods |
| QAC-5 | The paper presents the study results |

Specifically, we employ a binary scale (Yes/No) to evaluate each IC, EC, and QAC for every paper. Papers failing to meet any criteria are excluded from our study. We initially review the title and abstract of each of the 1,683 papers exported from our automated search process after deduplication. However, in some cases, we need to assess the full text to make a decision. Despite the presence of our search string terms in the title, abstract, or keywords of certain papers, it remains unclear how their content relates to the focus of our SLR. For example, while some papers appear to develop an automated security approach for a specific step in DevOps based on their title and abstract, a full-paper inspection reveals that the proposed approach is not relevant to either machine learning (ML) or deep learning (DL). Thus, to ensure that a paper is adequately aligned with the focus of our review (i.e., ML or DL-based security approaches for DevSecOps), we conducted a comprehensive full-text review following the initial assessment of titles and abstracts. By adhering to these steps, we can effectively filter out papers that do not address ML and DL for DevSecOps. Specifically, we excluded 1,584 irrelevant papers based on our IC, EC, and QAC, resulting in a collection of 88 papers selected for this study.

### 3.4 Snowballing Search

To expand our search for potentially relevant primary studies, we employed a snowballing approach. This method involves not only examining the reference lists and citations of papers but also systematically tracking where papers are referenced and cited. This dual approach, known as backward and forward snowballing, allows us to thoroughly explore relevant literature beyond the initial set of papers.

Before initiating the snowballing procedure, it is essential to curate a collection of initial papers. In this study, the initial paper collection includes 88 papers after the quality assessment. We performed forward and backward snowballing with deduplication and the full study selection process. Consequently, we obtained an additional 11 papers via our snowballing search.

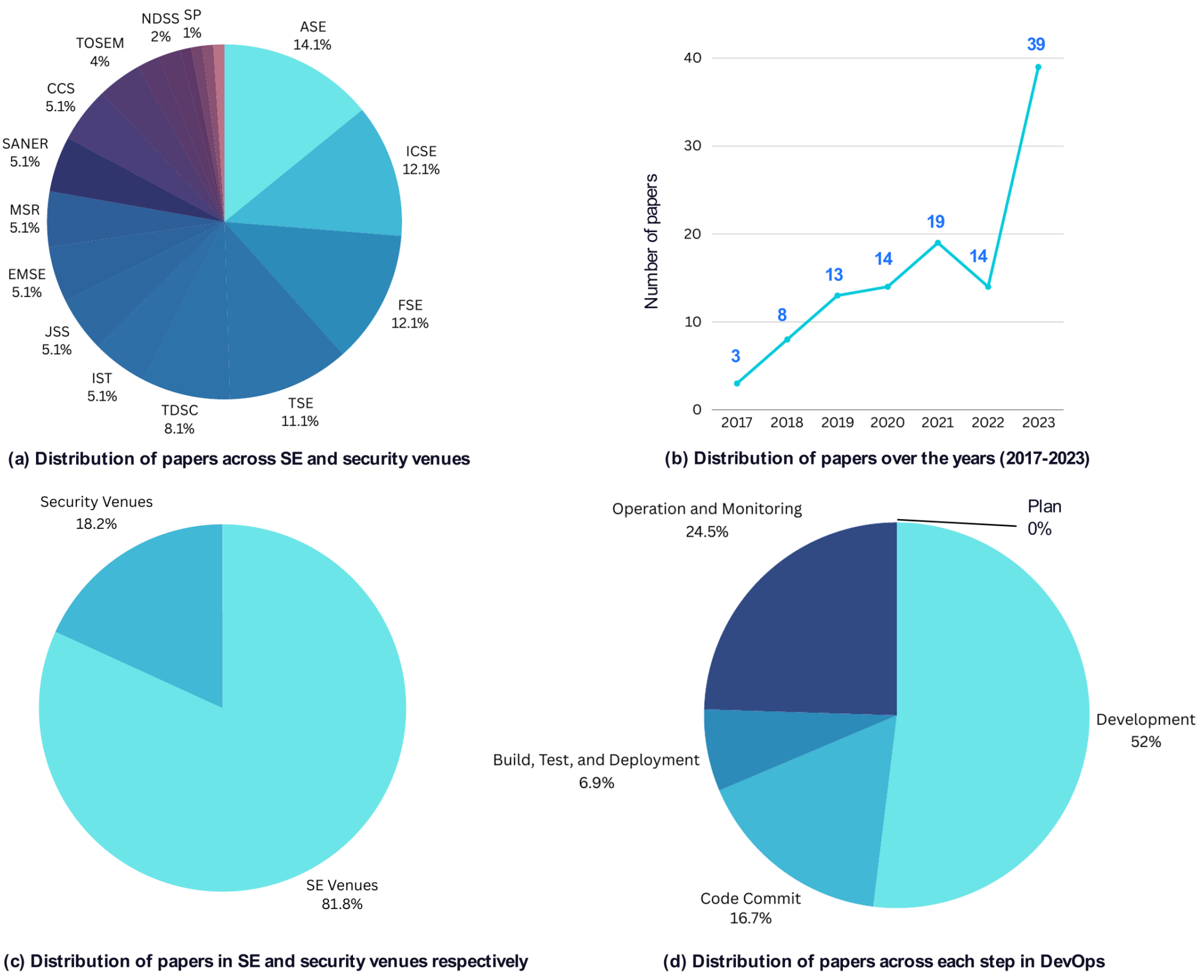

Fig. 3. The overview of the distributions of the selected 99 papers.

### 3.5 Data Extraction and Analysis

We obtained 99 relevant papers after searching, filtering, and snowballing. Figure 3 presents an overview of the distribution of our selected papers. All of the selected papers are peer-reviewed by the venues listed in Table 2. Figure 3 (a) illustrates the distribution of papers across various venues, revealing ASE as the most prevalent venue with a contribution of 14% of the total, followed closely by ICSE and FSE, each accounting for 12%. TDSC follows with 8%, while IST, JSS, EMSE,

MSR, SANER, and CCS each contribute 5% respectively. TOSEM contributes 4%, whereas NDSS and SP have the smallest contributions, at 2% and 1% respectively.

In Figure 3 (b), we present the paper trend over the years. Notably, the number of papers annually shows a steady linear increase from 2017 to 2022. However, there is a significant jump from 14 papers in 2022 to 39 papers in 2023. This sudden surge could be attributed to the recent advancements in generative AI and the escalating concerns surrounding software security. Figure 3 (c) presents the paper distributions of SE venues versus security venues where SE venues account for the majority of 82% of papers while security venues account for the remaining 18%.

Figure 3 (d) illustrates the distribution of papers across each step in DevOps. Notably, we found no relevant papers discussing an AI-driven security approach in the planning step of DevOps. This could be attributed to the nature of activities involved in this stage, such as threat modeling and impact analysis, which may require a higher degree of expertise and human intervention rather than relying on AI algorithms. The majority of studies, accounting for 52%, focus on the Development step in DevOps. In this step, AI-driven security approaches can directly assist software developers by detecting vulnerabilities in their source code, providing explanations, and suggesting repairs. Following this, 24% of the studies concentrate on the Operation and Monitoring step in DevOps. During this phase, AI-driven security approaches can learn from historical data and help monitor and detect anomalies occurring in software systems. Additionally, 17% of studies focus on developing AI-driven approaches for securing the Code Commit step in DevOps, while the remaining 7% concentrate on the Build, Test, and Deployment step in DevOps.

During the full-text review, we conducted data extraction to gather all relevant information necessary for a comprehensive and insightful response to our two research questions outlined in Section 3.1. This extraction phase involved collecting data on various AI-driven approaches proposed for security tasks associated with each step in DevOps, as outlined in Figure 1. With this compiled data, we systematically analyzed the relevant aspects of AI-driven approaches aimed at enhancing the security aspect of DevOps.

### 3.6 Data Synthesis and Mapping

To explore the landscape of AI-driven security approaches in DevSecOps, we conducted a thematic analysis, following the process outlined by Braun and Clarke [19] to synthesize the extracted data. Our analysis encompassed 99 papers relevant to our research questions (RQs), and we systematically categorized key aspects of AI-driven security within the DevOps process presented in Figure 1. Over a span of approximately three months, the first author led the primary open coding and mapping for both research questions (RQs). Throughout this period, the other two authors provided feedback, and all three authors engaged in regular discussions to refine the analysis. This feedback was instrumental in refining the codes and ensuring that they accurately represented the data. Through this collaborative process, the three authors gradually reached a consensus on the content and themes involved in both RQs, with any discrepancies resolved through discussion and iterative refinement.

**Data Synthesis For RQ1.** The data were imported into Google Sheets, where we conducted an open coding process [189]. This involved systematically breaking down the key information extracted from each paper—such as the security tasks addressed, the problems solved, the neural network architectures employed, and the benchmarks used—into smaller, meaningful components. These components were then labeled with *codes*, representing specific categories or themes that emerged from the data [182]. The coding process was iterative, with initial codes refined through multiple rounds of analysis to ensure they accurately captured the essence of the data. Finally, we summarized existing approaches and benchmarks into overview tables for each security task in DevOps to provide a comprehensive comparison of the methodologies and evaluation benchmarks used across the studies.

**Data Synthesis and Mapping For RQ2.** We adopted an identical open coding process as in RQ1 to systematically identify and categorize the challenges encountered by AI-driven security approaches. These challenges were derived directly from the coded data of the reviewed papers, ensuring that our findings were firmly grounded in the existing literature rather than subjective interpretations. By grouping these challenges into distinct thematic categories, we identified overarching themes that reflect the current state of the field. We also thematically mapped the challenges to the proposed potential avenues for future research. In doing the mapping, we first considered how the proposed research directions addressed specific challenges, as evidenced by the existing literature. Next, we evaluated the feasibility and relevance of these proposed directions by examining how they aligned with the identified challenges and the current gaps in the literature. This involved a careful review of existing studies to ensure that each proposed direction was well-grounded in the challenges it aimed to address. Finally, we summarized the challenge themes along with their proposed directions into an overview table, providing a clear and concise reference that highlights the relationship between current challenges and future research opportunities.

Through this multi-layered coding structure in Google Sheets in RQ1 and RQ2, we were able to systematically synthesize the data, leading to meaningful insights into both the present landscape and future opportunities in AI for DevSecOps.

## 4 RQ1: AI METHODS OVERVIEW FOR DEVSECOPS

In this section, we introduce AI-driven security methodologies and tools for DevSecOps. We derived insights from 99 collected literature sources to address our RQ1. Each subsection focuses on AI-driven approaches for specific steps in DevOps. These steps may encompass multiple security-related tasks. In each subsection, we begin with a summary table that encapsulates the results related to our RQ1 for that particular step. Furthermore, given the importance of benchmarks in evaluating the effectiveness of AI-driven approaches, we present tables summarizing these security benchmarks before delving into the specific security tasks. This provides an overview of the security benchmarks previously used and offers readers detailed information such as which benchmarks are most commonly used and whether each benchmark is synthetic or real-world.

### 4.1 Plan

*4.1.1 Threat Modeling.* Threat modeling is an engineering technique employed to identify threats, attacks, vulnerabilities, and countermeasures that may impact an application. It aids in shaping the application's design, meeting the organization's security objectives, and minimizing risk [150]. In the planning step of DevSecOps, threat modeling proactively identifies security concerns and integrates security measures into the development process from the outset. One widely adopted framework for threat modeling is STRIDE proposed by Howard and Lipner [88], which categorizes threats into six main types: Spoofing, Tampering, Repudiation, Information Disclosure, Denial of Service (DoS), and Elevation of Privilege. On the other hand, the DREAD framework proposed by [112] provides a structured approach to assess the severity of identified threats. DREAD stands for Damage potential, Reproducibility, Exploitability, Affected users, and Discoverability. This framework allows teams to assign scores to each criterion, typically on a scale from 0 to 10, to evaluate the potential impact of a threat. By considering factors such as the potential damage caused, the ease of exploitation, and the number of users affected, teams can prioritize their security efforts and focus on addressing the most critical vulnerabilities first. Both STRIDE and DREAD threat modeling is typically conducted by human security experts, often in collaboration with developers, architects, and other stakeholders involved in the software development process. These collaborative processes ensure that security considerations are integrated into DevOps planning throughout the development lifecycle and that potential threats are identified, assessed, and mitigated at an early stage of development.

Table 4. An overview of AI-driven approaches for security-related tasks in the development step.

| DevOps Step | Identified Security Task | AI Method | ML | DL | Reference |
|---|---|---|---|---|---|
| Development | Software Vulnerability Detection | RNN | | ✓ | [46, 122, 123, 185, 234] |
| | | TextCNN | | ✓ | [21] |
| | | GNN | | ✓ | [23, 25, 51, 86, 121, 151, 229, 241, 266, 275, 279] |
| | | Node2Vec | | ✓ | [263] |
| | | LM for code | | ✓ | [49, 64, 268] |
| | | LM + GNN | | ✓ | [49, 207] |
| | Software Vulnerability Classification | ML algorithms | ✓ | | [7, 231] |
| | | RNN | | ✓ | [123, 280] |
| | | TextRCNN | | ✓ | [50] |
| | | Transformer | | ✓ | [228] |
| | | LM | | ✓ | [47, 119] |
| | | LM for code | | ✓ | [62, 65] |
| | | LM for code + RNN | | ✓ | [161] |
| | Automated Vulnerability Repair | ML algorithms | ✓ | | [200] |
| | | CNN | | ✓ | [131] |
| | | RNN | | ✓ | [31, 156] |
| | | Tree-based RNN | | ✓ | [120] |
| | | GNN | | ✓ | [48] |
| | | Transformer | | ✓ | [30, 35, 276, 277] |
| | | LM for code | | ✓ | [14, 61, 66, 82, 95, 97, 137, 157, 164, 269, 278] |
| | Security Tools in IDEs | LM-based security tool | | ✓ | [65] |

*4.1.2 Software Impact Analysis.* Software impact analysis is a crucial process in DevSecOps that analyzes, predicts, and estimates the potential consequences before a change in the deployed product [105]. Impact analysis integrates security considerations into the planning phase of DevOps by analyzing potential unexpected side effects of decisions or changes within a system and identifying potentially affected areas. This process starts by identifying impacted modules and functionality, describing proposed changes, and delineating affected areas. Risk assessment is used to evaluate potential risks associated with each change, such as performance changes, security vulnerabilities, and compatibility issues, often using a qualitative scale or numerical scoring system [219]. In particular, Turver and Munro [221] introduced a technique for the early detection of ripple effects based on a simple graph-theoretic model of documentation and the themes within the documentation. This technique aims to provide a more accessible and effective approach to assessing the impact of changes, particularly during the early stages of a project when source code understanding may be limited. Gethers et al. [67] introduced an adaptive approach for conducting impact analysis from a change request to the source code. The approach begins with a textual change request, such as a bug report. It uses a single snapshot (release) of the source code which is indexed using Latent Semantic Indexing to estimate the impact set. In particular, the analysis encompasses information retrieval, dynamic analysis, and data mining of previous source code commits. They evaluated their approaches using open-source software systems and showed significant improvement in their combined approach over standalone approaches.

Despite the critical role of threat modeling and impact analysis, no relevant literature on AI-driven security approaches was found during our literature search. Thus, we have presented common approaches used in these areas. In RQ2, we will highlight AI-driven approaches found outside of academic literature, such as those developed by Aribot, Validata, and Copilot for Security. Additionally, we will outline future research directions, suggesting how researchers can build upon these industry practices to advance AI-driven methods for threat modeling and impact analysis within DevSecOps.

Table 5. Benchmarks used in evaluating AI-driven software vulnerability detection.

| Benchmark | Year | Granularity | Programming Language | Real-World | Synthesis | Evaluation Studies |
|---|---|---|---|---|---|---|
| Firefox [196] | 2013 | File | C, C++ | ✓ | | [46] |
| Android [190] | 2014 | File | Java | ✓ | | [46] |
| Draper [185] | 2018 | Function | C, C++ | ✓ | ✓ | [185, 241] |
| Vuldeepecker [123] | 2018 | Code Gadget | C, C++ | ✓ | ✓ | [123] |
| Du et al. [54] | 2019 | Function | C, C++ | ✓ | | [54] |
| Devign [275] | 2019 | Function | C, C++ | ✓ | | [121, 241, 263, 275] |
| FUNDED [227] | 2020 | Function | C, Java, Swift, PHP | ✓ | ✓ | [266] |
| Big-Vul [58] | 2020 | Function/Line | C, C++ | ✓ | | [64, 86, 121, 207, 241, 268, 279] |
| Reveal [25] | 2021 | Function | C, C++ | ✓ | | [25, 121, 263] |
| Cao et al. [23] | 2021 | Function | C, C++ | ✓ | | [23] |
| D2A [273] | 2021 | Function | C, C++ | ✓ | | [49, 241, 266] |
| Deepwukong [34] | 2021 | Function | C, C++ | ✓ | ✓ | [21, 34] |
| Vuldeelocator [122] | 2021 | Line | C, C++ | ✓ | ✓ | [122] |
| VulCNN [244] | 2022 | Function | C, C++ | ✓ | ✓ | [21, 244] |
| VUDENC [234] | 2022 | Token | Python | ✓ | | [234] |
| DeepVD [229] | 2023 | Function | C, C++ | ✓ | | [229] |
| VulChecker [151] | 2023 | Instruction | C, C++ | ✓ | | [151] |

## 4.2 Development

We provide an overview of current AI-driven security approaches for the development phase of DevSecOps in Table 4. Below, we delve into each associated security task in detail.

*4.2.1 Software Vulnerability Detection.* Machine learning and deep learning-based vulnerability detection (VD) have been proposed to predict potential vulnerabilities in developers' source code. These detection approaches achieve improved accuracy from traditional static analysis methods without requiring the compilation of developers' code [42]. By leveraging these VD methods during development, developers can proactively identify vulnerabilities, facilitating a "shift-left" in security testing—from the testing/deployment phase to the development phase. This proactive integration aligns with the principles of DevSecOps, embodying the paradigm during the "Development" step. We summarize the benchmarks used to evaluate the AI-driven VD in Table 5. In what follows, we present current AI-driven vulnerability detection methods, explore the problems they address, and discuss how each approach differs or is similar to the others.

**Existing AI methods**. Various AI-driven VD methods have been proposed to predict vulnerabilities on different granularities (i.e., file, function, and line levels) and programming languages (i.e., C, C++, Java, Python, Swift, Php). Below, we introduce the model architecture employed by existing VD approaches, the granularities and programming languages they target, the problems these approaches aim to solve, and how each approach either differs from or complements the others.

Dam et al. [46] focused on the challenges inherent in manually designing code features, such as complexity metrics or frequencies of code tokens, which represent characteristics of potentially vulnerable code. These manually designed features often struggle to sufficiently capture both the semantic and syntactic representations of source code, a crucial capability for building accurate prediction models. To overcome these limitations, Dam et al. [46] leverages a Long Short Term Memory (LSTM) model to sequentially learn both semantic and syntactic features of the code. This model aims to capture intricate patterns in source code files that indicate vulnerabilities. The LSTM-based approach targets file-level VD and has been evaluated using benchmarks from Android applications [190] and the Firefox application [196], highlighting a focus on Java, C, and C++ programming languages.

Most recent work focuses on function-level granularity in VD, as it can save more manual effort compared to file-level analysis, where each file may still contain large code blocks requiring inspection. This finer granularity enhances efficiency and precision in identifying vulnerabilities. For instance, Du et al. [54] proposed the LEOPARD framework which uses program metrics and avoids the need for prior knowledge of known vulnerabilities. This framework seeks to overcome the limitations of machine learning-based methods, which struggle with imbalanced data, and pattern-based methods, which require prior knowledge. LEOPARD was evaluated on a benchmark comprising 11 real-world open-source projects, including BIND, FFmpeg, Linux, OpenSSL, and Wireshark written in C and C++. LEOPARD's metric-based grouping and ranking can serve as a pre-filtering step, potentially enhancing the efficiency and accuracy of DL models. Similar to Dam et al. [46], Russell et al. [185] leveraged an RNN-based representation learning approach that treated a code function as a sequence of tokens to detect vulnerabilities on the function level. To enhance existing labeled datasets, the authors compiled a vast dataset of millions of open-source functions in C and C++, labeling them with findings from three different static analyzers that indicate potential exploits. They then evaluated the proposed approach using a benchmark parsed from real-world projects, including the SATE IV Juliet Test Suite, the Debian Linux distribution, and public Git repositories on GitHub.

Another mainstream approach involves considering the graph structures of source code, such as data flow graphs (DFG), control flow graphs (CFG), or abstract syntax trees (AST). In this context, a code function is treated as a graph with nodes and edges, and graph neural networks (GNNs) are employed to learn the graph representations for making vulnerability predictions.

In particular, Zhou et al. [275] proposed the Devign method to use DFG, CFG, and AST from the code property graph (CPG) [255]. This method leverages GNNs to analyze rich code semantic representations and includes a novel convolutional (Conv) module to improve graph-level classification. Evaluated on the C/C++ dataset from four large-scale open-source projects (i.e., FFmpeg, Wireshark, QEMU, and Linux Kernel), Devign significantly outperforms existing methods, marking one of the first applications of GNNs for VD and highlighting the benefits of graph-based approaches. Furthermore, Chakraborty et al. [25] investigated the significant drop in performance of state-of-the-art DL techniques for automated VD in real-world scenarios, highlighting issues with training data and model design that often lead to high false positives and negatives. To address these challenges, they proposed the Reveal approach, which integrates code property graphs and GNNs similar to Devign [275]. Additionally, Reveal employs techniques such as SMOTE [27] for class imbalance mitigation and triplet loss [134] to enhance the separation between vulnerable and neutral examples. This approach was evaluated on a C/C++ benchmark including Chromium and Debian source code repository. Similarly, Zhang et al. [266] leveraged code property graphs and GNNs to explore cross-domain VD, mitigating the challenge of tedious and time-consuming annotation of large-scale software code, which requires significant effort from domain experts. The proposed method was designed for C/C++ languages and was evaluated on several benchmarks, including Devign [275], Reveal [25], D2A [273], and FUNDED [227].

Cao et al. [23] found that existing GNNs often struggle with incorporating both incoming and outgoing edges into a node simultaneously. Thus, they proposed a bidirectional GNN-based approach that integrates code property graphs to enhance VD. This approach was evaluated on a benchmark including C/C++ open-source projects such as Linux Kernel, FFmpeg, Wireshark and Libav. Notably, Mirsky et al. [151] found that Existing VD approaches often fail to classify the type of vulnerability found, which leaves developers without crucial information on what specific vulnerabilities are present. Thus, they proposed VulChecker, a tool that can precisely locate vulnerabilities in source code as well as classify their type (CWE). To accomplish this, we propose a new program representation named enriched program dependency graph (ePDG), program slicing strategy, and applied gated graph recurrent neural networks (GRNN) to utilize all of the code's semantics and improve the reach between a vulnerability's root cause and manifestation points. This approach was evaluated on an augmented Juliet C/C++ benchmark. While these graph-based representations effectively capture program dependencies and semantics, Wang et al. [229] found that they often include numerous irrelevant entities and dependencies that do not enhance the model's ability to identify vulnerabilities. To mitigate this, Wang et al. [229] proposed, DeepVD, a graph-based neural network VD model that aims to leverage features emphasizing class separation between vulnerability and benign categories. This approach was evaluated on their collected C/C++ benchmark including 303 different projects and 13,130 vulnerable functions.

Yuan et al. [263] found that existing DL-based approaches are limited to single functions, failing to leverage inter-function information. To address this, they proposed the Behavior Graph Model, which extracts abstract behaviors and relationships between functions to enhance DL-based VD. This method was evaluated on two C/C++ benchmarks: Devign [275] and Reveal [25]. In addition, Zhang et al. [268] found DL models struggle with long code snippets due to input length limitations, which impede their ability to fully represent extensive vulnerable code. To mitigate this, they proposed to decompose long code snippets into multiple execution paths from the Control Flow Graph (CFG). By leveraging a pre-trained code model and convolutional neural network with intra- and inter-path attention, the method focuses on learning representations of these shorter paths. This strategy allows the model to better handle long sequences by processing each path individually, thus improving its ability to detect vulnerabilities in lengthy code. This approach was evaluated using three C/C++ benchmarks: Devign [275], Reveal [25], and Big-Vul [58].

Cai et al. [21] based their method on complex network analysis theory to convert the CPG into an image-like matrix and used the TextCNN model to mitigate the scalability issue. The proposed approach was evaluated on two benchmarks (i.e., VulCNN [244] and Deepwukong [34]) derived from the Software Assurance Reference Dataset (SARD) and CVE vulnerability databases. Moreover, Wu et al. [241] employed vulnerability-specific inter-procedural slicing algorithms to capture the semantics of various types of vulnerabilities and used GNNs to learn and understand these vulnerability semantics. They evaluate their approach on common C/C++ vulnerability benchmarks such as Devign [275], Draper [185], Big-Vul [58], and D2A [273]. Finally, Steenhoek et al. [207] combined the recently advanced large language models (LLMs) with data flow graph (DFG) to further improve the prediction accuracy on the Big-Vul [58] benchmark for C/C++ languages.

Nevertheless, these VD methods still operate on the function level, which may still consist of multiple lines of code that need to be manually inspected by developers. To address this issue, prior works have proposed various line-level VD approaches. Notably, the VulDeePecker proposed by Li et al. [123] marks the initial stride towards fine-grained VD. This approach introduces the concept of code gadgets, aiming to encompass a finer-grained code representation beyond program or function levels. VulDeePecker was evaluated on their curated benchmark including 19 popular C/C++ open-source projects. Li et al. [121] proposed IVDetect which leverages GNNs to predict on function level and used GNNExplainer [262] to interpret model predictions to locate fine-grained vulnerabilities for C/C++ languages. IVDetect was evaluated on Devign [275], Reveal [25], and Big-Vul [58] benchmarks. Wartschinski et al. [234] proposed VUDENC which uses Word2Vec embedding in conjunction with an RNN-based model to predict vulnerabilities within a few lines of code. VUDENC was evaluated on their Python benchmark collected from 1,009 vulnerability-fixing commits. In addition, Zou et al. [279] proposed a multi-granularity VD that can predict function and slice-level C/C++ vulnerabilities, which was evaluated on the Big-Vul [58] benchmark.

Recent works have proposed AI-driven methods that pinpoint vulnerable lines in source code, identifying not just vulnerable functions but also the specific code lines associated with vulnerabilities, aiming to reduce the effort required for manual security code review. Li et al. [122] proposed

Table 6. Benchmarks used in evaluating AI-driven vulnerability type classification.

| Benchmark | Year | Granularity | Programming Language | Real-World | Synthesis | Evaluation Studies |
|---|---|---|---|---|---|---|
| $\mu$VulDeePecker [280] | 2019 | Code Gadget | C, C++ | ✓ | ✓ | [280] |
| TreeVul [161] | 2023 | Commit | C, C++, Java, and Python | ✓ | | [161] |

VulDeeLocator which relies on intermediate code to accommodate extra semantic information with the BiRNN model to locate vulnerable lines for C/C++ languages. VulDeeLocator was evaluated on their curated benchmark involving real-world projects and synthetic and academic programs parsed from SARD. However, most fine-grained VD approaches rely on RNN-based models, which have limited capability to learn long-term code dependencies in lengthy source code functions. To address this challenge, Fu and Tantithamthavorn [64] proposed LineVul which uses a transformer architecture for VD, which retains better long-term dependencies than RNN-based models and utilizes intrinsic self-attention scores to identify line-level vulnerabilities. Alternatively, Hin et al. [86] employed CodeBERT embeddings [59] with GNNs, treating line-level VD as a node classification task. Both approaches focus on C/C++ languages and were evaluated on the Big-Vul benchmark. Similarly, Ding et al. [49] combined a pre-trained transformer model with GNNs to locate C/C++ vulnerabilities at the line level. This approach was evaluated on the synthetic Juliet test suite v1.3 and real-world D2A benchmarks [273]. Dong et al. [51] advanced the use of GNNs for line-level vulnerability detection by incorporating sub-graph embeddings. This approach, evaluated on both real-world projects from the National Vulnerability Database (NVD) and synthetic projects from SARD, builds upon previous GNN-based methods that leveraged graph representations to enhance VD.

In summary, recent advancements in fine-grained, line-level VD showcase a diverse array of methodologies, each addressing specific limitations of earlier approaches. Early works like VulDeePecker [123] introduced the concept of code gadgets to provide a finer-grained representation beyond function levels, while IVDetect [121] used GNNs and GNNExplainer for function-level predictions and fine-grained vulnerability localization. VUDENC's [234] multi-granularity approach further refined this by focusing on vulnerabilities within smaller code segments or slices. More recent approaches have shifted towards leveraging transformers and advanced embeddings: LineVul Fu and Tantithamthavorn [64] employed transformers to capture long-term dependencies, while CodeBERT and GNNs were combined by Hin et al. [86] for improved line-level predictions. Similarly, Ding et al. [49] integrated transformers with GNNs to enhance vulnerability localization, and Dong et al. [51] advanced GNN-based methods by incorporating sub-graph embeddings. Collectively, these methods highlight a progressive refinement in VD, moving from broad file and function-level predictions to more precise line-level detections, each contributing unique insights into improving vulnerability identification and reducing manual review efforts.

*4.2.2 Software Vulnerability Classification.* AI-driven methods hold the promise of predicting vulnerability types by analyzing the given vulnerable source code. These predictions explain detected vulnerable source code, furnishing developers with valuable insights. By employing this approach, developers can prioritize addressing critical vulnerability types promptly. In practical terms, integrating these AI-driven automation methods directly into developers' Integrated Development Environments (IDEs) has the potential to furnish real-time vulnerability insights during the development stage. This integration aligns with the DevSecOps concept, reflecting the principle of incorporating security into the "Development" step. We summarize the benchmarks used to evaluate the AI-driven vulnerability type classification in Table 6.

**Existing AI methods**. We observed that a notable number of studies utilizing AI-driven methods concentrated on categorizing vulnerability types and characteristics by analyzing the input from vulnerability descriptions [7, 47, 50, 119, 228]. Nevertheless, these studies fall outside the scope of our review because information regarding vulnerability descriptions might not be accessible in the initial phases of software development.

Instead, our focus centers on AI-driven approaches that use plain source code as input to predict vulnerability types, which can explain vulnerability types by scanning developers' source code. Despite efforts, the data imbalance challenge in vulnerability classification persists. In particular, some vulnerabilities such as buffer-related errors are common while other vulnerabilities rarely occur. While Das et al. [47] incorporated data augmentation [237], the performance of their transformer model showed no significant improvement. Similarly, Wang et al. [231] addressed data imbalance by focusing on the top 10 frequency CWE-IDs, yet this approach limited the model's ability to identify rare vulnerability types.

To mitigate the data imbalance issue, Fu et al. [62] proposed a hierarchical knowledge distillation framework named VulExplainer. VulExplainer involves grouping the imbalanced dataset into subsets based on CWE abstract types, creating more balanced subsets consisting of similar CWE-IDs. Separate TextCNN teacher models are trained for each subset, but they can only predict specific CWE-IDs within their subset. To address this limitation, a comprehensive transformer student model predicting all CWE-IDs is then developed through knowledge distillation. VulExplainer was evaluated using the Big-Vul [58] benchmark consisting of real-world C/C++ projects.

On the other hand, Fu et al. [65] proposed leveraging the pre-trained language model CodeBERT [59] combined with multi-objective optimization (MOO) for CWE-ID classification. Unlike traditional multi-task learning methods, which simply aggregate losses, MOO aims to optimize multiple objectives concurrently, such as CVSS severity score estimation, thereby enhancing the model's overall performance. The proposed MOO-based model was evaluated using the C/C++ Big-Vul benchmark [58], demonstrating improvements over conventional multi-task learning by more effectively handling correlated tasks.

Pan et al. [161] observed that many current methods for identifying security patches in open-source software (OSS) merely confirm the existence of a patch without classifying the specific type of vulnerability it addresses. To address this gap, they introduced TreeVul, a hierarchical and chained architecture designed to leverage the structured nature of CWE-IDs for more nuanced vulnerability classification. TreeVul utilizes CodeBERT for source code representation and multiple LSTM encoders to generate predictions across different levels of the CWE hierarchy. This approach allows TreeVul to offer developers detailed predictions for both high-level and low-level CWE-IDs. Evaluated on a benchmark encompassing C/C++, Java, and Python, TreeVul enhances the ability to pinpoint and address specific vulnerabilities, thus improving the precision of patch management and supporting more targeted security fixes.

Existing deep learning methods for vulnerability detection often only determine whether a program contains vulnerabilities but fail to classify the specific types. Thus, different from the approaches mentioned above, Zou et al. [280] combined vulnerability detection and type classification into an integrated multi-class detection task. Specifically, they introduced $\mu$VulDeePecker which introduces the concept of code attention and considers control-dependence when extracting code gadget [123]. The bidirectional LSTM was trained to detect and classify 40 different vulnerability types. They developed a new benchmark from SARD and NVD, featuring 116 CWE-IDs and 33,409 C/C++ programs, to evaluate $\mu$VulDeePecker. This benchmark demonstrated that $\mu$VulDeePecker significantly improves multiclass classification and detection capabilities by integrating control dependence in addition to data dependence.

Table 7. Benchmarks used in evaluating AI-driven program and vulnerability repairs.

| Benchmark | Year | Programming Language | Real-World | Synthesis | Evaluation Studies |
|---|---|---|---|---|---|
| Defects4J [98] | 2014 | Java | ✓ | | [31, 95, 120, 131, 131, 276, 277] |
| ManyBugs [111] | 2015 | C | ✓ | | [131] |
| BugAID [81] | 2016 | JavaScript | ✓ | | [131] |
| QuixBugs [124] | 2017 | Java, Python | ✓ | | [95, 131] |
| CodeFlaws [215] | 2017 | C | ✓ | | [131] |
| Bugs.jar [186] | 2018 | Java | ✓ | | [120] |
| SequenceR [31] | 2019 | Java | ✓ | | [31] |
| Bugs2Fix [220] | 2019 | Java | ✓ | | [82] |
| ManySStuBs4J [100] | 2020 | Java | ✓ | | [137] |
| Hoppity [48] | 2020 | JavaScript | ✓ | | [48] |
| CodeXGLUE [129] | 2021 | Java | ✓ | ✓ | [278] |
| TFix [14] | 2021 | JavaScript | ✓ | | [14, 278] |
| VRepair [30] | 2022 | C, C++ | ✓ | | [30, 61, 66] |
| Namavar et al. [156] | 2022 | JavaScript | ✓ | | [156] |
| Pearce et al. [164] | 2023 | C, C++ | ✓ | ✓ | [164] |
| Function-SStuBs4J [157] | 2023 | Java | ✓ | | [157] |
| InferFix [97] | 2023 | Java, C# | ✓ | | [97] |

In summary, the reviewed studies on AI-driven vulnerability classification address various aspects of the problem but differ in their approaches and focus areas. Some research [7, 47, 50, 119, 228] concentrates on leveraging vulnerability descriptions for classification, which may not always be available in early development stages. Other works such as [62], tackle the issue of data imbalance by introducing hierarchical knowledge distillation to better handle rare vulnerability types. This method creates balanced subsets of CWE-IDs to enhance classification accuracy. In contrast, Fu et al. [65] advances multi-task learning with multi-objective optimization (MOO) to improve CWE-ID classification by optimizing multiple objectives simultaneously, demonstrating its effectiveness over traditional multi-task learning methods. Meanwhile, Pan et al. [161] introduces TreeVul, which utilizes a hierarchical approach to classify vulnerabilities across different CWE tree levels, offering developers more detailed and actionable insights. Lastly, Zou et al. [280] integrates vulnerability detection and type classification into a unified multi-class task, using code attention and control-dependence to enhance both detection and classification. Collectively, these approaches complement each other by addressing distinct challenges in vulnerability classification, from data imbalance to granularity of vulnerability types, thus offering more practical and accurate automated vulnerability classification approaches for enhancing software security.

*4.2.3 Automated Vulnerability Repair.* The progress in sequence-to-sequence (seq2seq) learning within the realm of deep learning has facilitated significant advancements, particularly in the development of AI-driven automated programs and vulnerability repair approaches. These innovative solutions now offer the capability to automatically recommend fixes for vulnerable or buggy programs, addressing the time-consuming and labor-intensive nature of manual code repair. Program repair models take source code as input, which allows for potential integration with developers' IDEs, providing near real-time code repair suggestions during the development phase. This integration automates the vulnerable code repair process and incorporates security into the "Development" step. We summarize the benchmarks used to evaluate AI-driven vulnerability and program repair approaches in Table 7.

**Existing AI methods**. Several approaches have been suggested for program repair and vulnerability repair based on deep learning (DL), with program repair targeting general software defects and vulnerability repair addressing security-related weaknesses. Due to the inherent similarities between these tasks, transfer learning can be employed to enhance the model's ability to generalize across both as demonstrated by Chen et al. [30]. Consequently, the subsequent discussion encompasses a review of both DL-based program repair and vulnerability repair.

Automated program repair has traditionally relied on static and dynamic analysis techniques like Angelix [141] and CapGen [238], which, despite their progress, are limited to simple, small fixes and depend heavily on intelligent design and domain-specific knowledge. To address these limitations, Chen et al. [31] introduced SequenceR, a language-agnostic, DL-based approach for program repair, focusing specifically on one-line fixes. SequenceR employs an RNN-based sequence-to-sequence (seq2seq) network architecture with a copy mechanism [192] to handle rare or unseen tokens and constructs an abstract buggy context to capture long-range dependencies. It was evaluated using a benchmark curated from CodRep [32] and Bugs2Fix [220], which includes past Java commits from open-source projects and emphasizes the collection of one-line bug fixes.

Recently, considerable research attention has been directed towards employing transformer models for both program repair and vulnerability repair [14, 30, 35, 66, 82, 137, 164, 269, 278]. This shift from RNN-based to transformer-based models is driven by the limitations of RNNs in handling long-range dependencies and their sequential processing nature, which can lead to inefficiencies and difficulties in capturing complex code patterns. With their attention mechanisms, transformers [225] allow for the modeling of long-range dependencies more effectively by considering all tokens simultaneously rather than sequentially like RNNs. This parallel processing capability and the ability to capture contextual information across different parts of the code make transformers particularly well-suited for tasks like program repair and vulnerability detection, where understanding the broader context is crucial. Consequently, transformer-based models have demonstrated superior performance and scalability compared to their RNN-based predecessors.

However, transformers are more complex than RNNs in terms of the number of parameters and the architectural intricacies, requiring a greater amount of training data to achieve optimal performance. This requirement poses a challenge, as vulnerability data is significantly scarcer than general bug data. To address this, Chen et al. [30] explored the effectiveness of domain adaptation using a vanilla transformer model, demonstrating that pre-training on a larger general bug benchmark followed by fine-tuning on a vulnerability repair dataset can enhance performance. Their approach was evaluated on a curated C/C++ benchmark prepared from the CVEFixes [15] and Big-Vul [58] datasets. Similarly, Chi et al. [35] leveraged a vanilla transformer model with data-flow dependencies as input code sequences as input, which was evaluated on a Java benchmark [165].

Building on the concept of pre-training transformers on extensive datasets, another elegant solution to the data scarcity issue in vulnerability repair involves leveraging pre-trained transformers that have been trained on terabytes of code and natural language. For instance, Mashhadi and Hemmati [137] focused on fixing Java programs using the pre-trained CodeBERT model [59], evaluating their approach on a Java benchmark [100]. Similarly, Fu et al. [66] utilized the CodeT5 model [233] to repair C/C++ vulnerabilities, assessing the effectiveness of language models and different tokenizers on the C/C++ benchmark proposed by Chen et al. [30]. These studies illustrate the potential of pre-trained transformers in overcoming the limitations of data scarcity, providing a more robust and scalable solution for automated vulnerability repair and program repair tasks.

Nevertheless, Hao et al. [82] found that current pre-trained transformer code language models (CLMs) for program repairs still have a low success rate. This is primarily because these CLMs are typically developed for general coding purposes, and their potential for program repair applications has yet to be fully explored. To address this, they proposed APRFiT, a general curricular fine-tuning framework designed to improve the success rate of CLMs for program repair. This framework was evaluated on the Bugs2Fix benchmark [220]. On the other hand, Zirak and Hemmati [278] focused on using domain adaptation to enhance existing LM-based approaches such as TFix [14] and CodeXGLUE in cross-project scenarios, evaluating their effectiveness on JavaScript and Java benchmarks [129]. These efforts collectively highlight the need for specialized fine-tuning and domain adaptation strategies to fully realize the potential of pre-trained transformers in program and vulnerability repair tasks.

Fu et al. [61] argued that majority of vulnerable code consists of only a small portion that needs repair, suggesting that decoders should focus solely on the statements requiring fixes during decoding. Inspired by vision transformers, they introduced a vulnerability masking technique designed to guide the vulnerability repair model towards focusing on vulnerable code blocks during the decoding process. This approach was evaluated on the C/C++ vulnerability benchmark proposed by Chen et al. [30]. In a similar vein, Zhu et al. [276] proposed a syntax-guided edit decoder for program repair, which generates edits instead of modified code, offering an efficient representation of small modifications. This method was evaluated on the Defects4J v2.0 benchmark, which consists of Java bugs. These approaches emphasize the importance of targeted repairs and efficient representations for program and vulnerability repairs.

Namavar et al. [156] noted the lack of systematic studies on the effect of code representation on learning performance in program repair. To address this gap, they conducted a controlled experiment to examine the impact of various code representations on model accuracy and usefulness in deep learning-based program repair, using their curated JavaScript benchmark. Despite promising performance, Zhu et al. [277] identified a common limitation among deep learning models: the generation of numerous untypable patches. This issue arises because existing models often do not account for the constraints imposed by typing rules. To mitigate this problem, Zhu et al. [277] introduced Tare, a type-aware model for neural program repair that learns typing rules to generate more accurate patches. This approach was evaluated on the Defects4J [98] and QuixBugs [124] benchmarks, which include Java and Python code, respectively. These studies underscore the importance of code representation and type awareness in enhancing the effectiveness of automated program repair models.

Prior works have introduced context-aware program repair approaches to address the limitations in learning bug-fixing code changes and the context of the surrounding source code. For instance, Li et al. [120] proposed DLFix, which employs a tree-based RNN to learn the context surrounding code fixes. This approach was evaluated on the Defects4J [98] and Bugs.jar [186] Java benchmarks, demonstrating its effectiveness in capturing contextual information. Similarly, Lutellier et al. [131] leveraged CNN layers to extract hierarchical features, allowing for better modeling of source code at different granularity levels, such as statements and functions. This method was evaluated on five common benchmarks: Defects4J (Java) [98], QuixBugs (Java, Python) [124], ManyBugs (C) [111], BugAID (JavaScript) [81], and CodeFlaws (C) [215]. Another notable approach is CURE, proposed by Jiang et al. [95], which uses a code-aware search strategy to find correct fixes by focusing on compilable patches and those close in length to the buggy code, thus addressing issues related to the search space and syntax rules. CURE was evaluated on the Defects4J [98] and QuixBugs [124] Java benchmarks. On the other hand, Dinella et al. [48] represented code as a graph and leveraged GNNs to learn the context and fix bugs, showcasing their approach on a proposed JavaScript benchmark. These context-aware approaches collectively highlight the importance of incorporating surrounding code context to enhance the accuracy and effectiveness of automated program and vulnerability repair models.

Certain studies concentrate on addressing specific categories of software bugs or vulnerabilities and repairing warnings issued by static analysis tools automatically. For instance, Utture and Palsberg [222] focused on fixing resource leak warnings and proposed RLFixer, which was evaluated on warnings generated by popular Java resource-leak detectors. Similarly, Marcilio et al. [136] aimed to automatically generate fix suggestions for common warnings issued by static code analysis tools, providing a practical solution to common software issues identified by these tools. Siddiq et al. [200], on the other hand, focused on fixing SQL injection vulnerabilities, highlighting the importance of addressing specific security threats. In a different approach, Pearce et al. [164] investigated the effectiveness of zero-shot large language models (LLMs) for vulnerability repair and evaluated their approach on a collected C/C++ benchmark consisting of both synthetic and real-world vulnerabilities.

Finally, it is worth noting that some studies have developed end-to-end approaches for both vulnerability detection and repair [97, 145, 157]. Mesecan et al. [145] presented HyperGI which can detect, localize, and repair information leakage. Ni et al. [157] used multi-task learning to construct a comprehensive end-to-end model for both defect prediction and program repair. This approach was evaluated on their collected Java benchmark, Function-SStuBs4J. Jin et al. [97] proposed InferFix which relies on a static analysis tool, Infer, to detect and locate vulnerabilities, then leveraged large language models (LLMs) to generate corresponding repairs. They also curated a dataset of bugs extracted by executing the Infer static analyzer on the change histories of thousands of Java and C# repositories to evaluate their approach.

*4.2.4 Security Tools in IDEs.* Static analysis tools rely on predefined patterns to assist developers in identifying potential vulnerabilities in source code. Similarly, integrating deep learning (DL)-based vulnerability prediction approaches into security analysis tools and deploying them to developers' Integrated Development Environments (IDEs) is becoming increasingly feasible. Previous research has shown that DL models can surpass static analysis tools in terms of accuracy in detecting and locating vulnerabilities [42, 64]. Moreover, they exhibit the capability to identify specific vulnerability types and propose corresponding repairs. The incorporation of AI-driven security tools into IDEs streamlines developers' workflows by automating the identification of security issues in their code. Such automation integrates security considerations into the development stage, aligning with the principles of DevSecOps. Below, we explore existing AI-driven security tools within IDEs, examine the challenges they have encountered, and discuss potential avenues for future research.

**Existing AI methods**. We encountered a limited number of search results while exploring AI-driven software security tools. Notably, although deep learning-based approaches have demonstrated superior effectiveness compared to static analysis tools [42], the predominant tools in the current landscape continue to rely on static analysis and pre-defined patterns. Despite this trend, our investigation uncovered open-source and commercial AI-driven security tools (AIBugHunter [65] and Snyk IDE [202]) designed to assist developers in identifying security issues during development.

Fu et al. [65] introduced AIBugHunter, a deep learning-based security tool designed for C/C++ to assist developers in automating security aspects of their development process. This tool boasts several key capabilities, including the detection and pinpointing of vulnerabilities at the line level, explanation of detected vulnerability types, estimation of vulnerability severity, and suggestions for corresponding repairs. Each of these functionalities is powered by a dedicated language model. AIBugHunter was evaluated on the Big-Vul benchmark [58] and operates as an open-source tool which is accessible as a VSCode extension. According to the results of a user experiment conducted by Fu et al. [65], AIBugHunter can reduce the time spent on detecting, locating, estimating, explaining, and repairing vulnerabilities from 10-15 minutes to a mere 3-4 minutes.

Snyk IDE [202] is a commercial security tool supporting multiple programming languages such as Apex, C++, Go, Java, Kotlin, JavaScript, .NET, PHP, Python, Ruby, Scala, Swift, TypeScript, and VB NET. Through static code scans, it provides developers with crucial information, pinpointing vulnerabilities, offering explanations of their types, and suggesting actionable fixes. Snyk is powered by DeepCode AI, which consists of multiple AI models trained on security-specific data parsed from millions of open-source projects by security researchers. Moreover, Snyk IDE provides flexible plans, including both free and paid options. It seamlessly integrates as a security plugin across various popular IDEs like JetBrains, Visual Studio Code, Eclipse, and Visual Studio.MongoDB, a leading database technology company, has enhanced its security posture by integrating Snyk into its development workflow [109]. Snyk's dashboard and automated security features are embedded within MongoDB's existing GitHub, Slack, and Jira tools, allowing developers to efficiently manage open-source vulnerabilities. This integration streamlines the identification and remediation of security issues, significantly reducing manual checks and accelerating response times. By leveraging Snyk's automated solutions, MongoDB has achieved greater efficiency in securing third-party dependencies, illustrating the effectiveness of incorporating security tools directly into developer environments.

Table 8. An overview of AI-driven approaches for security-related tasks in the code commit step.

| DevOps Step | Identified Security Task | AI Method | ML | DL | Reference |
|---|---|---|---|---|---|
| Code Commit | CI/CD Secure Pipelines | ML algorithms | ✓ | | [20, 29, 117, 162, 167, 210, 211, 218, 256] |
| | | Explainable ML | ✓ | | [166] |
| | | RNN | | ✓ | [168, 184] |
| | | Tree-based RNN | | ✓ | [45] |
| | | Transformer | | ✓ | [171] |
| | | JIT-SDP tool | ✓ | | [102, 172] |
| | | Change analysis tool | ✓ | | [142] |
| | | LM | | ✓ | [108] |
| | | LM for code | | ✓ | [267] |

Last but not least, GitLab Duo [139] exemplifies an AI-driven approach for enhancing security from the development stage through the CI/CD process in an industrial context. GitLab Duo offers several practical features throughout the software development lifecycle, including automating comprehensive merge request descriptions, providing natural language explanations of code for QA testers, performing root cause analysis of pipeline errors, and facilitating vulnerability resolution. By leveraging generative AI, GitLab Duo helps engineers understand and address vulnerabilities within the development window, minimizing context-switching and ensuring that security measures are seamlessly integrated into the CI/CD process. A notable case study is Cube, a software development company based in the Netherlands, which adopted GitLab Duo to increase efficiency and speed in creating secure software [70]. Cube, serving diverse industries such as energy and real estate, uses GitLab Duo's features like Code Suggestions and AI-powered chat to save time and improve software security. By integrating these AI-driven tools, Cube has enhanced its ability to deliver secure software applications, demonstrating how industrial adoption of generative AI can complement academic approaches to achieve secure software development.

## 4.3 Code Commit

We provide an overview of current AI-driven security approaches for the code commit phase of DevSecOps in Table 8. Below, we delve into each associated security task in detail.

*4.3.1 Dependency Management.* The issue of vulnerable dependencies is widely recognized in software ecosystems due to the extensive interconnection of free and open-source software (FOSS) libraries [163]. Thus, dependency management is crucial in DevSecOps as it helps developers handle and organize the external components and libraries that a software project relies on to function correctly. This process involves identifying, tracking, and managing the dependencies between different modules, libraries, or packages within a software application. Effective dependency management ensures secure and resilient software development practices in DevSecOps, mitigating the risks associated with vulnerable dependencies. Dependency management includes package managers like npm (Node.js), pip (Python), Maven (Java), and NuGet (.NET), which automate the installation, configuration, and versioning of dependencies. In addition, there exist commercial tools that provide visibility into dependencies and help identify and mitigate security vulnerabilities. For instance, Sonatype Nexus [204] is a repository manager used for managing software components. This tool allows organizations to store and retrieve artifacts securely and efficiently. Nexus supports various package formats and integrates with popular build tools and CI/CD pipelines, providing dependency management, security scanning, and artifact promotion capabilities. Black Duck [214] is a software composition analysis (SCA) platform designed to manage risks linked to open-source and third-party software components. Black Duck scans codebases, detects open-source components, and assesses their compliance with licenses, security vulnerabilities, and overall quality. It integrates with development tools for proactive dependency management. Snyk [203] also offers a dependency management tool, which is a developer-centric approach to application security, seamlessly integrating into existing DevOps workflows. This tool offers a Dependency Tree View for identifying dependencies and their vulnerabilities, automatically updating them as they evolve. With features like automated scanning within IDEs, comprehensive vulnerability databases, and prioritized remediation, Snyk empowers developers to proactively manage their risk exposure and maintain the integrity of their software projects.

Table 9. Benchmarks used in evaluating AI-driven just-in-time (JIT) software defect prediction.

| Benchmark | Year | Granularity | Programming Language | Real-World | Synthesis | Evaluation Studies |
|---|---|---|---|---|---|---|
| PROMISE [17] | 2007 | Commit | Java | ✓ | | [45] |
| Kamei et al. [99] | 2012 | Commit | C, C++, Java, JavaScript, Perl | ✓ | | [29] |
| Qt & OpenStack [140] | 2018 | Commit/Line | C++, Python | ✓ | | [166, 167] |
| Cabral et al. [20] | 2019 | Commit/File | Java, JavaScript, Python | ✓ | | [20] |
| Yan et al. [256] | 2020 | Commit/File | Java | ✓ | | [171, 256] |
| Wattanakriengkrai et al. [236] | 2020 | Commit | Java | ✓ | | [168] |
| Suh [210] | 2020 | Commit/File | JavaScript, PHP | ✓ | | [210] |

Despite the critical role of effective dependency management, our review of AI-driven security approaches encountered challenges in finding relevant literature. Thus, we have introduced common dependency management practices. In RQ2, we will highlight AI-driven approaches found outside of academic literature followed by future research directions, suggesting how researchers can build upon these industry practices to advance AI-driven methods for dependency management within DevSecOps.

*4.3.2 CI/CD Secure Pipelines.* In the Code Commit step of DevSecOps, ensuring the security of the Continuous Integration/Continuous Deployment (CI/CD) pipeline is crucial. A Secure CI/CD Pipeline involves implementing security measures, starting when code is committed [179]. This may include integrating security checks, such as Just-in-Time (JIT) defect prediction approaches, to identify potential vulnerabilities in code changes as soon as they are submitted. The accuracy of these JIT defect prediction models can be influenced by the methods used to identify defects in the underlying datasets, as different approaches may yield varying predictive performance [261]. Additionally, establishing robust issue-report-to-fix-commit links enhances security by facilitating the swift resolution of reported issues, ensuring that security fixes are promptly applied to the codebase. By leveraging AI-driven techniques in JIT defect prediction and issue-report-to-fix-commit links, organizations can proactively address security concerns at the code commit stage, aligning with the principles of DevSecOps. We summarize the benchmarks used to evaluate the JIT defect prediction in Table 9.

**Existing AI methods**. With the increased interest in continuous deployment, a variant of software defect prediction called Just-in-Time (JIT) Software Defect Prediction (SDP) focuses on predicting whether each incremental software change (e.g., a commit) is defective [272]. JIT-SDP typically analyzes project metrics and code metrics, such as code complexity, lines of code, code churn, and the developer's experience, all collected from historical data of past commits. By leveraging these factors, teams can identify potential issues before they escalate. Integrating JIT-SDP into the CI/CD pipeline enables teams to proactively detect security vulnerabilities in code commits, such as improper input validation or insecure coding practices, allowing for timely mitigation and remediation.

A majority of studies have focused on building ML models for JIT-SDP. For instance, Chen et al. [29] proposed the supervised method MULTI, which aims to reduce manual effort from developers by maximizing the number of identified buggy changes while minimizing efforts in software quality assurance activities. MULTI was evaluated on a benchmark consisting of five programming languages (i.e., C, C++, Java, JavaScript, and Perl) from six open-source projects: Bugzilla, Columba, Eclipse JDT, Eclipse Platform, Mozilla, and PostgreSQL. Similarly, Li et al. [117] leveraged semi-supervised learning with a greedy strategy in unit code inspection effort to rank changes according to their tendency to be defect-prone. This approach was evaluated on the same benchmark as MULTI [29]. While both studies focus on the same task of JIT-SDP and aim to reduce manual effort, they explore different methodologies: MULTI uses multi-task supervised learning, whereas Li et al. [117] employs a semi-supervised approach.

Addressing challenges such as the class imbalance problem has also been a focus. Cabral et al. [20] proposed a class imbalance learning algorithm to improve existing ML models for JIT-SDP and evaluated on their collected benchmark with ten open-source GitHub projects including Java, JavaScript, and Python languages. Alternatively, Tessema and Abebe [218] suggested using change request information to further enhance ML-based JIT-SDP.

In the context of JIT-SDP, addressing challenges like the class imbalance problem is crucial for improving prediction accuracy, as highlighted by Cabral et al. [20] and Tessema and Abebe [218]. However, as ML models become increasingly complex, the need for explainability in these models has grown significantly. Explainable JIT-SDP is essential because it allows developers to understand the reasoning behind predictions, which can build trust in the models and facilitate the debugging and refinement process. Moreover, understanding the model's decision-making process can help in identifying and addressing biases or errors in the predictions, ultimately leading to more reliable software quality assurance. This connection to explainable AI is exemplified by the PyExplainer tool proposed by Pornprasit et al. [166], who explored rule-based explanations [60] to explain the predictions made by black-box ML models. PyExplainer was evaluated on a benchmark of two large-scale open-source software projects (i.e., Openstack and Qt) provided by McIntosh and Kamei [140]. The explainability of ML models not only enhances their usability but also aligns with the overarching goal of reducing manual effort in software quality assurance by making the models' outputs more interpretable and actionable for developers.

Fine-grained JIT-SDP models have emerged to address the limitations of traditional JIT-SDP by providing more precise bug localization, which aims to identify not just buggy commits but also the specific files or lines within a commit that are defective. These models enhance the efficiency of debugging and reduce the manual effort required by developers. For instance, Pascarella et al. [162] proposed a method to locate buggy files, while Pornprasit and Tantithamthavorn [167] and Yan et al. [256] focused on identifying buggy lines within a commit by ranking the prediction scores of ML models. Yan et al. [256] evaluated their bug localization method on a benchmark consisting of Java programs. In contrast, Pornprasit and Tantithamthavorn [167] assessed their JITLine approach using a benchmark that included projects from both the Qt and OpenStack ecosystems, as detailed by McIntosh and Kamei [140].

Furthermore, Qiu et al. [171] and Pornprasit and Tantithamthavorn [168] leveraged Recurrent Neural Networks (RNNs) to capture the semantic information of source code, aiming to locate buggy files within a commit and subsequently localize defective lines by ranking prediction scores. These approaches were evaluated on the Java benchmark prepared by Yan et al. [256] and the benchmark by Wattanakriengkrai et al. [236], respectively. Dam et al. [45] utilized an abstract syntax tree (AST) to represent source code and employed tree-based LSTM networks to effectively capture both the syntax and semantics at multiple levels. This method was evaluated on the PROMISE benchmark [17], which consists of Java projects, demonstrating the potential for more accurate prediction models through a deeper understanding of source code structure and semantics.

Some prior works have developed JIT-SDP tools for integration into CI/CD pipelines. For example, Qiu et al. [172] proposed JITO, integrated into Intellij IDEA, to detect defective code changes and locate exact defect lines. Similarly, Khanan et al. [102] developed JITBot, integrated into CI/CD pipelines, providing defect prediction in GitHub commits along with explanations to clarify reasons and mitigation plans. In addition, Mehta et al. [142] introduced Rex, a change analysis tool that leverages machine learning (ML) models and program analysis. Rex learns change rules that capture dependencies between different regions of code or configuration, based on patterns observed in commit logs spanning several months. When an engineer modifies a subset of files within a change rule, Rex suggests additional changes to ensure consistency and completeness. The tool has been effectively implemented within services such as Office 365 and Azure, impacting over 5,000 changes in the system.

On the other hand, issue-report-to-fix-commit links are critical for security, aiding in understanding code changes and assessing security implications. In particular, establishing robust links between issues and fixes helps in tracking which code changes address specific vulnerabilities or bugs. This traceability ensures that identified issues are addressed systematically and that fixes are correctly applied, which is essential for maintaining security [108]. Manual maintenance of these links is error-prone, potentially leading to vulnerabilities. Automatic link recovery methods have been proposed, but traditional classifiers such as Relink [242] may struggle due to limited positive links and dependency on their number for generating negative links. To address this, Sun et al. [211] formulated the missing link problem as a model learning problem and trained a machine learning (ML) classifier. In contrast, Ruan et al. [184] proposed leveraging recurrent neural networks (RNNs) to learn the semantic representation of natural language descriptions and code in issues and commits, and the semantic correlation between issues and commits. Recently, Lan et al. [108] suggested using language models such as BERT to further enhance performance. However, Zhang et al. [267] argued that the substantial size of language models like CodeBERT poses challenges in terms of computational resources and efficiency. To address this, they proposed a knowledge distillation approach to transfer the knowledge from the large CodeBERT model to a more compact model. This distilled model aims to retain competitive performance while significantly reducing the computational demands associated with training and deploying large-scale language models.

Prior works have proposed multiple AI-driven approaches to automatically identify security-related commits. For instance, Suh [210] used machine learning models such as Support Vector Machine (SVM) to predict whether a commit is likely to be reverted based on features extracted from the revision history of a codebase. This approach was motivated by the need to address the challenges associated with continuous integration and deployment at scale, where quickly identifying time-sensitive bugs is crucial. Evaluated on a benchmark collected from the Wayfair website repositories, which include code written in PHP, JavaScript, and Mustache templates, this method helps streamline the detection of problematic commits and enhances the overall efficiency and security of the development process.

Table 10. An overview of AI-driven approaches for security-related tasks in build, test, and deployment steps.

| DevOps Step | Identified Security Task | AI Method | ML | DL | Reference |
|---|---|---|---|---|---|
| Build, Test, and Deployment | Configuration Validation | ML algorithms | ✓ | | [33, 83, 247] |
| | | FFNN | | ✓ | [75] |
| | | GAN | | ✓ | [11, 199] |
| | Infrastructure Scanning | ML algorithms | ✓ | | [44, 175, 176] |
| | | Word2Vec-CBOW | | ✓ | [18] |

Table 11. Benchmarks used in evaluating AI-driven configuration validation.

| Benchmark | Year | Real-World | Synthesis | Evaluation Studies |
|---|---|---|---|---|
| SPLConqueror [201] | 2015 | ✓ | | [75, 199, 247] |
| ACTGAN [11] | 2019 | ✓ | | [11] |

### 4.4 Build, Test, and Deployment

We provide an overview of current AI-driven security approaches for the build, test, and deployment phase of DevSecOps in Table 10. Below, we delve into each associated security task in detail.

*4.4.1 Configuration Validation.* Configuration validation is a critical aspect of DevSecOps. It ensures that the configurations of software systems, including parameters and settings, are accurate, optimal, and secure. Common approaches to configuration validation include manual inspection, automated testing, and performance estimation models. The importance of configuration validation lies in its direct impact on system reliability, performance, and security. Misconfigurations can lead to vulnerabilities and system failures, hence making robust validation processes essential for safeguarding software systems against potential threats [252]. We summarize the benchmarks used to evaluate the AI-driven configuration validation in Table 11.

**Existing AI methods**. Manual configuration tuning in complex software systems presents a significant security challenge due to the extensive array of parameters for users to configure. However, understanding the intricacies of the software required for effective tuning often exceeds typical user capabilities [11]. This knowledge gap increases the risk of misconfigurations, leaving systems vulnerable to security breaches. To address this challenge, leveraging AI-driven performance estimation can provide valuable insights into the impact of various configuration settings on system performance and security, empowering users to make informed decisions.

In particular, deep learning-based approaches have been proposed. Ha and Zhang [75] proposed DeepPerf, which uses a deep feedforward neural network with sparsity regularization and efficient hyperparameter tuning to predict performance with high accuracy from a small set of configurations. DeepPerf was evaluated on a real-world benchmark [201] including systems such as Dune MGS, HIPAcc, and HSMGP, among others. These systems span various domains, are written in different programming languages, and have diverse configurations, both binary and numeric, which support configuration at either compile time or load time. Shu et al. [199] proposed Perf-AL, which employs generative networks with adversarial learning, comprising a generator and a discriminator. These networks iteratively refine the prediction model through competition until the predicted values converge towards the ground truth distribution. Perf-AL was evaluated on the benchmark presented by Siegmund et al. [201]. Additionally, Cheng et al. [33] proposed to combine multi-objective optimization and a performance prediction model to search for an optimal configuration for Spark deployed in the public cloud.

Table 12. Benchmarks used in evaluating AI-driven infrastructure security scanning.

| Benchmark | Year | Real-World | Synthesis | Evaluation Studies |
|---|---|---|---|---|
| Rahman and Williams [175] | 2018 | ✓ | | [175] |
| Rahman and Williams [176] | 2019 | ✓ | | [176] |
| Dalla Palma et al. [44] | 2021 | ✓ | | [44] |
| Borovits et al. [18] | 2022 | | ✓ | [18] |

Recently, Xia et al. [247] introduced CoMSA, a Modeling-driven Sampling Approach based on XGBoost, which prioritizes configurations with uncertain performance predictions for further training. CoMSA is adaptable to scenarios with or without historical performance testing results because not all software projects maintain such records. CoMSA was evaluated on a similar benchmark as DeepPerf including software systems such as LRZIP, LLVM, x264, and SQLite. Many existing approaches struggle with performance estimation due to the challenge of limited sample sizes, which impacts model accuracy, especially under tight tuning-time constraints. Instead of solely focusing on improving the performance estimation models, Bao et al. [11] proposed ACTGAN which aims to capture hidden structures within good configurations and use this knowledge to generate potentially better configurations. ACTGAN was evaluated on their collected benchmark including eight widely used software systems like Kafka and Spark.

On the other hand, container orchestrators (CO) are crucial for managing container clusters in virtualized infrastructures. Securing CO is challenging due to numerous configurable options and manual configuration is prone to errors and time-consuming. Thus, Haque et al. [83] proposed KGSecConfig, a machine learning-based approach for automating the security configuration of Container Orchestrators (CO), such as Kubernetes, Docker, Azure, and VMWare. It leverages knowledge graphs to systematically capture, link, and correlate heterogeneous and multi-vendor configuration options into a unified structure. By employing keyword and learning models, KGSecConfig enables the automated extraction of secured configuration options and concepts, aiding in the mitigation of misconfigurations in CO environments.

*4.4.2 Infrastructure Scanning.* In DevSecOps, Infrastructure Scanning plays a crucial role in ensuring the security and compliance of software systems. With the rise of Infrastructure as Code (IaC) tools like Ansible, Chef, and Puppet, the process of provisioning and configuring infrastructure has become more automated and scalable. These tools allow developers and operations teams to define infrastructure configurations as machine-readable code, enabling consistent, repeatable deployments across environments. Thus, IaC is a key DevOps practice and a component of continuous delivery [147]. However, during the development of IaC scripts, practitioners might unknowingly introduce security smells (e.g., hard-coded passwords). These recurring coding patterns signal security weaknesses that could lead to security breaches [174]. By integrating AI-driven IaC methods, organizations may proactively identify and mitigate security risks in their IaC scripts. This saves developers' manual security inspection efforts and strengthens the overall security posture of their systems. We summarize the benchmarks used to evaluate the AI-driven IaC scanning in Table 12.

**Existing AI methods**. Similar to other source code artifacts, Infrastructure as Code (IaC) scripts may contain defects that hinder their proper functionality. To automate the defect prediction process and reduce manual inspection, Rahman and Williams [175] leveraged text-mining techniques, such as Bag-of-Words (BoW) and TF-IDF, to extract features from IaC scripts and predict defective ones using machine learning (ML) models. They evaluated their method on their collected benchmark including Puppet IaC scripts. Rahman and Williams [176] further conducted qualitative analysis on defect-related commits extracted from open-source software repositories to identify source code characteristics correlated to defective IaC scripts. They evaluated their IaC defect prediction model on their collected benchmark including open-source repositories maintained by Mirantis, Mozilla, OpenStack, and Wikimedia Commons. They then surveyed practitioners to gauge their agreement with the identified characteristics, using them as features to construct their ML IaC defect prediction model. Similarly, Dalla Palma et al. [44] also suggested an ML-based technique to predict defects in IaC scripts. Their models rely on various metrics, such as lines of code, IaC-specific metrics like the number of configuration tasks, and process metrics such as the number of commits to a file, computed from the collected IaC scripts to predict their proneness to failure. The study was particularly implemented and targeted for their collected Ansible-based benchmark.

Table 13. An overview of AI-driven approaches for security-related tasks in the operation and monitoring step.

| DevOps Step | Identified Security Task | AI Method | ML | DL | Reference |
|---|---|---|---|---|---|
| Operation and Monitoring | Log Analysis and Anomaly Detection | ML algorithms | ✓ | | [38, 80, 104, 188, 243, 257] |
| | | RNN | | ✓ | [52, 53, 118, 143, 209, 230, 243, 271, 274] |
| | | RNN-based AE | | ✓ | [52, 264] |
| | | GNN | | ✓ | [116] |
| | | Transformer | | ✓ | [110] |
| | | Explainable DL | | ✓ | [3, 79] |
| | | Diffusion Model | | ✓ | [113] |
| | Cyber-Physical Systems | ML algorithms | ✓ | | [125, 245] |
| | | RNN + GNN | | | [250] |
| | | GAN | | ✓ | [249] |
| | | VAE | | ✓ | [248] |
| | | Transformer | | ✓ | [90] |
| | | LM + RNN | | ✓ | [251] |

Table 14. Benchmarks used in evaluating AI-driven log analysis and anomaly detection.

| Benchmark | Year | Real-World | Synthesis | Evaluation Studies |
|---|---|---|---|---|
| Yahoo! Webscope [254] | 2006 | ✓ | ✓ | [38] |
| BGL [159] | 2007 | ✓ | | [110, 118, 230, 243, 274] |
| HDFS [253] | 2009 | ✓ | | [52, 53, 116, 118, 143, 230, 243, 264, 271, 274] |
| ADFA-LD [40] | 2013 | | ✓ | [104] |
| SDS [37] | 2015 | | ✓ | [38] |
| UNSW-NB15 [153] | 2015 | | ✓ | [264] |
| OpenStack [53] | 2017 | ✓ | | [53, 274] |
| Microsoft [271] | 2019 | ✓ | | [271] |
| LogHub [85] | 2020 | ✓ | | [80, 118] |
| Studiawan et al. [209] | 2020 | ✓ | | [209] |
| Yang et al. [257] | 2023 | ✓ | | [257] |

On the other hand, Borovits et al. [18] focused on the linguistic anti-patterns in IaC. Linguistic anti-patterns are recurring poor practices concerning inconsistencies in the naming, documentation, and implementation of an entity. They impede the readability, understandability, and maintainability of source code. In particular, they proposed FINDICI, a deep learning (DL)-based approach for anti-pattern detection in IaC. They build and use the abstract syntax tree of IaC code units to create code embeddings used by DL models to detect inconsistent IaC code units. They also evaluated their approach using Ansible-based synthesis benchmark.

## 4.5 Operation and Monitoring

We provide an overview of current AI-driven security approaches for the operation and monitoring phase of DevSecOps in Table 13. Below, we delve into each associated security task in detail.

*4.5.1 Log Analysis and Anomaly Detection.* AI-driven approaches are pivotal in the detection and mitigation of anomalies in system logs. These techniques enable organizations to effectively identify irregularities and address potential issues. Additionally, explainable AI (XAI) contributes significantly by elucidating the underlying causes of anomalies, thus supporting informed decision-making. The scope of AI-driven anomaly detection extends beyond system logs to include cloud services, enhancing operational resilience and aligning with the principles of the DevSecOps paradigm. We summarize the benchmarks used to evaluate the AI-driven log analysis and anomaly detection in Table 14. The following sections present a comprehensive literature review on these topics.

**Existing AI methods**. Some studies have concentrated on machine learning (ML) models such as support vector machines (SVM) due to their lower computation and time requirements compared to deep learning (DL) models. For instance, Khreich et al. [104] integrated frequency and temporal information from system call traces using a one-class SVM, which preserves temporal dependencies among these events. The proposed approach was evaluated on ADFALD (Australian Defence Force Academy Linux Dataset) benchmark with synthesis data based on real-world attacking scenarios.

Moreover, Cid-Fuentes et al. [38] observed that certain anomaly detectors rely on historical failure data and cannot adapt to changes in system behavior at runtime. To address this limitation, they developed a model of system behavior at runtime using SVM, thereby eliminating the need for historical failure data and enabling adaptation to behavior changes. They evaluated their approach on the SDS (Synthetic Distributed System) [37] and Yahoo! Webscope benchmarks [254] consisting of both synthetic and real-world data. Han et al. [80] found that many machine learning algorithms, such as SVM and Logistic Regression, assume clean data and suffer from high training times. Thus, they proposed the Robust Online Evolving Anomaly Detection (ROEAD) framework, which incorporates a Robust Feature Extractor (RFE) to adapt to noise and an Online Evolving Anomaly Detection (OEAD) component for dynamic parameter updates. Their proposal was evaluated on the LogHub [85] benchmark. More recently, Yang et al. [257] used traditional Principal Component Analysis (PCA) and achieved comparable effectiveness to advanced supervised/semi-supervised DL-based techniques while demonstrating better stability under insufficient training data. Their approach was evaluated on their collected benchmark consist of HDFS logs from Hadoop jobs [253], BGL logs from the Blue Gene/L supercomputer [159], Spirit logs from the Spirit supercomputer [206], NC logs from a university network center, and MC logs from a motor corporation's online service system.

Due to recent advancements in deep learning (DL), numerous studies on log-based anomaly detection have introduced various DL-based approaches. For instance, Du et al. [53] proposed DeepLog, one of the pioneering Recurrent Neural Network (RNN) models that treat system logs as natural language sequences to autonomously learn patterns from normal execution log files. DeepLog was evaluated on HDFS [253] and their curated OpenStack log benchmarks. Following this, several other RNN-based detectors emerged, including those by Zhang et al. [271], who leveraged attention-based Bi-LSTM, and Meng et al. [143], who used the same architecture with their proposed template2Vec to extract semantic and syntax information from log templates. The former was evaluated on the HDFS [253] benchmark and log data collected from Microsoft's system, while the latter was evaluated on the HDFS benchmark. Studiawan et al. [209] employed Gated Recurrent Units (GRU) alongside sentiment analysis to identify negative sentiment indicative of anomalous activities in operating system (OS) logs. They evaluated their approach on the collected benchmark from four public datasets of OS and system logs: nps-2009-casper-rw, DFRWS 2009, Honeynet Forensic Challenge 7 2011, and BlueGene/L. Furthermore, Zhou et al. [274] proposed to use sentence embedding in their DeepSyslog method to capture the contextual and semantic relationships between log events, addressing the limitations of traditional methods that rely on incomplete and unstructured log data, which often leads to missed anomalies and false alarms. DeepSyslog was evaluated on HDFS [253], BGL [159] , and OpenStack [53] benchmarks. Wang et al. [230] introduced an online learning paradigm and used LSTM to handle incoming new and unstable log data and evaluated it on HDFS [253] and BGL [159] benchmarks. In addition, Yuan et al. [264] leveraged LSTM-based autoencoder to reconstruct discrete event logs and showed that their approach can detect not only sequences that include unseen or rare events but also structurally abnormal sequences, addressing the fundamental limitation of predicting upcoming events which often results in high false positives due to an inability to fully exploit sequence characteristics. They evaluated their approach on UNSW-NB15 [153] traffic logs and HDFS console logs [253].

Du et al. [52] focused on the challenge of continuously updating anomaly detection systems with new information over time. They introduced a lifelong learning framework called unlearning, which adjusts the model upon labeling false negatives or false positives post-deployment. This framework addressed both the challenge of exploding loss in anomaly detection and catastrophic forgetting in lifelong learning. They evaluated their approach on multiple benchmarks such as HDFS [253] and Yahoo Network Traffic [180]. Furthermore, Li et al. [118] identified challenges in analyzing interleaved logs in modern distributed systems. They proposed SwissLog as a solution to these challenges. The issues include the absence of log dependency mining, variability in log formats, and difficulty in non-intrusive performance issue detection. SwissLog tackles these by constructing ID relation graphs, grouping log messages by IDs, using an online data-driven log parser, and applying an attention-based Bi-LSTM model and heuristic searching algorithm for anomaly detection and localization. SwissLog was evaluated on common log benchmarks such as HDFS [253] and LogHub [85].

Some existing approaches, as discovered by Le and Zhang [110], rely on a log parser to convert log messages into log events, which are then used to create log sequences. However, errors in log parsing can negatively impact the performance of anomaly detectors based on unsupervised or supervised machine learning models. Thus, they introduced NeuralLog, which employs a transformer architecture and eliminates the need for log parsing. They used subword tokenization to address the out-of-vocabulary (OOV) issue. NeuralLog was evaluated on BGL [159] and LogHub [85] benchmarks. Instead of treating system logs as natural language sequences which might reduce anomaly detectors' sensitivity to the log flaws and the concurrency of multiple anomalies, Li et al. [116] transformed log record sequences into log event graphs using event semantic embedding and event adjacency matrix. An attention-based Gated Graph Neural Network (GGNN) model was then used to capture semantic information for anomaly identification. The proposed approach was evaluated on the HDFS [253] benchmark. Finally, Wu et al. [243] presented a comprehensive study investigating the effectiveness of different representations used in machine learning and deep learning models for log-based anomaly detection using common benchmarks such as HDFS [253] and BGL [159].

Some studies have delved into explainable AI (XAI) for anomaly detection in securing software systems. For instance, Han et al. [79] introduced an interpretation methodology tailored for unsupervised deep learning models specifically designed for security systems. Their approach formulates anomaly interpretation as an optimization problem, seeking to identify the most significant differences between anomalies and a normal reference. Furthermore, the interpretations underwent validation through feedback from human security experts. Additionally, Aguilar et al. [3] proposed a decision tree-based autoencoder aimed at anomaly detection, which offers insights into its decisions by exploring correlations among various attribute values.

Given the complexity of managing diverse services in cloud environments, there is a need for automated anomaly detection mechanisms that are easy to set up and operate without requiring extensive knowledge of individual services [188]. For instance, Sauvanaud et al. [188] used machine learning models to aid providers in diagnosing anomalous virtual machines (VMs). Recently, Lee et al. [113] proposed Maat, a framework for anticipating cloud service performance anomalies based on a conditional diffusion model [87]. Maat adopts a two-stage paradigm for anomaly anticipation, consisting of metric forecasting and anomaly detection on forecasts. It employs a conditional denoising diffusion model for multi-step forecasting and extracts anomaly-indicating features based on domain knowledge, followed by the application of isolation forest [127] with incremental learning to detect upcoming anomalies, thus uncovering anomalies that better conform to human expertise.

Table 15. Benchmarks used in evaluating AI-driven cyber-physical system (CPS) security approaches.

| Benchmark | Year | Real-World | Synthesis | Evaluation Studies |
|---|---|---|---|---|
| Gas Pipeline Dataset [152] | 2015 | ✓ | | [250] |
| SWaT [138] | 2016 | ✓ | | [125, 250] |
| WADI [4] | 2017 | ✓ | | [250] |
| BATADAL [216] | 2018 | | ✓ | [250] |
| MSDS [270] | 2023 | ✓ | | [90] |

*4.5.2 Cyber-Physical Systems.* Cyber-physical systems (CPS) integrate sensing, computation, control, and networking into physical objects and infrastructure, establishing connections among them and with the Internet to facilitate seamless interaction and automation [158]. Given the critical need for high security in CPS to ensure safe operation, anomaly detection, relying on data analysis and learning, emerges as a key security technology. The principles of DevSecOps, prioritizing security at every stage of software development, intersect with CPS, particularly when software interacts with physical systems or infrastructures. In such scenarios, the integration of AI-driven approaches for anomaly detection and security enhancement in CPS aligns with DevSecOps goals, aiming to seamlessly integrate security into the development and deployment processes. We summarize the benchmarks used to evaluate the AI-driven CPS security approaches in Table 15.

**Existing AI methods**. Prior works have proposed leveraging Digital Twins, which are digital replicas of physical entities [56], to train ML and DL models for anomaly detection. Digital twins are particularly useful for anomaly detection in cyber-physical systems (CPS) due to their ability to create virtual replicas of physical systems. For example, LATTICE [250] is a digital twin-based anomaly detection method employing deep curriculum learning. It assigns difficulty scores to each sample and uses a training scheduler to sample batches of training data based on these scores, facilitating learning from easy to difficult data. This approach has shown to be more effective than their previous DL-based proposal, ATTAIN [249]. LATTICE was evaluated on multiple CPS benchmarks including Secure Water Treatment (SWaT) [138], Water Distribution (WADI) [4], Battle of Attack Detection Algorithms (BATADAL) [216], Gas Pipeline Dataset [152]. Additionally, Xu et al. [251] identified challenges related to data complexity and insufficiency within CPS, particularly in train control and management systems. Consequently, they proposed employing a language model (LM) with an LSTM architecture to understand complex data, supplemented by a knowledge distillation technique to learn from out-of-domain datasets, addressing the issue of data insufficiency. They evaluated their approach using two TCMS (train control management system) benchmarks.

On the other hand, Xi et al. [245] found that existing anomaly detection methods in CPS, such as AutoEncoder (AE) [12] or Generative Adversarial Network (GAN) [191], often overlook implicit correlations between data points, like the relationship between vehicle speed and obstacle position in the Intelligent Cruise Control System (ICCS), resulting in suboptimal performance. Hence, they proposed an adaptive unsupervised learning method incorporating a Gaussian Mixture Model (GMM), dynamically constructing and updating data correlations via KNN and dynamic graph techniques. They evaluated their approach using benchmarks related to smart grids and smart home systems. Lin et al. [125] focused on Industrial Control Systems (ICS) such as water and power and leveraged the Bayesian network to discover dependencies between sensors and actuators and recognize irregular dependencies.Lin et al. [125] focused on Industrial Control Systems (ICS) like water and power systems, leveraging Bayesian networks to uncover dependencies between sensors and actuators and identify irregularities. They evaluated their approach on the SWaT [138] benchmark, demonstrating its effectiveness in recognizing abnormal dependencies.

Table 16. (RQ2) The overview of the 15 challenges and future research opportunities derived from previous studies.

| DevOps Step | Identified Security Task | Themes of Challenges | Research Opportunity |
|---|---|---|---|
| Plan | Threat Modeling | - | - |
| | Software Impact Analysis | - | - |
| Development | Software Vulnerability Detection | C1-1 - Data Imbalance<br>C4 - Cross Project<br>C5 - MBU Vulnerabilities<br>C6 - Data Quality | R1-1 - Data augmentation and logit adjustment<br>R4 - Evaluate cross-project SVD with diverse CWE-IDs<br>R5 - Evaluate SVD on MBU vulnerabilities<br>R6 - Address data inaccuracy from automatic data collection. |
| | Software Vulnerability Classification | C1-2 - Data Imbalance<br>C7 - Incompleted CWE Tree | R1-2 - Meta-learning and LLMs<br>R7 - Develop advanced tree-based SVC |
| | Automated Vulnerability Repair | C2-1 - Model Explainability<br>C8 - Sequence Length and Computing Resource<br>C9 - Loss of Pre-Trained Knowledge<br>C10 - Automated Repair on Real-World Scenarios | R2-1 - Evidence-based explainable AI (XAI)<br>R8 - Explore transformer variants that can process longer sequences<br>R9 - Explore different training paradigms during fine-tuning<br>R10 - Address limitations of LLMs |
| | Security Tools in IDEs | C3-1 - Lack of AI Security Tooling in IDEs | R3-1 - AI tool deployment and comprehensive tool evaluation |
| Code Commit | CI/CD Secure Pipelines | C2-2 - Model Explainability<br>C3-2 - Lack of AI Security Tooling in CI/CD<br>C11 - The Use of RNNs | R2-2 - Explainable AI (XAI) for DL Models<br>R3-2 - AI tool deployment in CI/CD pipelines<br>R11 - Explore LMs and LLMs |
| Build, Test, and Deployment | Configuration Validation | C12 - Complex Feature Space | R12 - Explore transformers for tabular data |
| | Infrastructure Scanning | C3-3 - Lack of AI Security Tooling for Infrastructure Scanning<br>C13 - Manual Feature Engineering | R3-3 - AI tool deployment and post-deployment evaluation<br>R13 - Explore DL-based techniques |
| Operation and Monitoring | Log Analysis and Anomaly Detection | C2-3 - Model Explainability<br>C14 - Normality Drift for Zero-Positive Anomaly Detection | R2-3 - Explainable AI (XAI) for ML Models<br>R14 - Enhance normality drift detection |
| | Cyber-Physical Systems | C15 - Monitoring Multiple Cyber-Attacks Simultaneously | R15 - Distributed anomaly detection and multi-agent systems |

On the other hand, microservice anomaly detection is vital for system reliability in a microservice architecture. Xie et al. [248] focused on anomaly detection in traces within a microservice architecture. Traces record inter-microservice invocations and are essential for diagnosing system failures. They suggested a group-wise trace anomaly detection algorithm, which categorizes traces based on shared sub-structures and employs a group-wise variational autoencoder (VAE) to obtain structural representations, effectively reducing system detection overhead and outperforming existing methods that analyze each trace individually without considering the structural relationships between them. They evaluated their proposed approach on a real-world benchmark from eBay's microservices system. Huang et al. [90] claimed that the main challenge arises from integrating multiple data modalities (e.g., metrics, logs, and traces) effectively. To address this, they proposed extracting and normalizing features from metrics, logs, and traces, integrating them using a graph representation called MST (Microservice System Twin) graph. A transformer architecture with spatial and temporal attention mechanisms is then employed to model inter-correlations and temporal dependencies, enabling accurate anomaly detection. They evaluated their approach on the MSDS benchmark [270], which includes distributed traces, application logs, and metrics from an OpenStack-based AI analytics system.

## 5 RQ2: CHALLENGES AND RESEARCH OPPORTUNITIES IN AI-DRIVEN DEVSECOPS

In the previous section, we reviewed existing AI-driven security methodologies and tools for DevSecOps to address our RQ1. Now, we will introduce themes of challenges encountered by prior studies and derive future research opportunities to answer our RQ2. To begin, we found that some of these challenges are shared across multiple security tasks related to the DevOps process. Thus, we will first illustrate these common challenges along with their associated research opportunities in the following section. After that, we will proceed to introduce challenges and opportunities specific to each security task in the DevOps process. For clarity, we will use "**C**" followed by a number to represent a challenge and "**R**" followed by a number to represent the corresponding research opportunity. Our answers to RQ2 are summarized in Table 16.

### 5.1 Common Challenges

We identified three common challenges: (C1) Data Imbalance, (C2) Model Explainability, and (C3) Lack of AI Security Tooling. Based on our investigation of previous literature, we found that these challenges are shared by multiple security tasks related to the DevOps process. In what follows, we introduce these common challenges and discuss the associated research opportunities.

**C1-1 - Data Imbalance in Software Vulnerability Detection (In DevOps Development).** Chakraborty et al. [25] observed that the performance of deep learning (DL)-based VD approaches could drop 73% of the F1-score due to the data imbalance issue. Thus, Yang et al. [260] further investigated the impact of data sampling on the effectiveness of existing state-of-the-art (SOTA) DL-based VD approaches. Their discovery revealed that, in DL-based VD, employing over-sampling proves more beneficial than under-sampling. Despite this observation, their experimental findings indicate a persistent challenge: a notable proportion of cases (ranging from 33% to 58%) where decisions were not determined by the presence of vulnerable statements. Consequently, the issue of data imbalance persists, with models continuing to prioritize non-vulnerable code statements when arriving at a vulnerable decision.

**R1-1 - Data Augmentation and Logit Adjustment.** Regarding the research opportunities of data imbalance, Yang et al. [260] suggested that future research explore data augmentation, emphasizing its potential value. Their results indicated that employing a straightforward repetition strategy could enhance the performance of models. Furthermore, the issue of data imbalance is also recognized in the computer vision domain, and proven methods like logit adjustment [144] have demonstrated effectiveness in addressing imbalances in image classification. The application of such methods holds the potential to improve the performance of vulnerability detection.

**C1-2 - Data Imbalance in Software Vulnerability Classification (In DevOps Development).** We found that current vulnerability classification methods still suffer from the data imbalance challenge where models have limited performance on the vulnerability types that infrequently occur. For instance, VulExplainer [62] can correctly identify 67%-69% for common CWE-IDs while the performance drops to 49%-56% for rare CWE-IDs. Moreover, the MOO-based vulnerability classification [65] could not correctly identify some of the infrequent vulnerabilities such as CWE-94 (Improper Control of Generation of Code).

**R1-2 - Meta-Learning and LLMs.** Fu et al. [62] showed that their VulExplainer approach outperformed the commonly used data imbalance techniques such as focal loss [126] and logit adjustment [144]. However, the performance on infrequent vulnerability types still has plenty of room to improve. Thus, future research should explore other techniques to address the data imbalance issue. For instance, meta-learning involves training models to learn a higher-level strategy or set of parameters that enable them to quickly adapt to new, unseen tasks with limited data. It might be suitable for imbalanced data because the exposure to diverse tasks during meta-training helps the model generalize effectively, allowing it to adapt to tasks with imbalanced class distributions. The transfer of knowledge across tasks and the few-shot learning nature of meta-learning contribute to improved performance in scenarios where certain classes have limited samples. Such an approach could also be integrated with few-shot learning of large language models (LLMs).

**C2-1 - Model Explainability in Automated Vulnerability Repair (In DevOps Development).** The advancement of language models has dramatically improved the accuracy of programs and vulnerability repair due to the substantial model size and training data. While recent studies focus on performance improvement, the predictions offered by those models are not explainable, posing challenges in establishing trust between the models and users. As highlighted by Winter et al. [240], trust in program repair is a crucial problem, in their empirical study involving Bloomberg developers, the sentiment conveyed was that an automated repair tool should demonstrate its reliability and foster trust with developers.

**R2-1 - Evidence-based XAI.** Given the complex structure of Large Language Models (LLMs), characterized by multiple hidden layers and a large number of parameters, developing intrinsically explainable AI to explain their repair predictions poses a significant challenge. Nonetheless, an avenue worth exploring involves leveraging the model's self-attention mechanism to ascertain if it can offer meaningful explanations. In addition, future studies could delve into evidence-based explainable AI (XAI), wherein the repair model not only presents its generated fix to end users but also showcases a similar repair case from its training data. This approach aims to establish trust with users, drawing inspiration from its successful application in explaining the story point estimation model in agile software development [63].

**C2-2 - Model Explainability in Software Defect Prediction (In DevOps Code Commit).** Our investigation revealed that the majority of just-in-time (JIT) software defect prediction (SDP) methods based on deep learning (DL) primarily emphasize enhancing performance and granularity [162, 167, 168, 256]. While more accurate and finer-grained SDP methods are beneficial for constructing robust and cost-effective DL-based SDP solutions, there is a noticeable lack of attention to the explainability of these DL-based SDPs. This lack of focus on explainability presents a challenge to the trustworthiness of DL-based SDPs.

**R2-2 - XAI for DL Models.** Given the complexity of DL models, explaining them is more challenging compared to machine learning (ML) models. Nonetheless, future studies could explore the use of extrinsic explanations such as Layer Integrated Gradient (LIG) [213], DeepLift [6, 198], DeepLiftSHAP [130], and GradientSHAP [130], which rely on gradients or their approximations to assess feature importance. Additionally, intrinsic explanations, such as the self-attention outputs from transformer architectures, could also highlight significant features contributing to model predictions. Finally, exploring case-based reasoning [1], which uses similar predictions retrieved from the training data as supporting evidence to explain model predictions, is another promising avenue to consider.

**C2-3 - Model Explainability in Log Analysis and Anomaly Detection (In DevOps Operation and Monitoring).** Our investigation reveals that while the majority of AI-driven anomaly detection approaches focused on performance improvement, only a few studies [3, 79] prioritized model explainability. Explainable AI (XAI) is crucial for anomaly detection models due to several reasons according to Han et al. [79]. Firstly, without detailed explanations for system decisions, security operators struggle to establish trust, leading to unreliable outputs. Secondly, the black-box nature of deep learning models makes it difficult to diagnose and address system mistakes, hindering effective troubleshooting. Additionally, integrating human expertise into security systems is challenging with opaque models, limiting human-in-the-loop capabilities and feedback incorporation. Thus, the lack of focus on explainability presents a significant challenge for AI-driven anomaly detection systems.

**R2-3 - XAI for ML Models.** To enhance the explainability of machine learning (ML) models in anomaly detection, future research can explore various interpretability methods. Approaches such as SHAP [130], LIME [181], and Breakdown [72] have demonstrated success in tasks like software defect prediction for understanding model decisions [96]. Additionally, recent advancements in explainable AI, such as the AIM framework proposed by Vo et al. [226], show potential for outperforming traditional XAI approaches like LIME and L2X, particularly in tasks like sentiment analysis. On the other hand, causal inference holds the potential to make machine learning models more interpretable [107]. By determining cause-and-effect relationships between variables, causal inference techniques offer insights into AI-driven anomaly detection. Overall, integrating causal inference methods into anomaly detection frameworks holds promise for enhancing model interpretability. By uncovering causal relationships between input features and detected anomalies, these techniques offer deeper insights into system behavior, enabling more informed decision-making by software developers and security analysts.

**C3-1 - Lack of AI Security Tooling in IDEs (In DevOps Development). Challenge - Lack of AI-driven Tools in IDEs** A significant hurdle in the current landscape is the limited availability of AI-driven security tools in developers' IDEs, posing a challenge to the widespread adoption of advanced security measures in software development. The scarcity of such tools restricts developers from harnessing the full potential of artificial intelligence in enhancing security practices. Additionally, there is a notable absence of a comprehensive user study that thoroughly evaluates the performance and effectiveness of existing AI-driven security tools, including prominent ones like AIBugHunter [65] and Snyk [202]. The lack of a comprehensive user study hinders a nuanced understanding of the strengths, weaknesses, and overall impact of these tools, impeding efforts to establish robust security practices in the evolving landscape of software development.

**R3-1 - AI Tool Deployment and Comprehensive Tool Evaluation.** We outline potential avenues for research in the domain of AI-driven security tools. Considering the extensive literature and varied techniques available for vulnerability detection, classification, and repair, future studies could focus on seamlessly integrating existing methods into developers' IDEs to enhance their practical applicability. For instance, commercial tools such as Code Scanning Autofix powered by GitHub Copilot are available in GitHub to help developers automatically fix vulnerable code in their pull requests. This tool is an expansion of Code Scanning that provides users with targeted recommendations to help them fix code scanning alerts in pull requests and avoid introducing new security vulnerabilities. The potential fixes are generated automatically by large language models (LLMs) using data from the codebase, the pull request, and from CodeQL analysis [217]. In particular, when a vulnerability is discovered in supported languages (i.e., JavaScript, TypeScript, Python, and Java), Autofix will generate a natural language explanation of the suggested fix, along with a preview of the code suggestion that the developer can accept, edit, or dismiss. Moreover, these code suggestions can include changes across multiple files and the dependencies that should be added to the project [68]. Furthermore, a comprehensive evaluation of these AI-driven security tools is imperative. This involves soliciting and analyzing user feedback, particularly the scenarios where the tools generate inaccurate predictions. Furthermore, it is crucial to investigate the robustness of AI-driven security tools. For example, a robust security tool should ideally predict the repaired program as benign. However, the existing literature does not definitively address whether a program continues to be predicted as vulnerable even after undergoing a successful repair. Exploring this aspect would contribute valuable insights into the efficacy of AI-driven security tools in practical scenarios.

**C3-2 - Lack of AI Security Tooling in CI/CD (In DevOps Code Commit).** Our investigation revealed that the majority of just-in-time (JIT) SDP tools used in CI/CD environments, such as JITO [172] and JITBot [102], primarily rely on machine learning models. Despite proposed deep learning (DL)-based SDP methodologies, there is a notable absence of DL-based tools integrated into CI/CD pipelines. This absence impedes the practical adoption of DL-based approaches.

**R3-2 - AI Tool Deployment in CI/CD Pipelines.** In contrast to machine learning (ML) models, which depend on manually predefined metrics for predicting software defects, DL models can learn source code representations directly from developers' code without the need for extensive feature engineering efforts. To enhance their practical adoption, future research should focus on integrating existing DL-based approaches for JIT SDP into CI/CD pipelines. Additionally, conducting a comprehensive evaluation of these DL-based tools is essential. This evaluation should encompass not only performance assessment post-deployment but also a large-scale user study to gather feedback from developers.

**C3-3 - Lack of AI Security Tooling for Infrastructure Scanning (In DevOps Build, Test, and Deployment).** Our investigation has uncovered a gap between AI-driven infrastructure scanning approaches and their practical adoption. Although several machine learning (ML)-based methods have been proposed to detect defective Infrastructure-as-Code (IaC) scripts [18, 44, 175, 176], they have yet to be deployed as software tools for developers. The lack of deployment hinders their practical adoption. Furthermore, the precision and practicality of these tools post-deployment remain unexplored.

**R3-3 - AI Tool Deployment and Post-Deployment Evaluation.** Future research should focus on bridging the gap between proposed AI-driven methods for detecting defective Infrastructure-as-Code (IaC) scripts and their practical deployment as software tools for developers. Exploring strategies to facilitate the deployment of these tools, such as developing user-friendly interfaces and integration into existing development workflows, could enhance their practical adoption. Additionally, investigating the precision and practicality of these tools post-deployment is crucial to assess their effectiveness in real-world scenarios. By addressing these research opportunities, advancements can be made towards empowering developers with effective and efficient tools for enhancing the security of Infrastructure-as-Code.

Below, we begin introducing the challenges and research opportunities specific to each security task in the DevOps process. Since we found no relevant literature discussing AI-driven approaches in the planning step of DevOps, we will start with the development step.

## 5.2 Plan

*5.2.1 Threat Modeling & Impact Analysis.* Since our literature review did not reveal relevant studies on Threat Modeling and Impact Analysis, we were unable to identify challenges directly from the literature. Instead, we examine existing AI-driven approaches used in industry and explore how future research could build upon these practices to advance the field.

**Existing AI-Driven Security Analysis Tools and Techniques in Industry.** The recent advancement of AI models presents promising opportunities for facilitating threat modeling and impact analysis in cybersecurity. For instance, an AI-driven commercial tool named Aribot for threat modeling has been introduced by Aristiun [8]. Aribot automates the threat modeling process through various functionalities. It automatically generates Infrastructure-as-Code templates, addressing public cloud-specific threats effectively. It ensures traceable security requirements throughout the lifecycle, facilitating comprehensive security coverage. Aribot also simplifies compliance adherence by mapping security requirements to frameworks such as the National Institute of Standards and Technology (NIST). Additionally, it enhances transparency and accountability by reporting and tracking records remediation efforts by development teams, providing real-time updates on implementation status without requiring manual intervention. Another AI-driven commercial tool for impact analysis has been introduced by Validata [224]. This AI-based solution automatically delivers crucial information for applications, expediting upgrades and patches while predicting the impact of new product versions before release. It empowers users to effortlessly analyze change impact on quality, performance, resource capacity, and costs within hours, automatically applying recommended changes through a corrective action plan.

Moreover, Microsoft [149] has recently introduced Copilot for Security. Informed by large-scale data and threat intelligence, including more than 78 trillion security signals processed by Microsoft each day. Copilot is coupled with large language models (LLMs) to deliver tailored security insights. It offers an interactive interface to support security practitioners during impact analysis and threat modeling with relevant security knowledge. Their randomized controlled trial indicated that experienced security analysts were 22% faster with Copilot, they were 7% more accurate across all tasks when using Copilot, and 97% said they want to use Copilot the next time they do the same task [55]. Copilot serves as the first critical step in leveraging generative AI to support security practitioners in their workflow.

In summary, the adoption of AI in threat modeling and impact analysis presents significant advantages for the planning step in DevSecOps. Tools like Aribot, Validata, and Copilot for Security showcase the potential of AI to automate these processes effectively. By leveraging the capabilities of AI to assess risks, identify vulnerabilities, and prioritize security measures, organizations could make informed decisions and allocate resources more efficiently.

**Future Research Directions.** To build on existing AI-driven tools like Aribot, Validata, and Copilot for Security, research can focus on a few key areas critical to enhancing DevSecOps planning. First, there is a need to refine AI models to enable dynamic updates in threat modeling and impact analysis as part of continuous integration/continuous deployment (CI/CD) pipelines. Given the iterative nature of DevSecOps, these updates would ensure that security assessments remain relevant and responsive to ongoing code changes. Additionally, the creation of benchmark datasets is essential for measuring the effectiveness of these AI-driven approaches. Such benchmarks would help identify gaps in current tools, guiding future improvements and providing a standardized basis for evaluating new solutions. Finally, research into the ethical implications of deploying AI in security is crucial. This includes ensuring that AI models are used responsibly, with attention to privacy, fairness, and transparency, which will be vital for the wider adoption of AI-driven security measures in DevSecOps environments.

## 5.3 Development

*5.3.1 Software Vulnerability Detection.* While the progress in code pre-trained language models enhances the F1-score of function-level vulnerability detection, reaching up to 96.5% [207], there remain several challenges that must be addressed in the current landscape of AI-driven vulnerability detection. We describe challenges and potential research directions in the following.

**C4 - Cross Project.** Most of the VD studies considered the mixed project (models are trained and tested on combined projects) scenario when evaluating their proposed approaches. However, in an empirical study evaluating SOTA DL-based VD approaches, Steenhoek et al. [208] observed a decline in model performance during detection under the cross-project scenario (models are trained on one set of projects and tested on completely different, non-overlapping projects). Specifically, the F1-score of function-level detection models experienced a reduction ranging from 11% to 32%. This underscores the challenge posed by cross-project vulnerability detection and emphasizes the limited generalizability of current methodologies. The findings emphasize the need for advancements in methodologies capable of generalizing effectively across various projects, especially in cross-project scenarios.

**R4 - Cross-Project SVD with Diverse CWE-IDs.** Zhang et al. [266] and Liu et al. [128] both proposed to use deep domain adaptation to address cross-project VD in C languages. Subsequent research avenues could delve into cross-programming language VD and explore cross-project VD for other programming languages. In particular, Liu et al. [128] mainly focused on CWE-119 and CWE-399 while other dangerous CWE-IDs [43] should be considered in future studies. Furthermore, the study concentrated on graph attention networks, yet there is potential for exploration of other graph neural networks (GNNs) and large language models (LLMs) for effective cross-project VD.

**C5 - MBU Vulnerabilities.** Sejfia et al. [193] pointed out that state-of-the-art deep learning-based vulnerability detection (VD) approaches concentrate on individual base units, assuming that vulnerabilities are confined to a single function. However, vulnerabilities may extend across multiple base units (MBU). They found that existing DL-based detectors do not work as well in detecting all comprising parts of MBU, which poses a challenge in predicting MBU vulnerabilities. Thus, future research should contemplate adapting their evaluation methods to incorporate the presence of MBU vulnerabilities.

**R5 - Evaluate SVD on MBU Vulnerabilities.** The majority of studies on function-level VD [25, 34, 64] have typically assessed performance based on the number of correctly detected vulnerable functions. However, it is important to note that a single vulnerability might encompass multiple vulnerable functions. Therefore, merely identifying one vulnerable function does not necessarily equate to the comprehensive detection of the entire vulnerability. Acknowledging this, Sejfia et al. [193] emphasized the need for future research to account for the scenario of Multiple Base Unit (MBU) vulnerabilities during the training and testing phases of DL-based VD. This requires refining evaluation metrics and methodologies to align with MBU vulnerabilities. Consequently, there is a critical call for an enhanced understanding and consideration of MBU vulnerabilities to further advance the field of vulnerability detection.

**C6 - Data Quality.** Finally, Croft et al. [41] noted data quality concerns related to accuracy, uniqueness, and consistency in widely used vulnerability datasets. Their findings revealed that a substantial portion, ranging from 20% to 71%, of labels in real-world datasets were inaccurately assigned. This inaccuracy had the potential to significantly impact the performance of resulting models by up to 65%.

**R6 - Addressing Data Inaccuracy from Automatic Data Collection.** Croft et al. [41] pointed out that automatic data collection often leads to data inaccuracy. Common vulnerability datasets [58, 273, 273] were constructed by using the code changes information to recover the vulnerable version of a method. However, the vulnerability fixing commits or vulnerable lines could be incorrectly selected. Thus, it is important for future research to devise semantic filters or heuristics—methods or rules designed to analyze and interpret the meaning of code changes. This is crucial for precisely pinpointing lines that signify vulnerability fixes, and addressing the data inaccuracies stemming from the automatic data collection process.

*5.3.2 Software Vulnerability Classification.* In our examination of AI-driven methods explaining vulnerability types to developers, we found that a significant amount of studies focus on vulnerability detection, with limited attention given to vulnerability classification. Below, we outline the challenge identified in current AI-driven vulnerability classification approaches followed by future research opportunities.

**C7 - Incompleted CWE Tree.** The TreeVul approach [161] currently relies exclusively on parent-child relations within the Common Weakness Enumeration (CWE) hierarchy for conducting top-down searches. This approach excludes the consideration of other potentially valuable relations, such as PeerOf, which could enhance the model's effectiveness. In addition, TreeVul only considers CWE categories with depth<=3 while some important categories may be located at depth>3. These incomplete considerations of CWE tree structure may hinder TreeVul's ability to generalize to additional CWE-IDs.

**R7 - Advanced Tree-based SVC.** According to Pan et al. [161], TreeVul did not encompass the entire CWE tree. Consequently, future investigations may broaden the TreeVul approach by incorporating extra relations, such as PeerOf. Nevertheless, this procedure views the CWE structure as a graph, introducing a higher level of complexity compared to the hierarchical tree structure considered by Pan et al. [161]. Thus, emphasis could be placed on streamlining the transformation process to reduce the complexities of converting the CWE tree into a graph. Furthermore, optimizing the TreeVul approach to dynamically determine the appropriate level for concluding top-down searches beyond the predefined depth-3 CWE categories offers an opportunity to enhance the model's adaptability. This optimization could involve developing a mechanism to assess confidence levels, allowing the model to automatically adjust its search depth based on contextual factors for each specific input. Such advancements could improve the precision and flexibility of AI-driven vulnerability classification.

*5.3.3 Automated Vulnerability Repair.* Recent advancements in transformer architecture and pre-trained language models for code have shown improvements over RNN-based models in vulnerability repair and program repair tasks. Nevertheless, beyond model architecture and performance, there are additional aspects that demand attention and further improvement, especially when considering their integration into real-world projects. Below, we present challenges derived from previous studies with potential research opportunities.

**C8 - Sequence Length and Computing Resource.** Most code pre-trained language models (LMs) have an input length limit of 512 subword tokens for base-size models and 1024 for large-size models. Thus, LMs may not fully comprehend long code programs due to the input length limit of LMs, which constrains the repair capability. Furthermore, the increase in output length not only impacts the performance of repair capability but also introduces a well-known challenge associated with long sequences in the NMT model. This difficulty arises from the limitations imposed by Markov chain assumptions and probabilistic constraints, where the model encounters challenges in maintaining coherence and capturing intricate dependencies over extended sequences. In particular, Fu et al. [66] observed that the repair model demonstrated a repair accuracy of 77% when both input and output lengths were below 100 and 10, respectively. However, the model's accuracy drastically dropped to 7% when the input and output lengths exceeded 500 and 50, respectively. In addition, Huang et al. [91] emphasized that the substantial size of language models can place a burden on computing resources, which hinders the generation of candidate patches. For example, during patch synthesis on the Defects4J dataset, the maximum beam size is set at 200, limiting the generation to only 200 patches for each bug.

**R8 - Transformer Variants That Process Longer Sequence.** Transformer architectures have led to remarkable progress in automated repair applications. However, despite their successes, modern transformers rely on the self-attention mechanism, whose time- and space-complexity is quadratic in the length of the input that requires substantial computing resources when encountering long sequences. Recently, many alternative architectures have been proposed to mitigate the long sequence challenge and computing burden of transformers. For example, Beltagy et al. [13] presented Longformer which uses global and local attention windows to mitigate the memory and time bottleneck of the self-attention mechanism from $O(ns \times ns)$ to $O(ns \times w)$, with ns being the sequence length and w being the average window size. In particular, the base-size Longformer can process up to 4,096 subword tokens. On the other hand, Sun et al. [212] proposed Retentive Network (RetNet) as a new foundation architecture for large language models. The chunkwise recurrent representation of RetNet facilitates efficient long-sequence modeling with linear complexity. Thus, future studies may explore their efficacy in the context of program and vulnerability repair.

**C9 - Loss of Pre-Trained Knowledge.** Most language model-based repair approaches follow a paradigm of taking the pre-trained checkpoint and further fine-tuning the model to fit the downstream program and vulnerability repair tasks. Nevertheless, according to Huang et al. [91], after fine-tuning, pre-trained models may experience a reduction in the knowledge gained during pre-training when compared to zero-shot learning (i.e., without fine-tuning). This could be attributed to conflicting training objectives between the unsupervised masked language modeling (MLM) of pre-training and the supervised neural machine translation (NMT) of fine-tuning.

**R9 - Explore Different Training Paradigms.** Huang et al. [91] suggested two research directions to address this challenge. Future research could investigate strategies to alleviate catastrophic forgetting [195] for program and vulnerability repair tasks. Additionally, exploring both neural machine translation (NMT) and masked language modeling (MLM) training paradigms during the fine-tuning stage is a potential direction. Notably, AlphaRepair [246] adopts a cloze task (MLM) approach instead of a translation task (NMT), predicting the token at the mask location based on contextual tokens. While employing MLM during fine-tuning could be advantageous as it aligns with the training objective of the model's pre-training stage, the differentiation in repair efficacy between the two paradigms (NMT and MLM) remains unclear.

**C10 - Automated Repair on Real-World Scenarios.** Pearce et al. [164] found that large language models (LLMs) can perfectly repair all of their synthetic and hand-crafted vulnerability scenarios. However, LLMs were not sufficiently reliable when producing automatic fixes for the real-world data in their qualitative analysis. Furthermore, they underscored the limitation of the current repair approach, which is confined to addressing issues within a single location in a single file.

**R10 - Addressing Limitations of LLMs.** Addressing the limitations of large language models (LLMs) in real-world vulnerability scenarios presents promising avenues for future research. Pearce et al. [164] identified the significant performance gap of LLMs in repairing synthetic and real-world vulnerability scenarios. Thus, understanding and mitigating the factors contributing to the discrepancy in performance between synthetic and real-world scenarios would be a valuable direction for further investigation, enabling the development of more robust automatic fixes in practical cybersecurity contexts. Furthermore, exploring multi-location and multi-file repair strategies and techniques could offer valuable insights.

## 5.4 Code Commit

*5.4.1 Dependency Management.* Since our literature review did not reveal relevant studies on Dependency Management, we were unable to identify challenges directly from the literature. Instead, we examine existing AI-driven approaches used in industry and explore how future research could build upon these practices to advance the field.

**Existing AI-Driven Security Analysis Tools and Techniques in Industry.** Recently, a startup company Infield introduced an AI-driven commercial tool designed to automate software dependency management and strengthen DevOps security [77]. The tool continuously monitors recommended updates of open-source components, provides step-by-step guidance for achieving the ideal status, and gathers unstructured information about open-source dependencies and their upgrades. This data is then structured to help users manage their backlog of upgrades efficiently, allowing them to prioritize upgrades based on risk and effort. For instance, the tool can be connected with users' codebase in GitHub, scans their code to determine the underlying dependencies, and recommends the steps needed to upgrade safely for their codebase. The tool offers a human-assisted approach to dependency management, helping organizations overcome the challenges of maintaining dependencies in a rapidly evolving ecosystem.

**Future Research Directions.** The introduction of AI-driven tools like Infield's dependency management solution marks a significant advancement in automating and enhancing security in DevSecOps. However, there remains substantial potential for further innovation, particularly in integrating Retrieval-Augmented Generation (RAG) techniques with Large Language Models (LLMs). One promising direction is to augment LLMs with a pre-built knowledge base containing comprehensive information on software dependencies, vulnerabilities, and best practices. In this setup, when a software update is identified, developers or security practitioners could use the RAG-based LLM system to query the knowledge base and receive tailored security recommendations for the specific dependencies involved. This approach would empower users to make more informed decisions, ensuring more effective and secure management of software dependencies. Additionally, developing a benchmark dataset specifically for evaluating AI-driven dependency management approaches is crucial. This dataset would allow researchers to systematically assess the effectiveness of RAG-based LLMs in real-world scenarios, identifying gaps and opportunities for improvement. By focusing on these areas, future research can build on existing AI applications and drive advancements in dependency management within the DevSecOps lifecycle.

*5.4.2 CI/CD Secure Pipelines.* AI-driven just-in-time (JIT) software defect prediction (SDP) approaches have been proposed and deployed into CI/CD pipelines for adoption [102, 172]. However, our investigation uncovers several challenges that must be addressed to further enhance these AI-driven approaches. Below, we present the challenge derived from previous JIT SDP studies with potential research opportunities.

**C11 - The Use of RNNs.** We discovered that many DL-based JIT SDP methods primarily rely on RNNs [45, 168]. However, pre-trained transformer architectures and language models (LMs) demonstrate promise for SDP, building on their successful track record in vulnerability detection [208], a closely related domain. RNNs excel at learning source code representations and predicting defects without relying on predefined metrics. Nonetheless, RNNs process input sequentially and struggle with long sequences, which can lead to suboptimal results compared to transformer-based models.

**R11 - Explore LMs and LLMs.** In light of the successful application of language models (LMs) for software vulnerability detection [64, 208], future research could explore their potential for JIT SDP. LMs like CodeBERT [59] and CodeT5 [232, 233] are transformer architectures pre-trained on source code, featuring self-attention mechanisms that excel in capturing semantic nuances and handling longer sequences compared to RNNs. Moreover, these LMs are pre-trained on extensive source code datasets encompassing various programming languages, enabling them to generate better source code representations compared to RNNs and enhancing SDP effectiveness. Furthermore, investigating the efficacy of large language models (LLMs) such as GPT-4 [2] and Code Llama [183] for SDP is a valuable research direction. These models are pre-trained extensively to understand source code. Thus, future research could explore strategies for leveraging these LLMs and deploying them to build secure CI/CD pipelines.

## 5.5 Build, Test, and Deployment

*5.5.1 Configuration Validation.* AI-driven methodologies have been suggested for automated verification of optimal and secure system configurations. Nevertheless, the complexity of software systems, characterized by an extensive array of configuration options, poses a challenge for machine learning (ML) and deep learning (DL) models. Below, we present the challenge derived from previous studies with potential research opportunities.

**C12 - Complex Feature Space.** Machine learning (ML) models for extracting insights from numeric and categorical features is a widely adopted strategy in configuration validation processes. Contemporary AI-driven performance prediction models frequently rely on conventional machine learning models like XGBoost [247] or basic feed-forward neural networks such as Deep-Perf [75] to analyze tabular configuration data and predict system performance. Nonetheless, software systems often encompass an extensive array of configuration options. For instance, one of the datasets used to evaluate Deep-Perf consists of 13,485 valid configurations [75]. The extensive configuration size results in intricate relationships among features and an expansive feature space. Consequently, traditional ML models and simple feed-forward neural networks may struggle to generalize to unseen configurations or capture subtle patterns and dependencies within the data, especially in high-dimensional spaces such as configuration datasets.

**R12 - Transformers for Tabular Data.** Addressing the challenge of capturing intricate relationships and dependencies within high-dimensional configuration datasets presents numerous future research opportunities in performance prediction. One promising avenue for exploration involves leveraging more advanced model architectures, such as transformer-based models [225], to enhance the predictive capabilities of performance prediction models. Transformer architectures, renowned for their success in natural language processing tasks, offer the potential to effectively capture complex patterns and dependencies within tabular data like configuration datasets. By adapting transformer models to handle tabular data, researchers can explore their ability to learn from the sequential and relational information present in configuration parameters and accurately predict system performance. In addition, researchers can explore semi-supervised learning techniques [160]. These methods leverage both labeled and unlabeled data, helping overcome the scarcity of labeled training instances in performance prediction. In summary, these research directions could improve the effectiveness of AI-driven configuration validation processes in complex software systems.

*5.5.2 Infrastructure Scanning.* AI-driven infrastructure scanning approaches have been proposed for detecting insecure Infrastructure-as-Code (IaC) scripts during the deployment phase of DevOps. Nevertheless, there remains room for improvement and identified gaps. Below, we outline the challenge derived from previous studies with potential research opportunities.

**C13 - Manual Feature Engineering.** We found that some AI-driven approaches still rely on manual feature engineering with machine learning (ML) models to predict insecure IaC scripts [44, 175, 176], which can demand substantial effort, especially as the project scales up. It is necessary to prepare a predetermined set of features before model training; for instance, Dalla Palma et al. [44] prepared a set of 108 features. Subsequently, the feature selection process is essential to filter out irrelevant features, and techniques such as normalization will be adopted to optimize ML model performance. This time-consuming process could hinder the practical adoption of AI-driven infrastructure scanning approaches.

**R13 - Explore DL-based Techniques.** Considering the recent advancement of deep learning (DL) and language models (LMs), it is feasible to use DL models for detecting defective IaC scripts. Unlike traditional machine learning (ML) models, DL models allow direct input of IaC scripts, thereby eliminating the need for manual feature engineering. DL models are capable of learning both semantic and syntactic features of IaC scripts. They can identify insecure patterns within IaC scripts based on historical data and predict defective scripts in future instances. Additionally, LMs have proven successful in various software security tasks, such as vulnerability detection [64, 208] and repairs [14, 66, 82]. Given that LMs are pre-trained on vast amounts of source code data, it is viable to fine-tune them for accurately detecting insecure IaC scripts. Hence, future research could explore the application of DL models to address the time-consuming challenges associated with manual feature engineering in detecting insecure IaC scripts.

## 5.6 Operation and Monitoring

*5.6.1 Log Analysis and Anomaly Detection.* Several AI-driven anomaly detection approaches have been developed to detect anomalies in software systems. However, our investigation reveals the challenge that requires attention and resolution. Below, we introduce the challenge derived from previous studies with potential research opportunities.

**C14 - Normality Drift for Zero-Positive Anomaly Detection.** In anomaly detection, zero-positive classification is commonly used because anomalies are typically rare and undefined. Zero-positive classification trains models on normal behavior, allowing them to generalize to unseen anomalies and adapt to changing data distributions. However, Han et al. [78] recognized a normality drift problem for the zero-positive classification used in anomaly detection. Normality drift refers to the phenomenon where the underlying distribution of normal (non-anomalous) data changes over time in a dataset used for training AI-driven anomaly detections. In other words, the characteristics of normal behavior exhibited by the system or environment being monitored may evolve or shift gradually or abruptly, leading to discrepancies between the training data and the data encountered during deployment or inference. This drift can adversely affect the performance of anomaly detection systems, as models trained on historical data may become less effective at identifying anomalies from the new normal behavior. Thus, addressing normality drift is crucial for maintaining the effectiveness and reliability of anomaly detection systems.

**R14 - Enhance Normality Drift Detection.** Han et al. [78] introduced OWAD as a solution to combat the challenge of normality drift encountered in deep learning-based anomaly detection for security applications. OWAD serves as a framework designed to detect, explain, and adapt to normality shifts at the distribution level, departing from sample-level explanations like CADE [258] which fail to comprehensively address the holistic shift in distribution, thereby limiting their applicability in understanding the overall normality drift. This development paves the way for numerous future research opportunities in anomaly detection. Particularly, there exists potential for exploring innovative adaptation strategies within OWAD aimed at enhancing its efficacy in adapting to evolving data distributions. This could entail investigating reinforcement learning approaches or meta-learning techniques to dynamically adjust model parameters in response to shifting distributions. Future investigations can also concentrate on refining OWAD's adaptation mechanisms to ensure effectiveness across diverse security environments.

*5.6.2 Cyber-Physical Systems.* AI-driven approaches have been proposed to address security concerns in cyber-physical systems (CPS) [12, 125, 191, 245, 249–251]. However, our investigation reveals that current approaches to CPS security primarily target individual cyber-attacks, neglecting the complex scenario of encountering multiple simultaneous threats across diverse CPS layers. Below, we describe this challenge and highlight potential avenues for future research.

**C15 - Monitoring Multiple Cyber-Attacks Simultaneously.** CPS consists of multiple layers representing distinct components and functionalities crucial for system operation. These layers typically include the physical layer encompassing hardware infrastructure, the communication layer facilitating data exchange between components, the control layer governing system behavior, and the data processing layer analyzing collected data. However, current approaches in CPS security often focus on addressing individual cyber-attacks, overlooking the reality of facing multiple simultaneous threats across various CPS layers [12, 125, 191, 245, 249–251]. For instance, attackers may simultaneously disrupt communication networks, manipulate control algorithms, and tamper with data analytics results in the smart grid CPS, posing complex challenges for system security and resilience. As cyber threats grow in sophistication and diversity, the need to coordinate responses and monitor ongoing attacks concurrently becomes paramount. However, achieving this in real time presents substantial challenges due to the intricate and interconnected nature of CPS environments.

**R15 - Distributed Anomaly Detection and Multi-Agent Systems.** Future research could focus on developing distributed detection and response mechanisms within CPS environments, empowering individual components to autonomously detect and respond to cyber threats in real time. This approach decentralizes the security architecture and reduces reliance on centralized control systems. By integrating AI-based anomaly detections, these mechanisms could effectively identify unusual patterns or behaviors indicative of cyber attacks across various CPS layers. Multi-agent systems (MAS) hold promise for addressing CPS attacks across different layers due to their decentralized and collaborative nature. MAS consists of multiple autonomous agents, each capable of independent decision-making and action. For example, Lyu and Brennan [132] introduced a two-layer architecture modeling framework for CPS, demonstrating how it enables real-time adaptation in dynamic industrial automation environments. Moreover, integrating AI techniques with MAS could facilitate collaborative decision-making and resource allocation among autonomous agents distributed across different CPS components or layers. By leveraging reinforcement learning (RL), agents can dynamically adapt their strategies based on feedback from the environment, allowing them to respond to emerging cyber threats with agility and precision. For instance, Ibrahim and Elhafiz [94] presented a case study of a smart grid and investigated CPS security using RL. In summary, AI-driven MAS could adaptively adjust defense strategies based on evolving threat landscapes and system conditions, enhancing the effectiveness of cyber defense operations within CPS environments.

## 6 THREATS TO VALIDITY

Our systematic literature review (SLR) was conducted in alignment with the guidelines outlined by Keele et al. [101], Kitchenham et al. [106]. However, like any SLR, our review also has certain limitations. Below, we provide a discussion of the external, internal, and construct threats to the validity of our SLR, along with corresponding mitigation strategies.

*Threats to construct validity* relate to the process of selecting the 12 security tasks for our review as presented in Figure 1. This process involves subjective judgment that could potentially influence the construct validity of our findings. The identification and refinement of tasks were based on the collective expertise of the three authors, which, while informed, inherently involved subjective elements. This subjectivity could affect how well the tasks represent the actual security concerns within the DevOps workflow. To mitigate this threat, we employed a rigorous approach involving multiple authors rather than relying on a single individual's perspective. The first author initially provided an overview of potential tasks, which was then critically discussed and refined through iterative brainstorming sessions and online meetings with the other two authors. This collaborative process ensured a broad range of insights and perspectives were considered, thereby enhancing the accuracy and relevance of the identified tasks.

*Threats to external validity* relate to the search string, the filtering process, and the selection of venues in this SLR aimed at identifying literature related to AI-driven methodologies and tools for DevSecOps. It is possible that our search string missed studies that should have been included in our review, potentially due to a missed term or a combination of terms that may have returned more significant results. Given that our study focuses on two areas—AI (specifically, machine learning and deep learning) and security methodologies and tools for integration into DevOps—we employ a systematic approach to testing different combinations of AI-related terms and security tasks as presented in Section 3.2. This involves experimenting with variations in search strings, including synonyms, related terms, and alternative phrasings. By doing so, we aim to increase the likelihood of identifying relevant studies that may have been overlooked initially. Furthermore, we leverage our understanding of the domain to refine and optimize our search strategy iteratively to mitigate this threat.

While conducting this systematic literature review (SLR), we acknowledge the potential for selection bias in the studies included. To mitigate this threat, we employ rigorous inclusion and exclusion criteria predefined before initiating the filtering process as illustrated in Table 3. Furthermore, we implement a snowballing search strategy to supplement our initial searches as presented in Section 3.4, thereby capturing papers that may have been overlooked initially.

The selection of venues for this SLR plays a crucial role in ensuring the reliability of our findings. In our approach, we deliberately include top-tier SE and security conferences and journals, focusing on venues with CORE A or CORE A* rankings. By prioritizing these esteemed venues, known for their rigorous peer-review processes and high academic standards, we aim to uphold the quality and credibility of the literature surveyed. While we acknowledge the possibility that our exclusion of lower-ranking venues may have resulted in the omission of relevant studies, our deliberate focus on top SE and security venues allows us to capture emerging trends and insights from reputable sources. To provide transparency and accountability in our venue selection process, we disclose the selected venues in Table 2, enabling readers to assess our review methodology.

*Threats to internal validity* relate to the potential absence of literature discussing AI-driven methods or tools for specific security tasks within the DevSecOps process. Notably, despite our comprehensive search efforts, we encountered a scarcity of studies addressing AI applications for threat modeling and impact analysis in the plan step, as well as dependency management in the code commit step. This gap in the literature poses a potential limitation to the comprehensiveness of our review. To mitigate this, we have examined common approaches for these three security tasks within the context of RQ1 and explored AI-driven applications in industry along with future research directions in RQ2. This integrated approach provides a comprehensive overview of current practices and underscores their relevance to DevSecOps.

Another threat to internal validity in our study arises from the fact that the themes and challenges were derived directly from the reviewed literature. This methodology means that some universally important challenges, such as data quality and model explainability, may only be associated with specific security tasks in DevSecOps if they were explicitly discussed in the literature we analyzed. Consequently, while our study highlights key challenges within the context of specific security tasks, there may be broader applicability of these challenges across other tasks that are not discussed in this paper.

## 7 CONCLUSION

AI-driven security approaches are revolutionizing software security automation, presenting opportunities to seamlessly integrate security practices into the DevOps workflow and realize the DevSecOps paradigm efficiently. Throughout this systematic literature review, we initially collected 1,897 papers from high-impact software engineering and software security venues, filtering down to 99 papers for in-depth analysis, all focusing on AI-driven security approaches for DevSecOps. In RQ1, we identified 12 critical security tasks in DevSecOps, examined existing AI-driven approaches and tools, and identified 65 benchmarks used to evaluate these methods. Subsequently, we identified 15 significant challenges confronting state-of-the-art AI-driven security methodologies and derived 15 corresponding future research directions in RQ2. In conclusion, this paper sheds light on the transformative potential of AI-driven security techniques in realizing the DevSecOps paradigm, while identifying critical challenges and emphasizing the need for future endeavors to address them in this crucial intersection of AI and DevSecOps.

## 8 ACKNOWLEDGMENTS

C. Tantithamthavorn was partially supported by the Australian Research Council's Discovery Early Career Researcher Award (DECRA) funding scheme (DE200100941).

## REFERENCES

[1] Agnar Aamodt and Enric Plaza. 1994. Case-based reasoning: Foundational issues, methodological variations, and system approaches. *AI communications* 7, 1 (1994), 39–59.

[2] Josh Achiam, Steven Adler, Sandhini Agarwal, Lama Ahmad, Ilge Akkaya, Florencia Leoni Aleman, Diogo Almeida, Janko Altenschmidt, Sam Altman, Shyamal Anadkat, et al. 2023. Gpt-4 technical report. *arXiv preprint arXiv:2303.08774* (2023).

[3] Diana Laura Aguilar, Miguel Angel Medina-Perez, Octavio Loyola-Gonzalez, Kim-Kwang Raymond Choo, and Edoardo Bucheli-Susarrey. 2022. Towards an interpretable autoencoder: A decision-tree-based autoencoder and its application in anomaly detection. *IEEE transactions on dependable and secure computing* 20, 2 (2022), 1048–1059.

[4] Chuadhry Mujeeb Ahmed, Venkata Reddy Palleti, and Aditya P Mathur. 2017. WADI: a water distribution testbed for research in the design of secure cyber physical systems. In *Proceedings of the 3rd international workshop on*

*cyber-physical systems for smart water networks*. 25–28.

[5] Muhammad Azeem Akbar, Kari Smolander, Sajjad Mahmood, and Ahmed Alsanad. 2022. Toward successful DevSecOps in software development organizations: A decision-making framework. *Information and Software Technology* 147 (2022), 106894.

[6] Marco Ancona, Enea Ceolini, Cengiz Öztireli, and Markus Gross. 2018. Towards better understanding of gradient-based attribution methods for Deep Neural Networks. In *6th International Conference on Learning Representations (ICLR)*. Arxiv-Computer Science.

[7] Masaki Aota, Hideaki Kanehara, Masaki Kubo, Noboru Murata, Bo Sun, and Takeshi Takahashi. 2020. Automation of vulnerability classification from its description using machine learning. In *2020 IEEE Symposium on Computers and Communications (ISCC)*. IEEE, 1–7.

[8] Aristiun. 2024. Automated Threat Modeling using AI. https://www.aristiun.com/automated-threat-modeling-using-ai.

[9] Robert S Arnold and Shawn A Bohner. 1993. Impact analysis-towards a framework for comparison. In *1993 conference on software maintenance*. IEEE, 292–301.

[10] Ahmed Bahaa, Ahmed Abdelaziz, Abdalla Sayed, Laila Elfangary, and Hanan Fahmy. 2021. Monitoring real time security attacks for IoT systems using DevSecOps: a systematic literature review. *Information* 12, 4 (2021), 154.

[11] Liang Bao, Xin Liu, Fangzheng Wang, and Baoyin Fang. 2019. ACTGAN: automatic configuration tuning for software systems with generative adversarial networks. In *2019 34th IEEE/ACM International Conference on Automated Software Engineering (ASE)*. IEEE, 465–476.

[12] Yuequan Bao, Zhiyi Tang, Hui Li, and Yufeng Zhang. 2019. Computer vision and deep learning–based data anomaly detection method for structural health monitoring. *Structural Health Monitoring* 18, 2 (2019), 401–421.

[13] Iz Beltagy, Matthew E Peters, and Arman Cohan. 2020. Longformer: The long-document transformer. *arXiv preprint arXiv:2004.05150* (2020).

[14] Berkay Berabi, Jingxuan He, Veselin Raychev, and Martin Vechev. 2021. Tfix: Learning to fix coding errors with a text-to-text transformer. In *International Conference on Machine Learning*. PMLR, 780–791.

[15] Guru Bhandari, Amara Naseer, and Leon Moonen. 2021. CVEfixes: automated collection of vulnerabilities and their fixes from open-source software. In *Proceedings of the 17th International Conference on Predictive Models and Data Analytics in Software Engineering*. 30–39.

[16] Ane Blázquez-García, Angel Conde, Usue Mori, and Jose A Lozano. 2021. A review on outlier/anomaly detection in time series data. *ACM Computing Surveys (CSUR)* 54, 3 (2021), 1–33.

[17] G Boetticher. 2007. The PROMISE repository of empirical software engineering data. *http://promisedata. org/repository* (2007).

[18] Nemania Borovits, Indika Kumara, Dario Di Nucci, Parvathy Krishnan, Stefano Dalla Palma, Fabio Palomba, Damian A Tamburri, and Willem-Jan van den Heuvel. 2022. FindICI: Using machine learning to detect linguistic inconsistencies between code and natural language descriptions in infrastructure-as-code. *Empirical Software Engineering* 27, 7 (2022), 178.

[19] Virginia Braun and Victoria Clarke. 2006. Using thematic analysis in psychology. *Qualitative research in psychology* 3, 2 (2006), 77–101.

[20] George G Cabral, Leandro L Minku, Emad Shihab, and Suhaib Mujahid. 2019. Class imbalance evolution and verification latency in just-in-time software defect prediction. In *2019 IEEE/ACM 41st International Conference on Software Engineering (ICSE)*. IEEE, 666–676.

[21] Wenjing Cai, Junlin Chen, Jiaping Yu, and Lipeng Gao. 2023. A software vulnerability detection method based on deep learning with complex network analysis and subgraph partition. *Information and Software Technology* 164 (2023), 107328.

[22] Jeanderson Cândido, Maurício Aniche, and Arie van Deursen. 2021. Log-based software monitoring: a systematic mapping study. *PeerJ Computer Science* 7 (2021), e489.

[23] Sicong Cao, Xiaobing Sun, Lili Bo, Ying Wei, and Bin Li. 2021. Bgnn4vd: Constructing bidirectional graph neural-network for vulnerability detection. *Information and Software Technology* 136 (2021), 106576.

[24] Antonio Carzaniga, Alfonso Fuggetta, Richard S Hall, Dennis Heimbigner, André Van Der Hoek, and Alexander L Wolf. 1998. *A characterization framework for software deployment technologies*. Technical Report. Technical Report CU-CS-857-98, Dept. of Computer Science, University of Colorado.

[25] Saikat Chakraborty, Rahul Krishna, Yangruibo Ding, and Baishakhi Ray. 2021. Deep learning based vulnerability detection: Are we there yet. *IEEE Transactions on Software Engineering* (2021).

[26] Wachiraphan Charoenwet, Patanamon Thongtanunam, Van-Thuan Pham, and Christoph Treude. 2024. An Empirical Study of Static Analysis Tools for Secure Code Review. *arXiv preprint arXiv:2407.12241* (2024).

[27] Nitesh V Chawla, Kevin W Bowyer, Lawrence O Hall, and W Philip Kegelmeyer. 2002. SMOTE: synthetic minority over-sampling technique. *Journal of artificial intelligence research* 16 (2002), 321–357.

[28] Checkmarx. 2006. Checkmarx. https://checkmarx.com/.
[29] Xiang Chen, Yingquan Zhao, Qiuping Wang, and Zhidan Yuan. 2018. MULTI: Multi-objective effort-aware just-in-time software defect prediction. *Information and Software Technology* 93 (2018), 1–13.
[30] Zimin Chen, Steve Kommrusch, and Martin Monperrus. 2022. Neural transfer learning for repairing security vulnerabilities in c code. *IEEE Transactions on Software Engineering* 49, 1 (2022), 147–165.
[31] Zimin Chen, Steve Kommrusch, Michele Tufano, Louis-Noël Pouchet, Denys Poshyvanyk, and Martin Monperrus. 2019. Sequencer: Sequence-to-sequence learning for end-to-end program repair. *IEEE Transactions on Software Engineering* 47, 9 (2019), 1943–1959.
[32] Zimin Chen and Martin Monperrus. 2018. The codrep machine learning on source code competition. *arXiv preprint arXiv:1807.03200* (2018).
[33] Guoli Cheng, Shi Ying, and Bingming Wang. 2021. Tuning configuration of apache spark on public clouds by combining multi-objective optimization and performance prediction model. *Journal of Systems and Software* 180 (2021), 111028.
[34] Xiao Cheng, Haoyu Wang, Jiayi Hua, Guoai Xu, and Yulei Sui. 2021. Deepwukong: Statically detecting software vulnerabilities using deep graph neural network. *ACM Transactions on Software Engineering and Methodology (TOSEM)* 30, 3 (2021), 1–33.
[35] Jianlei Chi, Yu Qu, Ting Liu, Qinghua Zheng, and Heng Yin. 2022. Seqtrans: automatic vulnerability fix via sequence to sequence learning. *IEEE Transactions on Software Engineering* 49, 2 (2022), 564–585.
[36] E Chickowski. 2018. Seven Winning DevSecOps Metrics Security Should Track. *Bitdefender) Retrieved March* 25 (2018), 2019.
[37] Javier Alvarez Cid-Fuentes, Claudia Szabo, and Katrina Falkner. 2015. Online behavior identification in distributed systems. In *2015 IEEE 34th Symposium on Reliable Distributed Systems (SRDS)*. IEEE, 202–211.
[38] Javier Alvarez Cid-Fuentes, Claudia Szabo, and Katrina Falkner. 2018. Adaptive performance anomaly detection in distributed systems using online svms. *IEEE Transactions on Dependable and Secure Computing* 17, 5 (2018), 928–941.
[39] CORE. 2023. International CORE Conference Rankings: ICORE. https://www.core.edu.au/icore-portal.
[40] Gideon Creech and Jiankun Hu. 2013. Generation of a new IDS test dataset: Time to retire the KDD collection. In *2013 IEEE wireless communications and networking conference (WCNC)*. IEEE, 4487–4492.
[41] Roland Croft, M Ali Babar, and M Mehdi Kholoosi. 2023. Data quality for software vulnerability datasets. In *2023 IEEE/ACM 45th International Conference on Software Engineering (ICSE)*. IEEE, 121–133.
[42] Roland Croft, Dominic Newlands, Ziyu Chen, and M Ali Babar. 2021. An empirical study of rule-based and learning-based approaches for static application security testing. In *Proceedings of the 15th ACM/IEEE international symposium on empirical software engineering and measurement (ESEM)*. 1–12.
[43] CWE. 2023. 2023 CWE Top 25 Most Dangerous Software Weaknesses. https://cwe.mitre.org/top25/archive/2023/2023_top25_list.html.
[44] Stefano Dalla Palma, Dario Di Nucci, Fabio Palomba, and Damian A Tamburri. 2021. Within-project defect prediction of infrastructure-as-code using product and process metrics. *IEEE Transactions on Software Engineering* 48, 6 (2021), 2086–2104.
[45] Hoa Khanh Dam, Trang Pham, Shien Wee Ng, Truyen Tran, John Grundy, Aditya Ghose, Taeksu Kim, and Chul-Joo Kim. 2019. Lessons learned from using a deep tree-based model for software defect prediction in practice. In *2019 IEEE/ACM 16th international conference on mining software repositories (MSR)*. IEEE, 46–57.
[46] Hoa Khanh Dam, Truyen Tran, Trang Pham, Shien Wee Ng, John Grundy, and Aditya Ghose. 2018. Automatic feature learning for predicting vulnerable software components. *IEEE Transactions on Software Engineering* 47, 1 (2018), 67–85.
[47] Siddhartha Shankar Das, Edoardo Serra, Mahantesh Halappanavar, Alex Pothen, and Ehab Al-Shaer. 2021. V2w-bert: A framework for effective hierarchical multiclass classification of software vulnerabilities. In *2021 IEEE 8th International Conference on Data Science and Advanced Analytics (DSAA)*. IEEE, 1–12.
[48] Elizabeth Dinella, Hanjun Dai, Ziyang Li, Mayur Naik, Le Song, and Ke Wang. 2020. Hoppity: Learning graph transformations to detect and fix bugs in programs. In *International Conference on Learning Representations (ICLR)*.
[49] Yangruibo Ding, Sahil Suneja, Yunhui Zheng, Jim Laredo, Alessandro Morari, Gail Kaiser, and Baishakhi Ray. 2022. VELVET: a noVel Ensemble Learning approach to automatically locate VulnErable sTatements. In *2022 IEEE International Conference on Software Analysis, Evolution and Reengineering (SANER)*. IEEE, 959–970.
[50] Yukun Dong, Yeer Tang, Xiaotong Cheng, and Yufei Yang. 2023. DeKeDVer: A deep learning-based multi-type software vulnerability classification framework using vulnerability description and source code. *Information and Software Technology* (2023), 107290.
[51] Yukun Dong, Yeer Tang, Xiaotong Cheng, Yufei Yang, and Shuqi Wang. 2023. SedSVD: Statement-level software vulnerability detection based on Relational Graph Convolutional Network with subgraph embedding. *Information and Software Technology* 158 (2023), 107168.

[52] Min Du, Zhi Chen, Chang Liu, Rajvardhan Oak, and Dawn Song. 2019. Lifelong anomaly detection through unlearning. In *Proceedings of the 2019 ACM SIGSAC conference on computer and communications security*. 1283–1297.

[53] Min Du, Feifei Li, Guineng Zheng, and Vivek Srikumar. 2017. Deeplog: Anomaly detection and diagnosis from system logs through deep learning. In *Proceedings of the 2017 ACM SIGSAC conference on computer and communications security*. 1285–1298.

[54] Xiaoning Du, Bihuan Chen, Yuekang Li, Jianmin Guo, Yaqin Zhou, Yang Liu, and Yu Jiang. 2019. Leopard: Identifying vulnerable code for vulnerability assessment through program metrics. In *2019 IEEE/ACM 41st International Conference on Software Engineering (ICSE)*. IEEE, 60–71.

[55] Benjamin G Edelman, James Bono, Sida Peng, Roberto Rodriguez, and Sandra Ho. 2023. Randomized Controlled Trial for Microsoft Security Copilot. *Available at SSRN 4648700* (2023).

[56] Abdulmotaleb El Saddik. 2018. Digital twins: The convergence of multimedia technologies. *IEEE multimedia* 25, 2 (2018), 87–92.

[57] Floris MA Erich, Chintan Amrit, and Maya Daneva. 2017. A qualitative study of DevOps usage in practice. *Journal of software: Evolution and Process* 29, 6 (2017), e1885.

[58] Jiahao Fan, Yi Li, Shaohua Wang, and Tien N Nguyen. 2020. AC/C++ code vulnerability dataset with code changes and CVE summaries. In *Proceedings of the 17th International Conference on Mining Software Repositories*. 508–512.

[59] Zhangyin Feng, Daya Guo, Duyu Tang, Nan Duan, Xiaocheng Feng, Ming Gong, Linjun Shou, Bing Qin, Ting Liu, Daxin Jiang, et al. 2020. CodeBERT: A Pre-Trained Model for Programming and Natural Languages. In *Findings of the Association for Computational Linguistics: EMNLP 2020*. 1536–1547.

[60] Jerome H Friedman and Bogdan E Popescu. 2008. Predictive learning via rule ensembles. (2008).

[61] Michael Fu, Van Nguyen, Chakkrit Tantithamthavorn, Dinh Phung, and Trung Le. 2023. Vision Transformer-Inspired Automated Vulnerability Repair. *ACM Transactions on Software Engineering and Methodology* (2023).

[62] Michael Fu, Van Nguyen, Chakkrit Kla Tantithamthavorn, Trung Le, and Dinh Phung. 2023. VulExplainer: A Transformer-based Hierarchical Distillation for Explaining Vulnerability Types. *IEEE Transactions on Software Engineering* (2023).

[63] Michael Fu and Chakkrit Tantithamthavorn. 2022. GPT2SP: A transformer-based agile story point estimation approach. *IEEE Transactions on Software Engineering* 49, 2 (2022), 611–625.

[64] Michael Fu and Chakkrit Tantithamthavorn. 2022. Linevul: A transformer-based line-level vulnerability prediction. In *Proceedings of the 19th International Conference on Mining Software Repositories*. 608–620.

[65] Michael Fu, Chakkrit Tantithamthavorn, Trung Le, Yuki Kume, Van Nguyen, Dinh Phung, and John Grundy. 2024. AIBugHunter: A Practical tool for predicting, classifying and repairing software vulnerabilities. *Empirical Software Engineering* 29, 1 (2024), 4.

[66] Michael Fu, Chakkrit Tantithamthavorn, Trung Le, Van Nguyen, and Dinh Phung. 2022. VulRepair: a T5-based automated software vulnerability repair. In *Proceedings of the 30th ACM Joint European Software Engineering Conference and Symposium on the Foundations of Software Engineering*. 935–947.

[67] Malcom Gethers, Bogdan Dit, Huzefa Kagdi, and Denys Poshyvanyk. 2012. Integrated impact analysis for managing software changes. In *2012 34th International Conference on Software Engineering (ICSE)*. IEEE, 430–440.

[68] GitHub. 2024. About autofix for CodeQL code scanning. https://docs.github.com/en/code-security/code-scanning/managing-code-scanning-alerts/about-autofix-for-codeql-code-scanning#supported-languages.

[69] GitHub. 2024. Dependabot. https://github.com/dependabot.

[70] GitLab. 2024. European tech company Cube drives secure software with AI in GitLab Duo. https://about.gitlab.com/customers/cube/.

[71] Google. 2023. 2023 State of DevOps Report. https://cloud.google.com/devops/state-of-devops.

[72] Alicja Gosiewska and Przemyslaw Biecek. 2019. IBreakDown: Uncertainty of model explanations for non-additive predictive models. *arXiv preprint arXiv:1903.11420* (2019).

[73] GuardRails. 2023. From Security Last to DevSecOps: The Driving Forces Behind the Transformation. https://www.guardrails.io/blog/from-security-last-to-devsecops-the-driving-forces-behind-the-transformation/.

[74] Michele Guerriero, Martin Garriga, Damian A Tamburri, and Fabio Palomba. 2019. Adoption, support, and challenges of infrastructure-as-code: Insights from industry. In *2019 IEEE international conference on software maintenance and evolution (ICSME)*. IEEE, 580–589.

[75] Huong Ha and Hongyu Zhang. 2019. DeepPerf: Performance prediction for configurable software with deep sparse neural network. In *2019 IEEE/ACM 41st International Conference on Software Engineering (ICSE)*. IEEE, 1095–1106.

[76] Tanja Hagemann and Katerina Katsarou. 2020. A systematic review on anomaly detection for cloud computing environments. In *Proceedings of the 2020 3rd Artificial Intelligence and Cloud Computing Conference*. 83–96.

[77] Susan Hall. 2024. AI-Assisted Dependency Updates without Breaking Things. https://thenewstack.io/ai-assisted-dependency-updates-without-breaking-things/.

[78] Dongqi Han, Zhiliang Wang, Wenqi Chen, Kai Wang, Rui Yu, Su Wang, Han Zhang, Zhihua Wang, Minghui Jin, Jiahai Yang, et al. 2023. Anomaly Detection in the Open World: Normality Shift Detection, Explanation, and Adaptation. In *30th Annual Network and Distributed System Security Symposium (NDSS)*.

[79] Dongqi Han, Zhiliang Wang, Wenqi Chen, Ying Zhong, Su Wang, Han Zhang, Jiahai Yang, Xingang Shi, and Xia Yin. 2021. Deepaid: Interpreting and improving deep learning-based anomaly detection in security applications. In *Proceedings of the 2021 ACM SIGSAC Conference on Computer and Communications Security*. 3197–3217.

[80] Shangbin Han, Qianhong Wu, Han Zhang, Bo Qin, Jiankun Hu, Xingang Shi, Linfeng Liu, and Xia Yin. 2021. Log-based anomaly detection with robust feature extraction and online learning. *IEEE Transactions on Information Forensics and Security* 16 (2021), 2300–2311.

[81] Quinn Hanam, Fernando S de M Brito, and Ali Mesbah. 2016. Discovering bug patterns in JavaScript. In *Proceedings of the 2016 24th ACM SIGSOFT international symposium on foundations of software engineering*. 144–156.

[82] Sichong Hao, Xianjun Shi, Hongwei Liu, and Yanjun Shu. 2023. Enhancing Code Language Models for Program Repair by Curricular Fine-tuning Framework. In *2023 IEEE International Conference on Software Maintenance and Evolution (ICSME)*. IEEE, 136–146.

[83] Mubin Ul Haque, M Mehdi Kholoosi, and M Ali Babar. 2022. KGSecConfig: A Knowledge Graph Based Approach for Secured Container Orchestrator Configuration. In *2022 IEEE International Conference on Software Analysis, Evolution and Reengineering (SANER)*. IEEE, 420–431.

[84] A.W. Harzing. 2007. Publish or Perish. Available online at https://harzing.com/resources/publish-or-perish.

[85] Shilin He, Jieming Zhu, Pinjia He, and Michael R Lyu. 2020. Loghub: A large collection of system log datasets towards automated log analytics. *arXiv e-prints* (2020), arXiv–2008.

[86] David Hin, Andrey Kan, Huaming Chen, and M Ali Babar. 2022. LineVD: Statement-level vulnerability detection using graph neural networks. In *Proceedings of the 19th International Conference on Mining Software Repositories*. 596–607.

[87] Jonathan Ho, Ajay Jain, and Pieter Abbeel. 2020. Denoising diffusion probabilistic models. *Advances in neural information processing systems* 33 (2020), 6840–6851.

[88] Michael Howard and Steve Lipner. 2006. *The security development lifecycle*. Vol. 8. Microsoft Press Redmond.

[89] Michael A Howard. 2006. A process for performing security code reviews. *IEEE Security & privacy* 4, 4 (2006), 74–79.

[90] Jun Huang, Yang Yang, Hang Yu, Jianguo Li, and Xiao Zheng. 2023. Twin Graph-Based Anomaly Detection via Attentive Multi-Modal Learning for Microservice System. In *2023 38th IEEE/ACM International Conference on Automated Software Engineering (ASE)*. IEEE, 66–78.

[91] Kai Huang, Xiangxin Meng, Jian Zhang, Yang Liu, Wenjie Wang, Shuhao Li, and Yuqing Zhang. 2023. An Empirical Study on Fine-Tuning Large Language Models of Code for Automated Program Repair. In *2023 38th IEEE/ACM International Conference on Automated Software Engineering (ASE)*. IEEE, 1162–1174.

[92] Peng Huang, William J Bolosky, Abhishek Singh, and Yuanyuan Zhou. 2015. Confvalley: A systematic configuration validation framework for cloud services. In *Proceedings of the Tenth European Conference on Computer Systems*. 1–16.

[93] Jez Humble and David Farley. 2010. *Continuous delivery: reliable software releases through build, test, and deployment automation*. Pearson Education.

[94] Mariam Ibrahim and Ruba Elhafiz. 2023. Security analysis of cyber-physical systems using reinforcement learning. *Sensors* 23, 3 (2023), 1634.

[95] Nan Jiang, Thibaud Lutellier, and Lin Tan. 2021. Cure: Code-aware neural machine translation for automatic program repair. In *2021 IEEE/ACM 43rd International Conference on Software Engineering (ICSE)*. IEEE, 1161–1173.

[96] Jirayus Jiarpakdee, Chakkrit Kla Tantithamthavorn, Hoa Khanh Dam, and John Grundy. 2020. An empirical study of model-agnostic techniques for defect prediction models. *IEEE Transactions on Software Engineering* 48, 1 (2020), 166–185.

[97] Matthew Jin, Syed Shahriar, Michele Tufano, Xin Shi, Shuai Lu, Neel Sundaresan, and Alexey Svyatkovskiy. 2023. Inferfix: End-to-end program repair with llms. *arXiv preprint arXiv:2303.07263* (2023).

[98] René Just, Darioush Jalali, and Michael D Ernst. 2014. Defects4J: A database of existing faults to enable controlled testing studies for Java programs. In *Proceedings of the 2014 international symposium on software testing and analysis*. 437–440.

[99] Yasutaka Kamei, Emad Shihab, Bram Adams, Ahmed E Hassan, Audris Mockus, Anand Sinha, and Naoyasu Ubayashi. 2012. A large-scale empirical study of just-in-time quality assurance. *IEEE Transactions on Software Engineering* 39, 6 (2012), 757–773.

[100] Rafael-Michael Karampatsis and Charles Sutton. 2020. How often do single-statement bugs occur? the manysstubs4j dataset. In *Proceedings of the 17th International Conference on Mining Software Repositories*. 573–577.

[101] Staffs Keele et al. 2007. Guidelines for performing systematic literature reviews in software engineering.

[102] Chaiyakarn Khanan, Worawit Luewichana, Krissakorn Pruktharathikoon, Jirayus Jiarpakdee, Chakkrit Tantithamthavorn, Morakot Choetkiertikul, Chaiyong Ragkhitwetsagul, and Thanwadee Sunetnanta. 2020. JITBot: an explainable

just-in-time defect prediction bot. In *Proceedings of the 35th IEEE/ACM international conference on automated software engineering*. 1336–1339.

[103] M Mehdi Kholoosi, M Ali Babar, and Cemal Yilmaz. 2023. Empirical Analysis of Software Vulnerabilities Causing Timing Side Channels. In *2023 IEEE Conference on Communications and Network Security (CNS)*. IEEE, 1–9.

[104] Wael Khreich, Babak Khosravifar, Abdelwahab Hamou-Lhadj, and Chamseddine Talhi. 2017. An anomaly detection system based on variable N-gram features and one-class SVM. *Information and Software Technology* 91 (2017), 186–197.

[105] Anna Khrupa. 2022. What Is Software Impact Analysis. https://qarea.com/blog/what-is-software-impact-analysis.

[106] Barbara Kitchenham, Lech Madeyski, and David Budgen. 2022. SEGRESS: Software engineering guidelines for reporting secondary studies. *IEEE Transactions on Software Engineering* 49, 3 (2022), 1273–1298.

[107] Kun Kuang, Lian Li, Zhi Geng, Lei Xu, Kun Zhang, Beishui Liao, Huaxin Huang, Peng Ding, Wang Miao, and Zhichao Jiang. 2020. Causal inference. *Engineering* 6, 3 (2020), 253–263.

[108] Jinpeng Lan, Lina Gong, Jingxuan Zhang, and Haoxiang Zhang. 2023. BTLink: automatic link recovery between issues and commits based on pre-trained BERT model. *Empirical Software Engineering* 28, 4 (2023), 103.

[109] Stuart Larsen. 2024. MongoDB security team enables secure development with Snyk. https://snyk.io/case-studies/mongodb/.

[110] Van-Hoang Le and Hongyu Zhang. 2021. Log-based anomaly detection without log parsing. In *2021 36th IEEE/ACM International Conference on Automated Software Engineering (ASE)*. IEEE, 492–504.

[111] Claire Le Goues, Neal Holtschulte, Edward K Smith, Yuriy Brun, Premkumar Devanbu, Stephanie Forrest, and Westley Weimer. 2015. The ManyBugs and IntroClass benchmarks for automated repair of C programs. *IEEE Transactions on Software Engineering* 41, 12 (2015), 1236–1256.

[112] David LeBlanc and Michael Howard. 2002. *Writing secure code*. Pearson Education.

[113] Cheryl Lee, Tianyi Yang, Zhuangbin Chen, Yuxin Su, and Michael R Lyu. 2023. Maat: Performance Metric Anomaly Anticipation for Cloud Services with Conditional Diffusion. In *2023 38th IEEE/ACM International Conference on Automated Software Engineering (ASE)*. IEEE, 116–128.

[114] Tiina Leppänen, Anne Honkaranta, and Andrei Costin. 2022. Trends for the DevOps security. A systematic literature review. In *International Symposium on Business Modeling and Software Design*. Springer, 200–217.

[115] Jingyue Li, Reidar Conradi, Christian Bunse, Marco Torchiano, Odd Petter N Slyngstad, and Maurizio Morisio. 2009. Development with off-the-shelf components: 10 facts. *IEEE software* 26, 2 (2009), 80–87.

[116] Jiangming Li, Huasen He, Shuangwu Chen, and Dong Jin. 2023. LogGraph: Log Event Graph Learning Aided Robust Fine-Grained Anomaly Diagnosis. *IEEE Transactions on Dependable and Secure Computing* (2023).

[117] Weiwei Li, Wenzhou Zhang, Xiuyi Jia, and Zhiqiu Huang. 2020. Effort-aware semi-supervised just-in-time defect prediction. *Information and Software Technology* 126 (2020), 106364.

[118] Xiaoyun Li, Pengfei Chen, Linxiao Jing, Zilong He, and Guangba Yu. 2022. SwissLog: Robust anomaly detection and localization for interleaved unstructured logs. *IEEE Transactions on Dependable and Secure Computing* (2022).

[119] Xiangwei Li, Xiaoning Ren, Yinxing Xue, Zhenchang Xing, and Jiamou Sun. 2023. Prediction of Vulnerability Characteristics Based on Vulnerability Description and Prompt Learning. In *2023 IEEE International Conference on Software Analysis, Evolution and Reengineering (SANER)*. IEEE, 604–615.

[120] Yi Li, Shaohua Wang, and Tien N Nguyen. 2020. Dlfix: Context-based code transformation learning for automated program repair. In *Proceedings of the ACM/IEEE 42nd International Conference on Software Engineering*. 602–614.

[121] Yi Li, Shaohua Wang, and Tien N Nguyen. 2021. Vulnerability detection with fine-grained interpretations. In *Proceedings of the 29th ACM Joint Meeting on European Software Engineering Conference and Symposium on the Foundations of Software Engineering*. 292–303.

[122] Zhen Li, Deqing Zou, Shouhuai Xu, Zhaoxuan Chen, Yawei Zhu, and Hai Jin. 2021. Vuldeelocator: a deep learning-based fine-grained vulnerability detector. *IEEE Transactions on Dependable and Secure Computing* 19, 4 (2021), 2821–2837.

[123] Zhen Li, Deqing Zou, Shouhuai Xu, Xinyu Ou, Hai Jin, Sujuan Wang, Zhijun Deng, and Yuyi Zhong. 2018. Vuldeepecker: A deep learning-based system for vulnerability detection. *arXiv preprint arXiv:1801.01681* (2018).

[124] Derrick Lin, James Koppel, Angela Chen, and Armando Solar-Lezama. 2017. QuixBugs: A multi-lingual program repair benchmark set based on the Quixey Challenge. In *Proceedings Companion of the 2017 ACM SIGPLAN international conference on systems, programming, languages, and applications: software for humanity*. 55–56.

[125] Qin Lin, Sridha Adepu, Sicco Verwer, and Aditya Mathur. 2018. TABOR: A graphical model-based approach for anomaly detection in industrial control systems. In *Proceedings of the 2018 on asia conference on computer and communications security*. 525–536.

[126] Tsung-Yi Lin, Priya Goyal, Ross Girshick, Kaiming He, and Piotr Dollár. 2017. Focal loss for dense object detection. In *Proceedings of the IEEE international conference on computer vision*. 2980–2988.

[127] Fei Tony Liu, Kai Ming Ting, and Zhi-Hua Zhou. 2012. Isolation-based anomaly detection. *ACM Transactions on Knowledge Discovery from Data (TKDD)* 6, 1 (2012), 1–39.
[128] Shigang Liu, Guanjun Lin, Lizhen Qu, Jun Zhang, Olivier De Vel, Paul Montague, and Yang Xiang. 2020. CD-VulD: Cross-domain vulnerability discovery based on deep domain adaptation. *IEEE Transactions on Dependable and Secure Computing* 19, 1 (2020), 438–451.
[129] Shuai Lu, Daya Guo, Shuo Ren, Junjie Huang, Alexey Svyatkovskiy, Ambrosio Blanco, Colin Clement, Dawn Drain, Daxin Jiang, Duyu Tang, et al. 2021. Codexglue: A machine learning benchmark dataset for code understanding and generation. *arXiv preprint arXiv:2102.04664* (2021).
[130] Scott M Lundberg and Su-In Lee. 2017. A unified approach to interpreting model predictions. In *Proceedings of the 31st international conference on neural information processing systems*. 4768–4777.
[131] Thibaud Lutellier, Hung Viet Pham, Lawrence Pang, Yitong Li, Moshi Wei, and Lin Tan. 2020. Coconut: combining context-aware neural translation models using ensemble for program repair. In *Proceedings of the 29th ACM SIGSOFT international symposium on software testing and analysis*. 101–114.
[132] Guolin Lyu and Robert W Brennan. 2023. Multi-agent modelling of cyber-physical systems for IEC 61499-based distributed intelligent automation. *International Journal of Computer Integrated Manufacturing* (2023), 1–27.
[133] Andi Mann, Michael Stahnke, Alanna Brown, and Nigel Kersten. 2019. State of DevOps Report, 2019.
[134] Chengzhi Mao, Ziyuan Zhong, Junfeng Yang, Carl Vondrick, and Baishakhi Ray. 2019. Metric learning for adversarial robustness. *Advances in neural information processing systems* 32 (2019).
[135] Runfeng Mao, He Zhang, Qiming Dai, Huang Huang, Guoping Rong, Haifeng Shen, Lianping Chen, and Kaixiang Lu. 2020. Preliminary findings about devsecops from grey literature. In *2020 IEEE 20th international conference on software quality, reliability and security (QRS)*. IEEE, 450–457.
[136] Diego Marcilio, Carlo A Furia, Rodrigo Bonifácio, and Gustavo Pinto. 2020. SpongeBugs: Automatically generating fix suggestions in response to static code analysis warnings. *Journal of Systems and Software* 168 (2020), 110671.
[137] Ehsan Mashhadi and Hadi Hemmati. 2021. Applying codebert for automated program repair of java simple bugs. In *2021 IEEE/ACM 18th International Conference on Mining Software Repositories (MSR)*. IEEE, 505–509.
[138] Aditya P Mathur and Nils Ole Tippenhauer. 2016. SWaT: A water treatment testbed for research and training on ICS security. In *2016 international workshop on cyber-physical systems for smart water networks (CySWater)*. IEEE, 31–36.
[139] Taylor McCaslin. 2024. How to put generative AI to work in your DevSecOps environment. https://about.gitlab.com/blog/2024/03/07/how-to-put-generative-ai-to-work-in-your-devsecops-environment/.
[140] Shane McIntosh and Yasutaka Kamei. 2018. Are fix-inducing changes a moving target? a longitudinal case study of just-in-time defect prediction. In *Proceedings of the 40th international conference on software engineering*. 560–560.
[141] Sergey Mechtaev, Jooyong Yi, and Abhik Roychoudhury. 2016. Angelix: Scalable multiline program patch synthesis via symbolic analysis. In *Proceedings of the 38th international conference on software engineering*. 691–701.
[142] Sonu Mehta, Ranjita Bhagwan, Rahul Kumar, Chetan Bansal, Chandra Maddila, Balasubramanyan Ashok, Sumit Asthana, Christian Bird, and Aditya Kumar. 2020. Rex: Preventing bugs and misconfiguration in large services using correlated change analysis. In *17th USENIX Symposium on Networked Systems Design and Implementation (NSDI 20)*. 435–448.
[143] Weibin Meng, Ying Liu, Yichen Zhu, Shenglin Zhang, Dan Pei, Yuqing Liu, Yihao Chen, Ruizhi Zhang, Shimin Tao, Pei Sun, et al. 2019. Loganomaly: Unsupervised detection of sequential and quantitative anomalies in unstructured logs.. In *IJCAI*, Vol. 19. 4739–4745.
[144] Aditya Krishna Menon, Sadeep Jayasumana, Ankit Singh Rawat, Himanshu Jain, Andreas Veit, and Sanjiv Kumar. 2020. Long-tail learning via logit adjustment. *arXiv preprint arXiv:2007.07314* (2020).
[145] Ibrahim Mesecan, Daniel Blackwell, David Clark, Myra B Cohen, and Justyna Petke. 2021. HyperGI: Automated detection and repair of information flow leakage. In *2021 36th IEEE/ACM International Conference on Automated Software Engineering (ASE)*. IEEE, 1358–1362.
[146] Microsoft. 2022. DevSecOps controls. https://learn.microsoft.com/en-us/azure/cloud-adoption-framework/secure/devsecops-controls.
[147] Microsoft. 2022. What is infrastructure as code (IaC)? https://learn.microsoft.com/en-us/devops/deliver/what-is-infrastructure-as-code.
[148] Microsoft. 2023. What is DevOps? https://learn.microsoft.com/en-us/devops/what-is-devops.
[149] Microsoft. 2024. Microsoft Copilot for Security is generally available on April 1, 2024, with new capabilities. https://www.microsoft.com/en-us/security/blog/2024/03/13/microsoft-copilot-for-security-is-generally-available-on-april-1-2024-with-new-capabilities/.
[150] Microsoft. 2024. Threat Modeling. https://www.microsoft.com/en-us/securityengineering/sdl/threatmodeling.
[151] Yisroel Mirsky, George Macon, Michael Brown, Carter Yagemann, Matthew Pruett, Evan Downing, Sukarno Mertoguno, and Wenke Lee. 2023. VulChecker: Graph-based Vulnerability Localization in Source Code. In *31st USENIX Security Symposium, Security 2022*.

[152] Thomas H Morris, Zach Thornton, and Ian Turnipseed. 2015. Industrial control system simulation and data logging for intrusion detection system research. *7th annual southeastern cyber security summit* (2015), 3–4.
[153] Nour Moustafa and Jill Slay. 2015. UNSW-NB15: a comprehensive data set for network intrusion detection systems (UNSW-NB15 network data set). In *2015 military communications and information systems conference (MilCIS)*. IEEE, 1–6.
[154] Håvard Myrbakken and Ricardo Colomo-Palacios. 2017. DevSecOps: a multivocal literature review. In *Software Process Improvement and Capability Determination: 17th International Conference, SPICE 2017, Palma de Mallorca, Spain, October 4–5, 2017, Proceedings*. Springer, 17–29.
[155] Rennie Naidoo and Nicolaas Möller. 2022. Building Software Applications Securely With DevSecOps: A Socio-Technical Perspective. In *ECCWS 2022 21st European Conference on Cyber Warfare and Security*. Academic Conferences and publishing limited.
[156] Marjane Namavar, Noor Nashid, and Ali Mesbah. 2022. A controlled experiment of different code representations for learning-based program repair. *Empirical Software Engineering* 27, 7 (2022), 190.
[157] Chao Ni, Kaiwen Yang, Yan Zhu, Xiang Chen, and Xiaohu Yang. 2023. Unifying Defect Prediction, Categorization, and Repair by Multi-Task Deep Learning. In *2023 38th IEEE/ACM International Conference on Automated Software Engineering (ASE)*. IEEE, 1980–1992.
[158] NSF. 2024. Cyber-physical systems. https://www.nsf.gov/news/special_reports/cyber-physical/.
[159] Adam Oliner and Jon Stearley. 2007. What supercomputers say: A study of five system logs. In *37th annual IEEE/IFIP international conference on dependable systems and networks (DSN'07)*. IEEE, 575–584.
[160] Yassine Ouali, Céline Hudelot, and Myriam Tami. 2020. An overview of deep semi-supervised learning. *arXiv preprint arXiv:2006.05278* (2020).
[161] Shengyi Pan, Lingfeng Bao, Xin Xia, David Lo, and Shanping Li. 2023. Fine-grained commit-level vulnerability type prediction by CWE tree structure. In *2023 IEEE/ACM 45th International Conference on Software Engineering (ICSE)*. IEEE, 957–969.
[162] Luca Pascarella, Fabio Palomba, and Alberto Bacchelli. 2019. Fine-grained just-in-time defect prediction. *Journal of Systems and Software* 150 (2019), 22–36.
[163] Ivan Pashchenko, Henrik Plate, Serena Elisa Ponta, Antonino Sabetta, and Fabio Massacci. 2018. Vulnerable open source dependencies: Counting those that matter. In *Proceedings of the 12th ACM/IEEE international symposium on empirical software engineering and measurement*. 1–10.
[164] Hammond Pearce, Benjamin Tan, Baleegh Ahmad, Ramesh Karri, and Brendan Dolan-Gavitt. 2023. Examining zero-shot vulnerability repair with large language models. In *2023 IEEE Symposium on Security and Privacy (SP)*. IEEE, 2339–2356.
[165] Serena Elisa Ponta, Henrik Plate, Antonino Sabetta, Michele Bezzi, and Cédric Dangremont. 2019. A manually-curated dataset of fixes to vulnerabilities of open-source software. In *2019 IEEE/ACM 16th International Conference on Mining Software Repositories (MSR)*. IEEE, 383–387.
[166] Chanathip Pornprasit, Chakkrit Tantithamthavorn, Jirayus Jiarpakdee, Michael Fu, and Patanamon Thongtanunam. 2021. Pyexplainer: Explaining the predictions of just-in-time defect models. In *2021 36th IEEE/ACM International Conference on Automated Software Engineering (ASE)*. IEEE, 407–418.
[167] Chanathip Pornprasit and Chakkrit Kla Tantithamthavorn. 2021. JITLine: A simpler, better, faster, finer-grained just-in-time defect prediction. In *2021 IEEE/ACM 18th International Conference on Mining Software Repositories (MSR)*. IEEE, 369–379.
[168] Chanathip Pornprasit and Chakkrit Kla Tantithamthavorn. 2022. Deeplinedp: Towards a deep learning approach for line-level defect prediction. *IEEE Transactions on Software Engineering* 49, 1 (2022), 84–98.
[169] Gede Artha Azriadi Prana, Abhishek Sharma, Lwin Khin Shar, Darius Foo, Andrew E Santosa, Asankhaya Sharma, and David Lo. 2021. Out of sight, out of mind? How vulnerable dependencies affect open-source projects. *Empirical Software Engineering* 26 (2021), 1–34.
[170] Luís Prates, João Faustino, Miguel Silva, and Rúben Pereira. 2019. Devsecops metrics. In *Information Systems: Research, Development, Applications, Education: 12th SIGSAND/PLAIS EuroSymposium 2019, Gdansk, Poland, September 19, 2019, Proceedings 12*. Springer, 77–90.
[171] Fangcheng Qiu, Zhipeng Gao, Xin Xia, David Lo, John Grundy, and Xinyu Wang. 2021. Deep just-in-time defect localization. *IEEE Transactions on Software Engineering* 48, 12 (2021), 5068–5086.
[172] Fangcheng Qiu, Meng Yan, Xin Xia, Xinyu Wang, Yuanrui Fan, Ahmed E Hassan, and David Lo. 2020. JITO: a tool for just-in-time defect identification and localization. In *Proceedings of the 28th ACM joint meeting on european software engineering conference and symposium on the foundations of software engineering*. 1586–1590.
[173] Saima Rafi, Wu Yu, Muhammad Azeem Akbar, Ahmed Alsanad, and Abdu Gumaei. 2020. Prioritization based taxonomy of DevOps security challenges using PROMETHEE. *IEEE Access* 8 (2020), 105426–105446.

[174] Akond Rahman, Chris Parnin, and Laurie Williams. 2019. The seven sins: Security smells in infrastructure as code scripts. In *2019 IEEE/ACM 41st International Conference on Software Engineering (ICSE)*. IEEE, 164–175.
[175] Akond Rahman and Laurie Williams. 2018. Characterizing defective configuration scripts used for continuous deployment. In *2018 IEEE 11th International conference on software testing, verification and validation (ICST)*. IEEE, 34–45.
[176] Akond Rahman and Laurie Williams. 2019. Source code properties of defective infrastructure as code scripts. *Information and Software Technology* 112 (2019), 148–163.
[177] Roshan N Rajapakse, Mansooreh Zahedi, M Ali Babar, and Haifeng Shen. 2022. Challenges and solutions when adopting DevSecOps: A systematic review. *Information and software technology* 141 (2022), 106700.
[178] RedHat. 2022. What is Infrastructure as Code (IaC)? https://www.redhat.com/en/topics/automation/what-is-infrastructure-as-code-iac.
[179] RedHat. 2023. What is CI/CD security? https://www.redhat.com/en/topics/security/what-is-cicd-security.
[180] Yahoo Research. 2015. A Benchmark Dataset for Time Series Anomaly Detection. https://yahooresearch.tumblr.com/post/114590420346/a-benchmark-dataset-for-time-series-anomaly.
[181] Marco Tulio Ribeiro, Sameer Singh, and Carlos Guestrin. 2016. " Why should i trust you?" Explaining the predictions of any classifier. In *Proceedings of the 22nd ACM SIGKDD international conference on knowledge discovery and data mining*. 1135–1144.
[182] Maria Rosala. 2019. How to analyze qualitative data from UX research: Thematic analysis. *NN-Nielsen Norman Group* (2019).
[183] Baptiste Roziere, Jonas Gehring, Fabian Gloeckle, Sten Sootla, Itai Gat, Xiaoqing Ellen Tan, Yossi Adi, Jingyu Liu, Tal Remez, Jérémy Rapin, et al. 2023. Code llama: Open foundation models for code. *arXiv preprint arXiv:2308.12950* (2023).
[184] Hang Ruan, Bihuan Chen, Xin Peng, and Wenyun Zhao. 2019. DeepLink: Recovering issue-commit links based on deep learning. *Journal of Systems and Software* 158 (2019), 110406.
[185] Rebecca Russell, Louis Kim, Lei Hamilton, Tomo Lazovich, Jacob Harer, Onur Ozdemir, Paul Ellingwood, and Marc McConley. 2018. Automated vulnerability detection in source code using deep representation learning. In *2018 17th IEEE international conference on machine learning and applications (ICMLA)*. IEEE, 757–762.
[186] Ripon K Saha, Yingjun Lyu, Wing Lam, Hiroaki Yoshida, and Mukul R Prasad. 2018. Bugs. jar: A large-scale, diverse dataset of real-world java bugs. In *Proceedings of the 15th international conference on mining software repositories*. 10–13.
[187] Mary Sánchez-Gordón and Ricardo Colomo-Palacios. 2020. Security as culture: a systematic literature review of DevSecOps. In *Proceedings of the IEEE/ACM 42nd International Conference on Software Engineering Workshops*. 266–269.
[188] Carla Sauvanaud, Mohamed Kaâniche, Karama Kanoun, Kahina Lazri, and Guthemberg Da Silva Silvestre. 2018. Anomaly detection and diagnosis for cloud services: Practical experiments and lessons learned. *Journal of Systems and Software* 139 (2018), 84–106.
[189] Alexandra Sbaraini, Stacy M Carter, R Wendell Evans, and Anthony Blinkhorn. 2011. How to do a grounded theory study: a worked example of a study of dental practices. *BMC medical research methodology* 11 (2011), 1–10.
[190] Riccardo Scandariato, James Walden, Aram Hovsepyan, and Wouter Joosen. 2014. Predicting vulnerable software components via text mining. *IEEE Transactions on Software Engineering* 40, 10 (2014), 993–1006.
[191] Thomas Schlegl, Philipp Seeböck, Sebastian M Waldstein, Ursula Schmidt-Erfurth, and Georg Langs. 2017. Unsupervised anomaly detection with generative adversarial networks to guide marker discovery. In *International conference on information processing in medical imaging*. Springer, 146–157.
[192] Abigail See, Peter J Liu, and Christopher D Manning. 2017. Get to the point: Summarization with pointer-generator networks. *arXiv preprint arXiv:1704.04368* (2017).
[193] Adriana Sejfia, Satyaki Das, Saad Shafiq, and Nenad Medvidović. 2023. Toward Improved Deep Learning-based Vulnerability Detection. In *2024 IEEE/ACM 46th International Conference on Software Engineering (ICSE)*. IEEE Computer Society, 730–741.
[194] Mojtaba Shahin, Mansooreh Zahedi, Muhammad Ali Babar, and Liming Zhu. 2019. An empirical study of architecting for continuous delivery and deployment. *Empirical Software Engineering* 24 (2019), 1061–1108.
[195] Chenze Shao and Yang Feng. 2022. Overcoming catastrophic forgetting beyond continual learning: Balanced training for neural machine translation. *arXiv preprint arXiv:2203.03910* (2022).
[196] Yonghee Shin and Laurie Williams. 2013. Can traditional fault prediction models be used for vulnerability prediction? *Empirical Software Engineering* 18 (2013), 25–59.
[197] Adam Shostack. 2014. *Threat modeling: Designing for security.* John Wiley & Sons.
[198] Avanti Shrikumar, Peyton Greenside, and Anshul Kundaje. 2017. Learning important features through propagating activation differences. In *International Conference on Machine Learning*. PMLR, 3145–3153.

[199] Yangyang Shu, Yulei Sui, Hongyu Zhang, and Guandong Xu. 2020. Perf-AL: Performance prediction for configurable software through adversarial learning. In *Proceedings of the 14th ACM/IEEE International Symposium on Empirical Software Engineering and Measurement (ESEM)*. 1–11.

[200] Mohammed Latif Siddiq, Md Rezwanur Rahman Jahin, Mohammad Rafid Ul Islam, Rifat Shahriyar, and Anindya Iqbal. 2021. Sqlifix: Learning based approach to fix sql injection vulnerabilities in source code. In *2021 IEEE International Conference on Software Analysis, Evolution and Reengineering (SANER)*. IEEE, 354–364.

[201] Norbert Siegmund, Alexander Grebhahn, Sven Apel, and Christian Kästner. 2015. Performance-influence models for highly configurable systems. In *Proceedings of the 2015 10th Joint Meeting on Foundations of Software Engineering*. 284–294.

[202] Snyk. 2015. Snyk. https://snyk.io.

[203] Snyk. 2022. Software dependencies: How to manage dependencies at scale. https://snyk.io/series/open-source-security/software-dependencies/.

[204] Sonatype. 2024. Sonatype. https://www.sonatype.com/.

[205] Daniel Stahl, Torvald Martensson, and Jan Bosch. 2017. Continuous practices and devops: beyond the buzz, what does it all mean?. In *2017 43rd Euromicro Conference on Software Engineering and Advanced Applications (SEAA)*. IEEE, 440–448.

[206] Jon Stearley and Adam J Oliner. 2008. Bad words: Finding faults in spirit's syslogs. In *2008 Eighth IEEE International Symposium on Cluster Computing and the Grid (CCGRID)*. IEEE, 765–770.

[207] Benjamin Steenhoek, Hongyang Gao, and Wei Le. 2023. Dataflow Analysis-Inspired Deep Learning for Efficient Vulnerability Detection. In *2024 IEEE/ACM 46th International Conference on Software Engineering (ICSE)*. IEEE Computer Society, 166–178.

[208] Benjamin Steenhoek, Md Mahbubur Rahman, Richard Jiles, and Wei Le. 2023. An empirical study of deep learning models for vulnerability detection. In *2023 IEEE/ACM 45th International Conference on Software Engineering (ICSE)*. IEEE, 2237–2248.

[209] Hudan Studiawan, Ferdous Sohel, and Christian Payne. 2020. Anomaly detection in operating system logs with deep learning-based sentiment analysis. *IEEE Transactions on Dependable and Secure Computing* 18, 5 (2020), 2136–2148.

[210] Alexander Suh. 2020. Adapting bug prediction models to predict reverted commits at Wayfair. In *Proceedings of the 28th ACM Joint Meeting on European Software Engineering Conference and Symposium on the Foundations of Software Engineering*. 1251–1262.

[211] Yan Sun, Celia Chen, Qing Wang, and Barry Boehm. 2017. Improving missing issue-commit link recovery using positive and unlabeled data. In *2017 32nd IEEE/ACM International Conference on Automated Software Engineering (ASE)*. IEEE, 147–152.

[212] Yutao Sun, Li Dong, Shaohan Huang, Shuming Ma, Yuqing Xia, Jilong Xue, Jianyong Wang, and Furu Wei. 2023. Retentive Network: A Successor to Transformer for Large Language Models. arXiv:2307.08621 [cs.CL]

[213] Mukund Sundararajan, Ankur Taly, and Qiqi Yan. 2017. Axiomatic attribution for deep networks. In *International Conference on Machine Learning*. PMLR, 3319–3328.

[214] Synopsys. 2024. Black Duck Software Composition Analysis. https://www.synopsys.com/.

[215] Shin Hwei Tan, Jooyong Yi, Sergey Mechtaev, Abhik Roychoudhury, et al. 2017. Codeflaws: a programming competition benchmark for evaluating automated program repair tools. In *2017 IEEE/ACM 39th International Conference on Software Engineering Companion (ICSE-C)*. IEEE, 180–182.

[216] Riccardo Taormina, Stefano Galelli, Nils Ole Tippenhauer, Elad Salomons, Avi Ostfeld, Demetrios G Eliades, Mohsen Aghashahi, Raanju Sundararajan, Mohsen Pourahmadi, M Katherine Banks, et al. 2018. Battle of the attack detection algorithms: Disclosing cyber attacks on water distribution networks. *Journal of Water Resources Planning and Management* 144, 8 (2018), 04018048.

[217] Pierre Tempel and Eric Tooley. 2024. Found means fixed: Introducing code scanning autofix, powered by GitHub Copilot and CodeQL. https://github.blog/2024-03-20-found-means-fixed-introducing-code-scanning-autofix-powered-by-github-copilot-and-codeql/.

[218] Hailemelekot Demtse Tessema and Surafel Lemma Abebe. 2021. Enhancing just-in-time defect prediction using change request-based metrics. In *2021 IEEE International Conference on Software Analysis, Evolution and Reengineering (SANER)*. IEEE, 511–515.

[219] Testsigma. 2023. Impact Analysis In Software Testing- A Complete Overview. https://testsigma.com/blog/impact-analysis-in-testing/.

[220] Michele Tufano, Cody Watson, Gabriele Bavota, Massimiliano Di Penta, Martin White, and Denys Poshyvanyk. 2019. An empirical study on learning bug-fixing patches in the wild via neural machine translation. *ACM Transactions on Software Engineering and Methodology (TOSEM)* 28, 4 (2019), 1–29.

[221] Richard J Turver and Malcolm Munro. 1994. An early impact analysis technique for software maintenance. *Journal of Software Maintenance: Research and Practice* 6, 1 (1994), 35–52.

[222] Akshay Utture and Jens Palsberg. 2023. From Leaks to Fixes: Automated Repairs for Resource Leak Warnings. In *Proceedings of the 31st ACM Joint European Software Engineering Conference and Symposium on the Foundations of Software Engineering*. 159–171.

[223] Yolanda Valdés-Rodríguez, Jorge Hochstetter-Diez, Jaime Díaz-Arancibia, and Rodrigo Cadena-Martínez. 2023. Towards the Integration of Security Practices in Agile Software Development: A Systematic Mapping Review. *Applied Sciences* 13, 7 (2023), 4578.

[224] Validata. 2024. AI-powered Live Impact Analysis. https://www.validata-software.com/products/validata-sense-ai/ai-powered-live-impact-analysis.

[225] Ashish Vaswani, Noam Shazeer, Niki Parmar, Jakob Uszkoreit, Llion Jones, Aidan N Gomez, Łukasz Kaiser, and Illia Polosukhin. 2017. Attention is all you need. *Advances in neural information processing systems* 30 (2017).

[226] Vy Vo, Van Nguyen, Trung Le, Quan Hung Tran, Gholamreza Haffari, Seyit Camtepe, and Dinh Phung. 2022. An additive instance-wise approach to multi-class model interpretation. *arXiv preprint arXiv:2207.03113* (2022).

[227] Huanting Wang, Guixin Ye, Zhanyong Tang, Shin Hwei Tan, Songfang Huang, Dingyi Fang, Yansong Feng, Lizhong Bian, and Zheng Wang. 2020. Combining graph-based learning with automated data collection for code vulnerability detection. *IEEE Transactions on Information Forensics and Security* 16 (2020), 1943–1958.

[228] Tianyi Wang, Shengzhi Qin, and Kam Pui Chow. 2021. Towards Vulnerability Types Classification Using Pure Self-Attention: A Common Weakness Enumeration Based Approach. In *2021 IEEE 24th International Conference on Computational Science and Engineering (CSE)*. IEEE, 146–153.

[229] Wenbo Wang, Tien N Nguyen, Shaohua Wang, Yi Li, Jiyuan Zhang, and Aashish Yadavally. 2023. DeepVD: Toward Class-Separation Features for Neural Network Vulnerability Detection. In *2023 IEEE/ACM 45th International Conference on Software Engineering (ICSE)*. IEEE, 2249–2261.

[230] Xuheng Wang, Jiaxing Song, Xu Zhang, Junshu Tang, Weihe Gao, and Qingwei Lin. 2023. LogOnline: A Semi-Supervised Log-Based Anomaly Detector Aided with Online Learning Mechanism. In *2023 38th IEEE/ACM International Conference on Automated Software Engineering (ASE)*. IEEE, 141–152.

[231] Xinda Wang, Shu Wang, Kun Sun, Archer Batcheller, and Sushil Jajodia. 2020. A machine learning approach to classify security patches into vulnerability types. In *2020 IEEE Conference on Communications and Network Security (CNS)*. IEEE, 1–9.

[232] Yue Wang, Hung Le, Akhilesh Deepak Gotmare, Nghi DQ Bui, Junnan Li, and Steven CH Hoi. 2023. Codet5+: Open code large language models for code understanding and generation. *arXiv preprint arXiv:2305.07922* (2023).

[233] Yue Wang, Weishi Wang, Shafiq Joty, and Steven CH Hoi. 2021. Codet5: Identifier-aware unified pre-trained encoder-decoder models for code understanding and generation. *arXiv preprint arXiv:2109.00859* (2021).

[234] Laura Wartschinski, Yannic Noller, Thomas Vogel, Timo Kehrer, and Lars Grunske. 2022. VUDENC: vulnerability detection with deep learning on a natural codebase for Python. *Information and Software Technology* 144 (2022), 106809.

[235] Cody Watson, Nathan Cooper, David Nader Palacio, Kevin Moran, and Denys Poshyvanyk. 2022. A systematic literature review on the use of deep learning in software engineering research. *ACM Transactions on Software Engineering and Methodology (TOSEM)* 31, 2 (2022), 1–58.

[236] Supatsara Wattanakriengkrai, Patanamon Thongtanunam, Chakkrit Tantithamthavorn, Hideaki Hata, and Kenichi Matsumoto. 2020. Predicting defective lines using a model-agnostic technique. *IEEE Transactions on Software Engineering* 48, 5 (2020), 1480–1496.

[237] Jason Wei and Kai Zou. 2019. EDA: Easy Data Augmentation Techniques for Boosting Performance on Text Classification Tasks. In *Proceedings of the 2019 Conference on Empirical Methods in Natural Language Processing and the 9th International Joint Conference on Natural Language Processing (EMNLP-IJCNLP)*. 6382–6388.

[238] Ming Wen, Junjie Chen, Rongxin Wu, Dan Hao, and Shing-Chi Cheung. 2018. Context-aware patch generation for better automated program repair. In *Proceedings of the 40th international conference on software engineering*. 1–11.

[239] David A. Wheeler. 2024. Flawfinder. https://dwheeler.com/flawfinder/.

[240] Emily Rowan Winter, Vesna Nowack, David Bowes, Steve Counsell, Tracy Hall, Sæmundur Haraldsson, John Woodward, Serkan Kirbas, Etienne Windels, Olayori McBello, et al. 2022. Towards developer-centered automatic program repair: findings from Bloomberg. In *Proceedings of the 30th ACM Joint European Software Engineering Conference and Symposium on the Foundations of Software Engineering*. 1578–1588.

[241] Bozhi Wu, Shangqing Liu, Yang Xiao, Zhiming Li, Jun Sun, and Shang-Wei Lin. 2023. Learning Program Semantics for Vulnerability Detection via Vulnerability-Specific Inter-procedural Slicing. In *Proceedings of the 31st ACM Joint European Software Engineering Conference and Symposium on the Foundations of Software Engineering*. 1371–1383.

[242] Rongxin Wu, Hongyu Zhang, Sunghun Kim, and Shing-Chi Cheung. 2011. Relink: recovering links between bugs and changes. In *Proceedings of the 19th ACM SIGSOFT symposium and the 13th European conference on Foundations of software engineering*. 15–25.

[243] Xingfang Wu, Heng Li, and Foutse Khomh. 2023. On the effectiveness of log representation for log-based anomaly detection. *Empirical Software Engineering* 28, 6 (2023), 137.

[244] Yueming Wu, Deqing Zou, Shihan Dou, Wei Yang, Duo Xu, and Hai Jin. 2022. Vulcnn: An image-inspired scalable vulnerability detection system. In *Proceedings of the 44th International Conference on Software Engineering*. 2365–2376.

[245] Liang Xi, Dehua Miao, Menghan Li, Ruidong Wang, Han Liu, and Xunhua Huang. 2023. Adaptive-Correlation-aware Unsupervised Deep Learning for Anomaly Detection in Cyber-physical Systems. *IEEE Transactions on Dependable and Secure Computing* (2023).

[246] Chunqiu Steven Xia and Lingming Zhang. 2022. Less training, more repairing please: revisiting automated program repair via zero-shot learning. In *Proceedings of the 30th ACM Joint European Software Engineering Conference and Symposium on the Foundations of Software Engineering*. 959–971.

[247] Yuanjie Xia, Zishuo Ding, and Weiyi Shang. 2023. CoMSA: A Modeling-Driven Sampling Approach for Configuration Performance Testing. In *2023 38th IEEE/ACM International Conference on Automated Software Engineering (ASE)*. IEEE, 1352–1363.

[248] Zhe Xie, Changhua Pei, Wanxue Li, Huai Jiang, Liangfei Su, Jianhui Li, Gaogang Xie, and Dan Pei. 2023. From Point-wise to Group-wise: A Fast and Accurate Microservice Trace Anomaly Detection Approach. In *Proceedings of the 31st ACM Joint European Software Engineering Conference and Symposium on the Foundations of Software Engineering*. 1739–1749.

[249] Qinghua Xu, Shaukat Ali, and Tao Yue. 2021. Digital twin-based anomaly detection in cyber-physical systems. In *2021 14th IEEE Conference on Software Testing, Verification and Validation (ICST)*. IEEE, 205–216.

[250] Qinghua Xu, Shaukat Ali, and Tao Yue. 2023. Digital Twin-based Anomaly Detection with Curriculum Learning in Cyber-physical Systems. *ACM Transactions on Software Engineering and Methodology* (2023).

[251] Qinghua Xu, Shaukat Ali, Tao Yue, Zaimovic Nedim, and Inderjeet Singh. 2023. KDDT: Knowledge Distillation-Empowered Digital Twin for Anomaly Detection. In *Proceedings of the 31st ACM Joint European Software Engineering Conference and Symposium on the Foundations of Software Engineering*. 1867–1878.

[252] Tianyin Xu and Yuanyuan Zhou. 2015. Systems approaches to tackling configuration errors: A survey. *ACM Computing Surveys (CSUR)* 47, 4 (2015), 1–41.

[253] Wei Xu, Ling Huang, Armando Fox, David Patterson, and Michael I Jordan. 2009. Detecting large-scale system problems by mining console logs. In *Proceedings of the ACM SIGOPS 22nd symposium on Operating systems principles*. 117–132.

[254] Yahoo. 2024. Yahoo! Webscope Dataset. https://webscope.sandbox.yahoo.com/.

[255] Fabian Yamaguchi, Nico Golde, Daniel Arp, and Konrad Rieck. 2014. Modeling and discovering vulnerabilities with code property graphs. In *2014 IEEE Symposium on Security and Privacy*. IEEE, 590–604.

[256] Meng Yan, Xin Xia, Yuanrui Fan, Ahmed E Hassan, David Lo, and Shanping Li. 2020. Just-in-time defect identification and localization: A two-phase framework. *IEEE Transactions on Software Engineering* 48, 1 (2020), 82–101.

[257] Lin Yang, Junjie Chen, Shutao Gao, Zhihao Gong, Hongyu Zhang, Yue Kang, and Huaan Li. 2023. Try with Simpler–An Evaluation of Improved Principal Component Analysis in Log-based Anomaly Detection. *ACM Transactions on Software Engineering and Methodology* (2023).

[258] Limin Yang, Wenbo Guo, Qingying Hao, Arridhana Ciptadi, Ali Ahmadzadeh, Xinyu Xing, and Gang Wang. 2021. {CADE}: Detecting and explaining concept drift samples for security applications. In *30th USENIX Security Symposium (USENIX Security 21)*. 2327–2344.

[259] Lanxin Yang, He Zhang, Haifeng Shen, Xin Huang, Xin Zhou, Guoping Rong, and Dong Shao. 2021. Quality assessment in systematic literature reviews: A software engineering perspective. *Information and Software Technology* 130 (2021), 106397.

[260] Xu Yang, Shaowei Wang, Yi Li, and Shaohua Wang. 2023. Does data sampling improve deep learning-based vulnerability detection? Yeas! and Nays!. In *2023 IEEE/ACM 45th International Conference on Software Engineering (ICSE)*. IEEE, 2287–2298.

[261] Suraj Yatish, Jirayus Jiarpakdee, Patanamon Thongtanunam, and Chakkrit Tantithamthavorn. 2019. Mining software defects: Should we consider affected releases?. In *2019 IEEE/ACM 41st International Conference on Software Engineering (ICSE)*. IEEE, 654–665.

[262] Zhitao Ying, Dylan Bourgeois, Jiaxuan You, Marinka Zitnik, and Jure Leskovec. 2019. Gnnexplainer: Generating explanations for graph neural networks. *Advances in neural information processing systems* 32 (2019).

[263] Bin Yuan, Yifan Lu, Yilin Fang, Yueming Wu, Deqing Zou, Zhen Li, Zhi Li, and Hai Jin. 2023. Enhancing Deep Learning-based Vulnerability Detection by Building Behavior Graph Model. In *2023 IEEE/ACM 45th International Conference on Software Engineering (ICSE)*. IEEE, 2262–2274.

[264] Lun-Pin Yuan, Peng Liu, and Sencun Zhu. 2021. Recompose event sequences vs. predict next events: A novel anomaly detection approach for discrete event logs. In *Proceedings of the 2021 ACM Asia Conference on Computer and Communications Security*. 336–348.

[265] Kev Zettler. 2022. The DevSecOp tools that secure DevOps workflows. https://www.redhat.com/en/topics/devops/what-is-devsecops.

[266] Chunyong Zhang, Bin Liu, Yang Xin, and Liangwei Yao. 2023. CPVD: Cross Project Vulnerability Detection Based On Graph Attention Network And Domain Adaptation. *IEEE Transactions on Software Engineering* (2023).

[267] Chenyuan Zhang, Yanlin Wang, Zhao Wei, Yong Xu, Juhong Wang, Hui Li, and Rongrong Ji. 2023. EALink: An Efficient and Accurate Pre-trained Framework for Issue-Commit Link Recovery. In *2023 38th IEEE/ACM International Conference on Automated Software Engineering (ASE)*. IEEE, 217–229.

[268] Junwei Zhang, Zhongxin Liu, Xing Hu, Xin Xia, and Shanping Li. 2023. Vulnerability Detection by Learning from Syntax-Based Execution Paths of Code. *IEEE Transactions on Software Engineering* (2023).

[269] Quanjun Zhang, Chunrong Fang, Bowen Yu, Weisong Sun, Tongke Zhang, and Zhenyu Chen. 2023. Pre-trained model-based automated software vulnerability repair: How far are we? *IEEE Transactions on Dependable and Secure Computing* (2023).

[270] Tianzhu Zhang, Han Qiu, Gabriele Castellano, Myriana Rifai, Chung Shue Chen, and Fabio Pianese. 2023. System log parsing: A survey. *IEEE Transactions on Knowledge and Data Engineering* 35, 8 (2023), 8596–8614.

[271] Xu Zhang, Yong Xu, Qingwei Lin, Bo Qiao, Hongyu Zhang, Yingnong Dang, Chunyu Xie, Xinsheng Yang, Qian Cheng, Ze Li, et al. 2019. Robust log-based anomaly detection on unstable log data. In *Proceedings of the 2019 27th ACM Joint Meeting on European Software Engineering Conference and Symposium on the Foundations of Software Engineering*. 807–817.

[272] Yunhua Zhao, Kostadin Damevski, and Hui Chen. 2023. A systematic survey of just-in-time software defect prediction. *Comput. Surveys* 55, 10 (2023), 1–35.

[273] Yunhui Zheng, Saurabh Pujar, Burn Lewis, Luca Buratti, Edward Epstein, Bo Yang, Jim Laredo, Alessandro Morari, and Zhong Su. 2021. D2a: A dataset built for ai-based vulnerability detection methods using differential analysis. In *2021 IEEE/ACM 43rd International Conference on Software Engineering: Software Engineering in Practice (ICSE-SEIP)*. IEEE, 111–120.

[274] Junwei Zhou, Yijia Qian, Qingtian Zou, Peng Liu, and Jianwen Xiang. 2022. DeepSyslog: Deep Anomaly Detection on Syslog Using Sentence Embedding and Metadata. *IEEE Transactions on Information Forensics and Security* 17 (2022), 3051–3061.

[275] Yaqin Zhou, Shangqing Liu, Jingkai Siow, Xiaoning Du, and Yang Liu. 2019. Devign: Effective vulnerability identification by learning comprehensive program semantics via graph neural networks. *Advances in neural information processing systems* 32 (2019).

[276] Qihao Zhu, Zeyu Sun, Yuan-an Xiao, Wenjie Zhang, Kang Yuan, Yingfei Xiong, and Lu Zhang. 2021. A syntax-guided edit decoder for neural program repair. In *Proceedings of the 29th ACM Joint Meeting on European Software Engineering Conference and Symposium on the Foundations of Software Engineering*. 341–353.

[277] Qihao Zhu, Zeyu Sun, Wenjie Zhang, Yingfei Xiong, and Lu Zhang. 2023. Tare: Type-aware neural program repair. In *2023 IEEE/ACM 45th International Conference on Software Engineering (ICSE)*. IEEE, 1443–1455.

[278] Armin Zirak and Hadi Hemmati. 2022. Improving automated program repair with domain adaptation. *ACM Transactions on Software Engineering and Methodology* (2022).

[279] Deqing Zou, Yutao Hu, Wenke Li, Yueming Wu, Haojun Zhao, and Hai Jin. 2022. mVulPreter: A Multi-Granularity Vulnerability Detection System With Interpretations. *IEEE Transactions on Dependable and Secure Computing* (2022).

[280] Deqing Zou, Sujuan Wang, Shouhuai Xu, Zhen Li, and Hai Jin. 2019. $\mu$VulDeePecker: A Deep Learning-Based System for Multiclass Vulnerability Detection. *IEEE Transactions on Dependable and Secure Computing* 18, 5 (2019), 2224–2236.